\documentclass[reprint,prb,aps,twocolumn,superscriptaddress,10pt]{revtex4-2}
\usepackage{amsmath}
\usepackage[english]{babel}
\usepackage{graphicx}
\usepackage{color}
\usepackage{sidecap}
\usepackage{siunitx}
\usepackage[dvipsnames]{xcolor}

\usepackage[colorlinks,
            citecolor = blue, 
						linkcolor = blue,
						urlcolor  = blue,
						anchorcolor = blue,
						breaklinks = true
						]{hyperref}

\usepackage{cleveref}

\usepackage{soul}
\usepackage{ amssymb }
\newif\ifdraft
\drafttrue

\def \NIMSKW{Research Center for Electronic and Optical Materials, National Institute for Materials Science, 1-1 Namiki, Tsukuba 305-0044, Japan}
\def \TUM{Technical University of Munich, TUM School of Natural Sciences, Physics Department, 85748 Garching, Germany}
\def \MCQST{Munich Center for Quantum Science and Technology (MCQST), Schellingstr. 4, 80799 M{\"u}nchen, Germany}
\def \NIMSTT{Research Center for Materials Nanoarchitectonics, National Institute for Materials Science,  1-1 Namiki, Tsukuba 305-0044, Japan}

\def \Basel{Department of Physics, University of Basel, Klingelbergstrasse 82, Basel CH-4056, Switzerland}
\def \Harvard{Department of Physics, Harvard University, Cambridge, Massachusetts 02138, USA\\$^\ddag$These authors contributed equally to this work}

\begin{document}


\title{Out-of-equilibrium spin-valley dynamics of ferromagnets in topological Chern bands}

\author{J. James$^\ddag$}
\affiliation{\Basel}

\author{I. Krastilevskiy$^\ddag$}
\affiliation{\Basel}

\author{F. Pichler$^\ddag$}
\affiliation{\TUM}
\affiliation{\MCQST}

\author{L. Wang}
\affiliation{\Basel}

\author{A. Iafarova}
\affiliation{\Basel}

\author{F. Menzel}
\affiliation{\Basel}

\author{K.~Watanabe}
\affiliation{\NIMSKW}

\author{T.~Taniguchi}
\affiliation{\NIMSTT}

\author{C. Kuhlenkamp}
\affiliation{\Harvard}

\author{M. Knap}
\affiliation{\TUM}
\affiliation{\MCQST}

\author{T. Smole\'nski}
\email{tomasz.smolenski@unibas.ch}
\affiliation{\Basel}

\maketitle

{\bf Understanding quantum matter far from equilibrium is a central goal of modern physics. Twisted MoTe$_2$ bilayers~\cite{anderson2023,cai2023,zeng2023,xu2023observation,Park2023,li2025,ji2024,redekop2024,Jia2026,Pan2026,xu2025SC} constitute a promising platform for exploring this frontier~\cite{pichler2026} by combining strong Coulomb interactions, nontrivial band geometry, and optical control~\cite{holtzmann2026optical,huber2026optical,cai2026optical}. Here, we exploit this setting to investigate the role of topology and many-body correlations in the out-of-equilibrium dynamics of ferromagnets in Chern bands. Using a focused circularly polarized light pulse, we create a local magnetic domain oriented opposite to an external magnetic field and directly image its subsequent spin-valley relaxation in spatially and time-resolved low-temperature experiments. We demonstrate that in the vicinity of both integer and fractional Chern insulating states, the dynamics is governed by qualitatively different mechanisms than in ferromagnetic metals. Whereas metallic domains collapse by shrinking, Chern domains melt via thermal activation, resulting in drastically different temporal spin evolution and orders-of-magnitude longer relaxation times. These findings demonstrate the influence of topology and strong correlations on far-from-equilibrium collective spin phases, opening new opportunities for dynamical control of ferromagnets in the quantum Hall regime.}

Many-body correlations, topology, and out-of-equilibrium dynamics are three cornerstones of condensed matter physics. Over the past decades, tremendous effort has been devoted to exploring the intersections of these frontiers. Strong electromagnetic driving has emerged as a powerful tool for non-equilibrium control of correlated quantum materials~\cite{fausti2011lightinduced,wang2013observation,stojchevska2014ultrafast,mciver2020Lightinduced,kobayashi2023floquet,mitra2024light,delatorre2021nonthermal,bao2022Lightinduced}. The combination of topology and interactions, in turn, gives rise to topologically ordered fractional quantum Hall states~\cite{tsui1982,laughlin1983} and their lattice analogues, zero-field fractional Chern insulators (FCIs)~\cite{Hafezi_fractional_2007,kapit_exact_2010,sheng2011,neupert2011,tang2011,regnault_fractional_2011}, which have recently been realized in charge-tunable moir\'e materials~\cite{cai2023,zeng2023,xu2023observation,lu2024}. Despite these advances, understanding the nature of far-from-equilibrium processes in a setting where all three of the above concepts are brought together remains an outstanding challenge.

Twisted MoTe$_2$ ($t$-MoTe$_2$) bilayers~\cite{anderson2023,cai2023,zeng2023,xu2023observation,Park2023,ji2024,redekop2024,li2025,xu2025SC,Jia2026,Pan2026} offer a promising route to address this problem by combining ultrafast optical access~\cite{wang2025hidden} with strong Coulomb interactions and nontrivial band topology~\cite{Wu_TopologicalInsulators_2019,Devakul2021,Reddy2023,Wang2024,Yu2024}. In particular, these bilayers uniquely host gate-tunable ferromagnetic metals intertwined with fractional Chern insulators (FCIs) and integer Chern insulators (ICIs)~\cite{cai2023,zeng2023,xu2023observation,Park2023}, whose spin-valley state can be controlled with circularly polarized light~\cite{holtzmann2026optical,huber2026optical,cai2026optical}.

Here, we demonstrate that this optical control can be exploited to drive ferromagnets in $t$-MoTe$_2$ out of their thermal equilibrium, and investigate the thus-far-unexplored dynamics of the resulting spin-valley relaxation. Inspired by the concept of false-vacuum decay~\cite{Coleman_FalseVacuum_1977,Callan_FalseVacuum_1977}, we use a short circularly polarized light pulse to prepare a metastable local magnetic domain embedded in an equilibrium ferromagnetic background with opposite magnetization. We then directly visualize the evolution of this domain using spatially and time-resolved magneto-optical microscopy. We show that the decay of these false-vacuum domains proceeds through qualitatively different pathways for ferromagnetic metals and nearby Chern insulators, leading to orders-of-magnitude longer spin lifetimes in the latter case. These findings demonstrate how the emergence of correlated topological insulators fundamentally modifies the dynamics of electrons in Chern bands, highlighting the role of topology and many-body correlations in shaping the far-from-equilibrium properties of quantum matter.

\begin{figure*}[]
    \includegraphics[width=\textwidth]{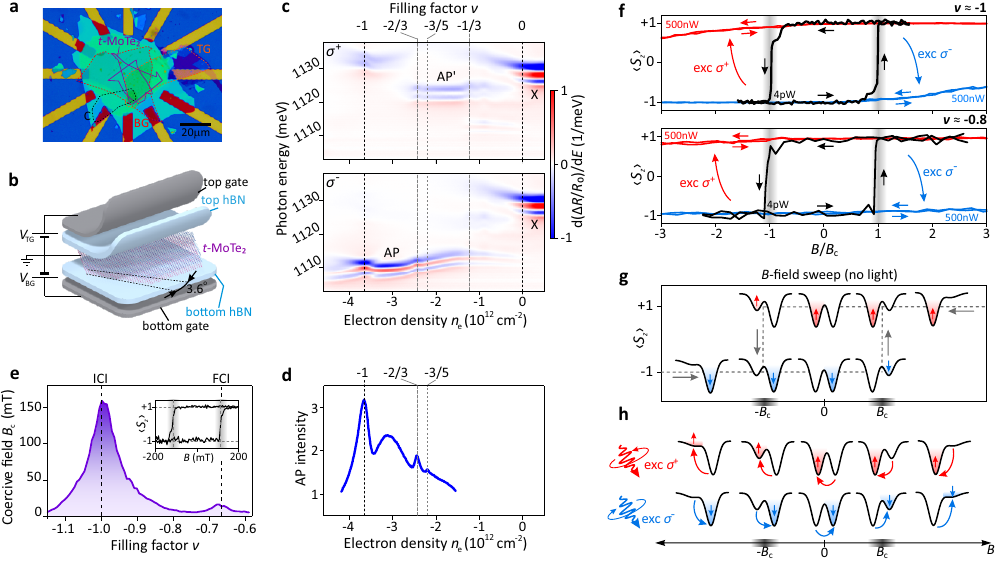}
	\caption{{\bf Optical control of collective magnetic states in $t$-MoTe$_2$.} ({\bf a}) Optical micrograph and ({\bf b}) schematic of the dual-gated $t$-MoTe$_2$ device A used in the main text. ({\bf c}) Filling-factor evolution of the reflectance-contrast spectra, $\Delta R/R_0$, differentiated with respect to photon energy $E$. The spectra were acquired in the two circular polarizations at a magnetic field of $B=0.2$~T. ({\bf d}) Intensity of the AP transition in $\sigma^-$ polarization, extracted by integrating $d(\Delta R/R_0)/dE$ over a few-meV-wide spectral window around the AP resonance. Local maxima arise from the formation of ICI and FCI phases. ({\bf e}) Filling-factor-dependent coercive field $B_\mathrm{c}$, extracted from low-power measurements of the magnetic hysteresis loop (see Methods Sec.~\ref{sec:coercive_field_determination} for details). Inset shows an example hysteresis loop measured at $\nu=-1.04$. ({\bf f}) Evolution of the hole spin polarization for the ICI (top) and metallic (bottom) phases while sweeping the magnetic field cyclically under resonant excitation of the AP with low (4~pW; black) and high (500~nW; red, $\sigma^+$, and blue, $\sigma^-$) excitation power. In the latter case, $\langle S_z\rangle$ becomes entirely determined by the helicity of the excitation light, demonstrating that the spin system can be oriented against the magnetic field (see Methods Sec.~\ref{sec:optical_orientation} for the details on $\langle S_z\rangle$ determination). ({\bf g}, {\bf h})~Cartoons illustrating the hole-spin configurations at various magnetic fields in the absence ({\bf g}) and presence ({\bf h}) of resonant AP excitation. In the former case, the system exhibits magnetic hysteresis, which disappears under $\sigma^\pm$ -polarized optical drive that sets $\langle S_z\rangle$ to $\sim\pm1$.\label{fig:Fig1}}
\end{figure*}

{\vspace{1.5mm}
\noindent{\bf \textsf{Optical control of collective magnetic states}}
\vspace{0.5mm}}

\noindent Our experiments are carried out at the base temperature $T\approx1.6$~K in a dry magneto-optical cryostat integrated with a confocal microscope setup (see Methods Sec.~\ref{sec:experimental_setup} for details). We obtain consistent results on multiple regions in two different $t$-MoTe$_2$ devices with twist angles of $\sim3.6^\circ$ (A) and $\sim4.1^\circ$ (B). In both devices, the thicknesses of hexagonal boron-nitride (hBN) layers encapsulating $t$-MoTe$_2$ were selected to ensure a large contrast of optical resonances in the reflectance spectra (see Methods Sec.~\ref{sec:device_fabrication} for details of device fabrication). In the main text, we focus on device A (Figs.~\ref{fig:Fig1}{\bf a,b}), which uniquely features a large $\sim10\mu\mathrm{m}^2$ homogeneous region where the twist angle varies by less than $\pm3$\% (see Methods Sec.~\ref{sec:homogeneity}), enabling us to perform spatially resolved experiments (see Methods Sec.~\ref{sec:2nd_device_data} for the results from device B). The moir\'e filling factor $\nu$, determined by the doping density $n_\mathrm{e}$, and the displacement field $D$ are independently controlled by two voltages applied to the top and bottom graphene gates (see Methods Sec.~\ref{sec:gates_and_filling}). Unless otherwise stated, we focus on the symmetric regime, $D\approx0$, where hybridization between the valence bands of the two layers---and thus the stability of the ferromagnetic phases---is strongest (see Extended Data Fig.~\ref{fig:muE} for the $(\nu,D)$ dependence). 

\begin{figure*}[]
    \includegraphics[width=\textwidth]{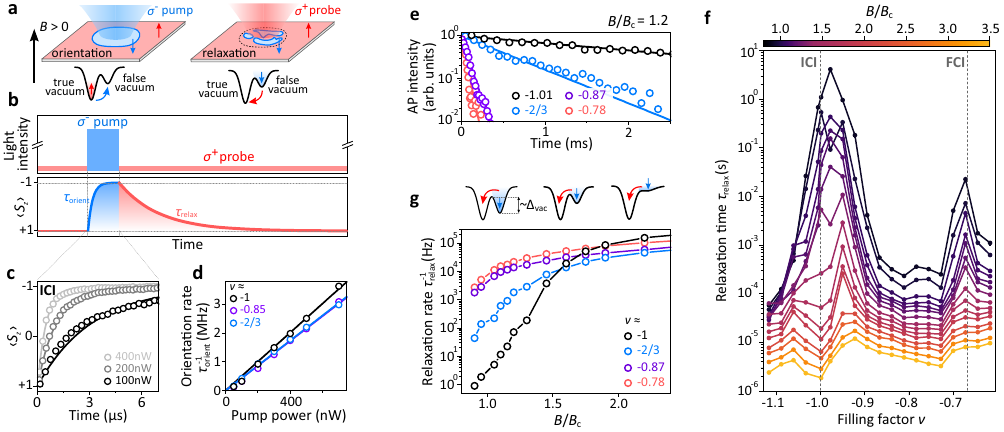}
    \caption{{\bf Out-of-equilibrium spin-valley dynamics of ferromagnets in $t$-MoTe$_2$}. ({\bf a,b}) Schematics of the pulse sequence ({\bf a}) and temporal light-intensity profiles ({\bf b}) used in time-resolved pump-probe spin relaxation experiments. A local magnetic domain is optically created using a $\sigma^-$-polarized pump beam, which populates the false-vacuum state oriented against the equilibrium magnetization. Its subsequent relaxation towards the true-vacuum ground state is monitored using a weaker $\sigma^+$-polarized probe pulse. The temporal evolution of the spin polarization during either pump or probe pulses is retrieved from the reflected light intensity measured with a fast single-photon detector (see Methods Sec.~\ref{sec:time_res_experiments} for details). ({\bf c}) Dynamics of optical spin orientation around $\nu=-1$ measured during the pump pulse for different pump powers, as indicated. Solid lines show exponential fits of the form $\langle S_z\rangle(t)=2\exp(-t/\tau_\mathrm{orient})-1$. ({\bf d}) Optical spin-orientation rates as a function of pump power for various filling factors at $B=0$. Solid lines represent linear fits. 
    ({\bf e}) Normalized AP intensity of the probe pulse as a function of time, reflecting the domain spin relaxation dynamics, measured at $B/B_\mathrm{c}=1.2$ for four different filling factors, as indicated. Solid lines show exponential fits of the form $\propto\exp(-t/\tau_\mathrm{relax})$. ({\bf f}) Spin-relaxation times $\tau_\mathrm{relax}$ extracted for various values of $B/B_\mathrm{c}$ as a function of $\nu$. ({\bf g}) Evolution of the relaxation rate with $B/B_\mathrm{c}$ for selected filling factors: near the coercive threshold, relaxation in Chern insulators is substantially slower, whereas at higher $B/B_\mathrm{c}$, ferromagnets at all filling factors exhibit similar and orders-of-magnitude faster relaxation dynamics.}
	\label{fig:Fig_relax_dynamics}
\end{figure*}

Fig.~\ref{fig:Fig1}{\bf c} shows the evolution of the differentiated white-light reflectance-contrast spectrum with $\nu$ (see Methods Sec.~\ref{sec:rc_analysis}). Upon hole doping, the exciton (X) transition disappears, and its spectral weight is transferred to a set of red-shifted exciton-polaron resonances~\cite{sidler2017fermi,huber2026optical}. For sufficiently large $|\nu|$, the hole-doped system becomes ferromagnetic, which manifests in a complete circular polarization of the lowest-energy attractive polaron (AP) resonance whose spectral weight in $\sigma^+$ ($\sigma^-$) polarization is directly proportional to the density of spin-down (spin-up) holes. Its polarization degree thus directly corresponds to the spin-valley polarization degree of the hole system, $\langle S_z\rangle=(n_\uparrow-n_\downarrow)/(n_\uparrow+n_\downarrow)$~\cite{anderson2023}, which exhibits a prominent magnetic hysteresis in a broad range of $-1.2<\nu<-0.5$ covering both metals and Chern insulators (see Fig.~\ref{fig:Fig1}{\bf e}). Consistent with previous studies~\cite{cai2023,zeng2023,xu2023observation,Park2023,li2025}, we find that the corresponding coercive field $B_\mathrm{c}$ is sizably enhanced around $\nu=-1$ and $\nu=-2/3$, where ICI and FCI phases emerge (see Methods Sec.~\ref{sec:coercive_field_determination} for details of $B_\mathrm{c}$ determination). The formation of these insulators---along with a more fragile FCI at $\nu=-3/5$---concurrently gives rise to sharp changes in the energy and intensity of the AP resonance (Fig.~\ref{fig:Fig1}{\bf d}). These features disperse with the magnetic field according to the St\v{r}eda formula~\cite{cai2023,zeng2023,xu2023observation,Park2023,li2025} (see Extended Data Fig.~\ref{fig:streda}), confirming the topological origin of the underlying insulating states.

Along with enabling optical spin readout at low excitation powers $\lesssim1$~nW, the AP transition can also be used to manipulate $\langle S_z\rangle$. As recently demonstrated in Refs.~\cite{holtzmann2026optical,huber2026optical,cai2026optical}, resonant AP excitation with $\sigma^\mp$-polarized light flips the hole spins from $\langle S_z\rangle=\pm1$ to $\mp1$ in the absence of a magnetic field. Here, we extend this approach to dynamically orient the spin system opposite to the direction of the $B$-field \textit{beyond the coercive threshold}. This is revealed in Fig.~\ref{fig:Fig1}{\bf f}, which shows the field evolution of $\langle S_z\rangle$ under resonant continuous-wave (CW) drive of the AP. By switching the light helicity to $\sigma^+$ ($\sigma^-$), we reorient the hole spins in both insulating and metallic phases against $B<0$ ($B>0$) over a wide range of $B$-fields, extending to several times $B_\mathrm{c}$ (see also schematic illustrations in Figs.~\ref{fig:Fig1}{\bf g,h}). The optically induced change in hole-spin polarization increases with excitation power (see Methods Sec.~\ref{sec:optical_orientation}). As a result, at powers of a few hundred nW, $\langle S_z\rangle$ no longer exhibits magnetic hysteresis and becomes reversed relative to its field-induced value.

{\vspace{1.5mm}
\noindent{\bf \textsf{Out-of-equilibrium spin-valley dynamics}}
\vspace{0.5mm}}

\noindent The above approach uniquely enables the non-thermal creation of a local, out-of-equilibrium ferromagnetic domain with a diameter of $\sim\mu$m determined by the optical spot size (Fig.~\ref{fig:Fig_relax_dynamics}{\bf a}). The holes within this domain are polarized against the $B$-field and therefore reside in a high-energy false-vacuum state. Combined with the gate-controlled tuning between metallic and Chern-insulating phases, this provides a unique setting for exploring the role of topology and correlations in the out-of-equilibrium relaxation towards the true-vacuum state. To examine this dynamics, we excite the sample with a train of synchronized, cross-circularly polarized light pulses (Fig.~\ref{fig:Fig_relax_dynamics}{\bf b}; see Methods Sec.~\ref{sec:time_res_experiments} for details). One cycle consists of a pump and a probe pulse, both resonant with the AP transition.

\begin{figure*}[]
    \includegraphics[width=\textwidth]{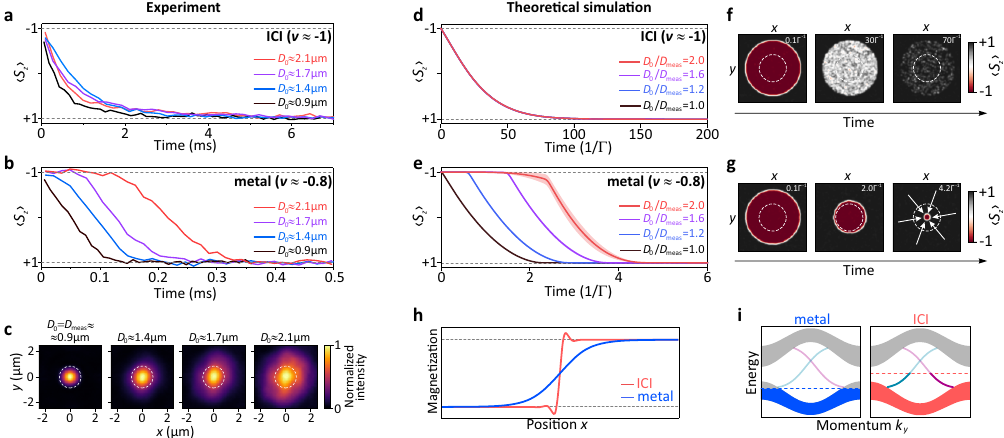}
    \caption{{\bf Spin-valley relaxation mechanisms of Chern insulators and ferromagnetic metals}. ({\bf a,b}) Temporal spin relaxation profiles of ICI ($\nu\approx-1$; {\bf a}) and metal ($\nu\approx-0.8$; {\bf b}) acquired at a fixed $B/B_\mathrm{c}=1.3$ for magnetic domains generated with pump beams of different diameters $D_0$, as indicated. In each case, the pump power was adjusted such that the average light intensity falling onto a central diffraction-limited probed area (marked with a dashed circle in panel {\bf c}) is similar, ensuring complete domain preparation during the pump pulse. ({\bf c}) Color-scale plots showing spatial intensity profiles of pump beams used for time-resolved measurements in panels {\bf a,b}. ({\bf d,e}) Theoretical relaxation profiles of magnetic domains in ICI ({\bf d}) and metal ({\bf e}) calculated at $B$ close to the coercive threshold $B_\mathrm{c}$ for different initial domain diameters $D_\mathrm{0}$ ($\Gamma$ sets the time scale in our simulations; see Methods Sec.~\ref{sec:theory_model} for details). The presented spin polarization is averaged over a central probed area with a fixed diameter $D_\mathrm{meas}$. Shaded regions around the relaxation curves mark the uncertainty of $\langle S_z\rangle(t)$ obtained for multiple simulations of the same domain relaxation. ({\bf f,g}) Snapshots of spatial spin configurations at different times (as indicated) calculated for ICI ({\bf f}) and metal ({\bf g}) assuming $D_\mathrm{0}=2D_\mathrm{meas}$ (the maps were smoothed by a convolution with a Gaussian of width $\approx10$~nm; see Methods Sec.~\ref{sec:theory_domain_stability}). Dashed circles mark the probed area. While the ICI domain decays through formation of smaller, homogeneously distributed internal domains, the metallic domain shrinks, giving rise to an initial plateau in the $\langle S_z\rangle(t)$ time traces in panels {\bf b} and {\bf e}, whose size increases with the initial domain size. ({\bf h, i}) Theoretical domain wall profiles ({\bf h}) and associated band structures ({\bf i}) in the ICI and the metal. Due to a sizably larger spin stiffness, the domain wall in the metal is much wider than in the ICI.} 
	\label{fig:Fig_spatial}
\end{figure*}

First, a short, strong $\sigma^-$-polarized pump pulse generates the domain. Its temporal spin evolution $\langle S_z\rangle(t)$ during the pulse is retrieved based on the time-dependent intensity of the reflected pump light averaged over multiple pump cycles using a fast superconducting nanowire single-photon detector (SNSPD; see Methods Sec.~\ref{sec:experimental_setup}). As expected for optical spin pumping, $\langle S_z\rangle(t)$ follows an exponential increase (Fig.~\ref{fig:Fig_relax_dynamics}{\bf c}), with a rate proportional to the pump intensity. The resulting spin-orientation timescale is only weakly dependent on the filling factor at a fixed $B$-field (Fig.~\ref{fig:Fig_relax_dynamics}{\bf d}; see Methods Sec.~\ref{sec:optical_orientation}) and remains comparatively short, enabling complete spin switching within a few~$\mu$s at pump powers of a few hundred~nW.

The relaxation of such a light-induced false-vacuum domain is probed in an analogous way using a second, $\sigma^+$-polarized pulse. Its power of $\sim4$~pW is orders of magnitude lower, ensuring non-destructive spin readout~\cite{Smolenski2022,ciorciaro2023} (see Methods Sec.~\ref{sec:time_res_experiments} for details), while its duration is sufficiently long to allow for complete domain collapse before the arrival of the next pump pulse. Fig.~\ref{fig:Fig_relax_dynamics}{\bf e} presents example spin relaxation traces acquired at selected filling factors for a fixed ratio of $B/B_\mathrm{c}=1.2$. Strikingly, the relaxation time $\tau_\mathrm{relax}$ strongly depends on the underlying many-body ground state. This is further confirmed in Fig.~\ref{fig:Fig_relax_dynamics}{\bf f}, which shows filling-dependent $\tau_\mathrm{relax}$ extracted for various $B/B_\mathrm{c}$. At low $B/B_\mathrm{c}\sim1$, the relaxation is orders of magnitude slower in the vicinity of both ICI and FCI phases as compared to metallic states. This difference correlates with the height of the energy barrier $\Delta_\mathrm{vac}$ separating the false- and true-vacuum states (see Fig.~\ref{fig:Fig_relax_dynamics}{\bf g}), which is enhanced for Chern insulators by many-body interactions. Upon increasing $B/B_\mathrm{c}$, the local minimum associated with the false-vacuum state becomes shallower, reducing $\Delta_\mathrm{vac}$ and thereby accelerating spin relaxation by multiple orders of magnitude for all states. Importantly, $\Delta_\mathrm{vac}$ remains finite even for $B>B_\mathrm{c}$, since the experimentally determined $B_\mathrm{c}$ at finite temperature is smaller than the field threshold required to close the barrier between the false- and true-vacuum states (see Methods Secs.~\ref{sec:coercive_field_determination} and~\ref{sec:theory_coercive_field}). This eventually occurs for $B/B_\mathrm{c}\gtrsim2$--3, resulting in similar, nearly $\nu$-independent relaxation dynamics. In this regime, $\tau_\mathrm{relax}$ depends only weakly on $B$ (cf. Fig.~\ref{fig:Fig_relax_dynamics}{\bf g}) and remains on the order of a few tens of $\mu$s. This timescale relatively closely corresponds to intervalley spin relaxation times previously reported for itinerant carriers in TMD monolayers~\cite{dey2017gate,goryca2019_relaxation,Li2021_relaxation,glazov2019intervalley}, thus suggesting that these processes determine $\tau_\mathrm{relax}$ in our experiments at high fields. 

\begin{figure*}
    \includegraphics[width=\textwidth]{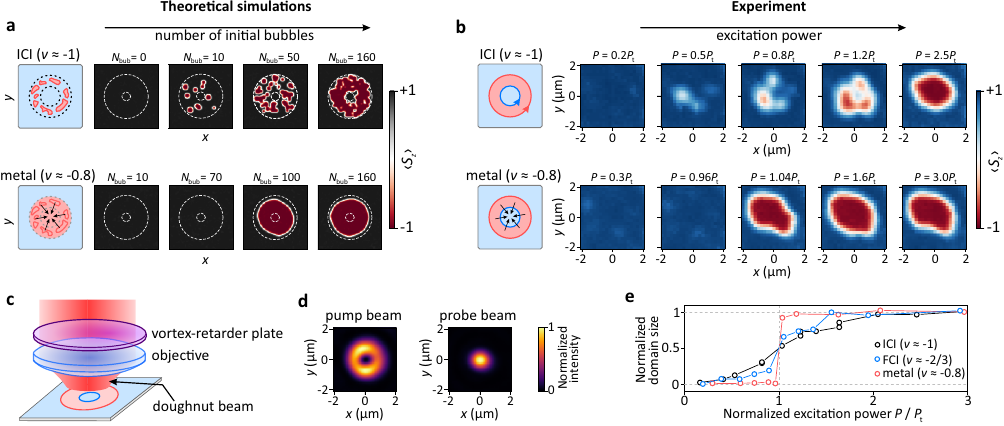}
    \caption{{\bf Stability of magnetic domains in Chern insulators and metals at $B=0$}. ({\bf a}) Theoretical simulation of steady-state spatial magnetization profiles in the ICI (top) and ferromagnetic metal (bottom), initialized with a large number $N_\mathrm{bub}$ of small seeds randomly distributed within a ring-shaped area (dashed lines). Similarly as in Fig.~\ref{fig:Fig_spatial}{\bf f,g}, the maps were smoothed by a convolution with a Gaussian of width $\approx10$~nm (see Methods Sec.~\ref{sec:theory_domain_stability}). ({\bf b}) Experimentally measured steady-state spin-polarization profiles of magnetic domains created by illuminating a spin-up-polarized ICI ($\nu\approx-1$; top) and ferromagnetic metal ($\nu\approx-0.8$; bottom) with a $\sigma^-$-polarized expanded doughnut-shaped pump beam at different powers. The domains were visualized by mapping the sample area with a diffraction-limited probe beam (see Methods Sec.~\ref{sec:domain_writing} for details). ({\bf c}) Schematic of the doughnut-shaped beam obtained using a vortex-retarder plate (see Methods Sec.~\ref{sec:experimental_setup} for details). ({\bf d}) Spatial intensity profiles of the pump and probe beams used for creating and visualizing the domains in {\bf b}. ({\bf e}) Normalized magnetic-domain area as a function of relative excitation power $P$, determined by integrating the spin polarization in maps similar to those shown in {\bf b} for the ferromagnetic metal and Chern insulators. Whereas the domain area in both ICI and FCI increases smoothly with $P$, metallic domains decay below a threshold power $P_\mathrm{t}$, but remain stable and merge into one large domain above this threshold, highlighting the enhanced mobility of domain walls in ferromagnetic metals.}
	\label{fig:Fig_domains}
\end{figure*}

{\vspace{1.5mm}
\noindent{\bf \textsf{Mechanisms of Spin-Valley Relaxation}}
\vspace{0.5mm}}

\noindent In addition to drastically different timescales, the relaxation of ferromagnetic domains in metals and nearby Chern insulators at $B$-fields slightly exceeding $B_\mathrm{c}$ is of qualitatively different nature. To demonstrate this, we perform spatially and time-resolved experiments in which the pump beam is expanded, allowing us to generate domains with a larger diameter $D_\mathrm{0}$. Their evolution is monitored using the same probe beam focused onto a co-centered, diffraction-limited spot with a fixed diameter $D_\mathrm{meas}\approx0.9\ \mu$m. Figs.~\ref{fig:Fig_spatial}{\bf a,b} show relaxation time traces measured in this configuration for the ICI and metallic states at $B/B_\mathrm{c}=1.3$, using pump beams with four different diameters, whose profiles are illustrated in Fig.~\ref{fig:Fig_spatial}{\bf c}. In each case, the pump power is chosen to ensure complete domain preparation within the pump-pulse duration. While the traces for the ICI remain nearly independent of domain size, the relaxation of metallic domains varies appreciably: for expanded beams, $\langle S_z\rangle(t)$ initially remains constant rather than decaying immediately. This gives rise to a prominent plateau, whose duration increases with pump-beam size and its power (see Extended Data Fig.~\ref{fig:method_expanded_vs_normal_spot} and Methods Sec.~\ref{sec:profile_dependence_on_area_and_power} for details as well as Methods Sec.~\ref{sec:relaxation_reproducibility} for reproducibility on a different device region), and is therefore proportional to~$D_\mathrm{0}$. These observations indicate that, unlike in Chern insulators, metallic domains do not decay through the nucleation of smaller internal domains. Instead, they collapse by shrinking from their perimeter, leaving the local magnetization in the central region unchanged until the domain edge approaches the probe-spot boundary.

This interpretation is supported by our theoretical analysis of the out-of-equilibrium magnetization dynamics, which we model by a stochastic differential equation describing a non-conserved order parameter~\cite{HohenbergHalperin77} (Methods Sec.~\ref{sec:theory_model}). We find that the key factor determining the magnetic-domain relaxation timescale is the structure of the domain wall, which is qualitatively distinct in metals and near Chern insulators due to their different spin stiffness~\cite{pichler2026}. As shown in Fig.~\ref{fig:Fig_spatial}{\bf h}, metallic domain walls are smooth and broad. By contrast, the occupation of topologically protected edge modes at domain walls near the ICI leads to very sharp domain walls, which exhibit a characteristic spatial profile with excess spin density around them (Fig.~\ref{fig:Fig_spatial}{\bf i}).

This qualitative difference in the domain-wall structure has crucial implications for the interplay between two major mechanisms governing spin relaxation: (1)~shrinking of the entire magnetic domain; and (2)~thermal fluctuations that locally flip the magnetization by overcoming the barrier $\Delta_\mathrm{vac}$ separating the false- and true-vacuum states. The former mechanism strongly depends on the velocity of the propagating domain edge, which increases with the domain-wall width (see Methods Sec.~\ref{sec:theory_dynamics} for details). Domain shrinking, therefore, remains efficient only in ferromagnetic metals with smooth domain walls, where it dominates over thermal-fluctuation-driven relaxation at low temperatures and $B$-fields close to the coercive threshold (where $\Delta_\mathrm{vac}$ is sufficiently large). This leads to a contraction of metallic domains (Fig.~\ref{fig:Fig_spatial}\textbf{g}), giving rise to a striking dependence of the $\langle S_z\rangle(t)$ decay profile on the initial domain size (Fig.~\ref{fig:Fig_spatial}\textbf{e}), with a characteristic initial plateau seen in the experiments.

By contrast, the sharp domain walls in the ICI strongly suppress domain shrinking. As a result, thermally driven local spin flips remain the only available relaxation channel even for $B\sim B_\mathrm{c}$, resulting in spatially homogeneous ICI-domain decay (Fig.~\ref{fig:Fig_spatial}\textbf{f}) with $\langle S_z\rangle(t)$ profiles independent of the initial domain size (Fig.~\ref{fig:Fig_spatial}\textbf{d}). 

At higher $B$-fields, where $\Delta_\mathrm{vac}$ decreases, thermally driven relaxation is accelerated and dominates over the domain-shrinking dynamics even for the ferromagnetic metal, as evidenced by the absence of the initial plateau in the $\langle S_z\rangle(t)$ decay profiles measured at $B\sim3.0B_\mathrm{c}$ with an expanded beam (see Extended Data Fig.~\ref{fig:method_power_dependance_expanded_spot} and Methods Sec.~\ref{sec:profile_dependence_on_B} for details). 

{\vspace{1.5mm}
\noindent{\bf \textsf{Visualizing magnetic domain formation}}
\vspace{0.5mm}}

\noindent A direct consequence of the above-discussed relaxation mechanisms is that small magnetic domains are inherently unstable in the metal, even in the absence of a magnetic field~\cite{bray1994theory}. To illustrate this effect, we theoretically simulate steady-state spatial profiles of magnetic domains initialized at $B=0$ by creating a large number $N_\mathrm{bub}$ of $\sim100$-nm seed bubbles distributed across a ring-shaped region with a $\sim\mu$m radius (Fig.~\ref{fig:Fig_domains}{\bf a}; see Methods Sec.~\ref{sec:theory_domain_stability} for details). Due to the sharp domain walls in the vicinity of the ICI, such bubbles remain stable, and the total domain area increases continuously with $N_\mathrm{bub}$, ultimately filling the ring-shaped region in the limit of large $N_\mathrm{bub}$. By contrast, in the ferromagnetic metal, the spin stiffness and hence domain-wall mobility are large. As a consequence, metallic bubbles collapse when their spatial density is low and aggregate into a fully magnetized domain above a critical initial bubble number.

This prediction is directly confirmed by our spatially resolved domain-writing measurements carried out using a doughnut-shaped pump beam, which is obtained by passing an expanded Gaussian beam through a vortex-retarder plate (Fig.~\ref{fig:Fig_domains}{\bf c}; see Methods Sec.~\ref{sec:experimental_setup} for details). First, the ferromagnet is prepared in a spin-up configuration with a positive magnetic field. After ramping the field back to zero, the sample is exposed for a few tens of seconds with a $\sigma^-$-polarized doughnut-shaped beam of a given power $P$. As a last step, the steady-state spatial spin profile $\langle S_z\rangle(\bf{r})$ of the resulting magnetic domain is mapped out using a weak, diffraction-limited probe beam (see Fig.~\ref{fig:Fig_domains}{\bf d} and Methods Sec.~\ref{sec:domain_writing}). As shown in Fig.~\ref{fig:Fig_domains}{\bf b}, for the ICI, the resulting $\langle S_z\rangle(\bf{r})$ grows gradually with $P$ and resembles the ring-shaped beam intensity profile. For the metal, no permanent domain is generated below a certain power level $P_\mathrm{t}$. Once $P$ exceeds this threshold, a large domain appears whose size no longer increases with $P$. This transition is very abrupt, in sharp contrast to both ICI and FCI, where the domain area increases smoothly with pump power (Fig.~\ref{fig:Fig_domains}{\bf e}). Given that optical spin orientation in $t$-MoTe$_2$ proceeds via local spin-flip events~\cite{holtzmann2026optical,huber2026optical,cai2026optical}, these findings confirm that small domains in ferromagnetic metals are unstable, in contrast to the case of Chern insulators. This is further corroborated by the shape of the resulting metallic domains: although the sample is illuminated with a doughnut-shaped beam, the domains emerge as filled circles rather than rings, consistent with the theoretical simulations (cf. Figs.~\ref{fig:Fig_domains}{\bf a,b}). This striking agreement between experiment and theory further highlights the key role of topology and electronic correlations in magnetic-domain relaxation.

{\vspace{1.5mm}
\noindent{\bf \textsf{Outlook}}
\vspace{0.5mm}}

\noindent The qualitatively different relaxation mechanisms of insulating and metallic ferromagnets in topological Chern bands that we uncover highlight the influence of band topology and many-body correlations on the out-of-equilibrium dynamics of collective spin phases. The fast optical control of metastable magnetic domains demonstrated here circumvents the necessity to change external magnetic fields, which can neither be done locally nor rapidly, thus offering dynamical access to phase diagrams of far-from-equilibrium ferromagnets in the quantum Hall regime. In a longer-term perspective, our approach may provide a route to study transport in the presence of dynamical edge modes or, when combined with scanning probe microscopy~\cite{park2026localspectroscopy,deng2026realspace}, enable ultrafast local access to collective excitations of topologically ordered phases~\cite{li2026}.

\vspace{-0.2cm}
\subsection*{Acknowledgments}\vspace{-0.1cm}
We thank Richard Warburton for letting us use their single-photon detectors. This work was supported by the European Commission through ERC grant OptoQuantTOP (Grant Number 101219354). The views and opinions expressed are those of the authors only and do not necessarily reflect those of the European Union. Neither the European Union nor the granting authorities can be held responsible for them. F.P. and M.K. acknowledge support from the Deutsche Forschungsgemeinschaft (DFG, German Research Foundation) under Germany's Excellence Strategy--EXC--2111--390814868, TRR 360 -- 492547816 and DFG grants No. KN1254/1-2, KN1254/2-1, the European Union (grant agreement No 101169765), as well as the Munich Quantum Valley, which is supported by the Bavarian state government with funds from the Hightech Agenda Bayern Plus. C.K. acknowledges support from NSF DMR-2220703. K.W. and T.T. acknowledge support from the JSPS KAKENHI (grant numbers 21H05233 and 23H02052), the CREST (JPMJCR24A5), JST and World Premier International Research Center Initiative (WPI), MEXT, Japan.

\vspace{-0.2cm}
\subsection*{Author contributions}\vspace{-0.1cm}
T.S. conceived the project. J.J., I.K., and T.S. performed magneto-optical and time-resolved experiments as well as analyzed the data, with assistance from L.W.. A.I. contributed to spatially resolved domain writing measurements and their analysis. J.J. fabricated device A, while F.M. prepared device B. K.W. and T.T. grew the hBN crystals. F.P. and C.K. developed the theoretical model and carried out the simulations under the guidance of M.K.. T.S., M.K., J.J, I.K., and F.P. wrote the manuscript. M.K. and T.S. supervised the project.

\section*{Methods}

\renewcommand{\figurename}{Extended Data Figure}
\renewcommand{\theHfigure}{M\arabic{figure}}
\renewcommand{\thesubsection}{\arabic{subsection}}
\setcounter{figure}{0}

\subsection{Experimental setup \label{sec:experimental_setup}}

\begin{figure*}[]
	\includegraphics[width = \textwidth]{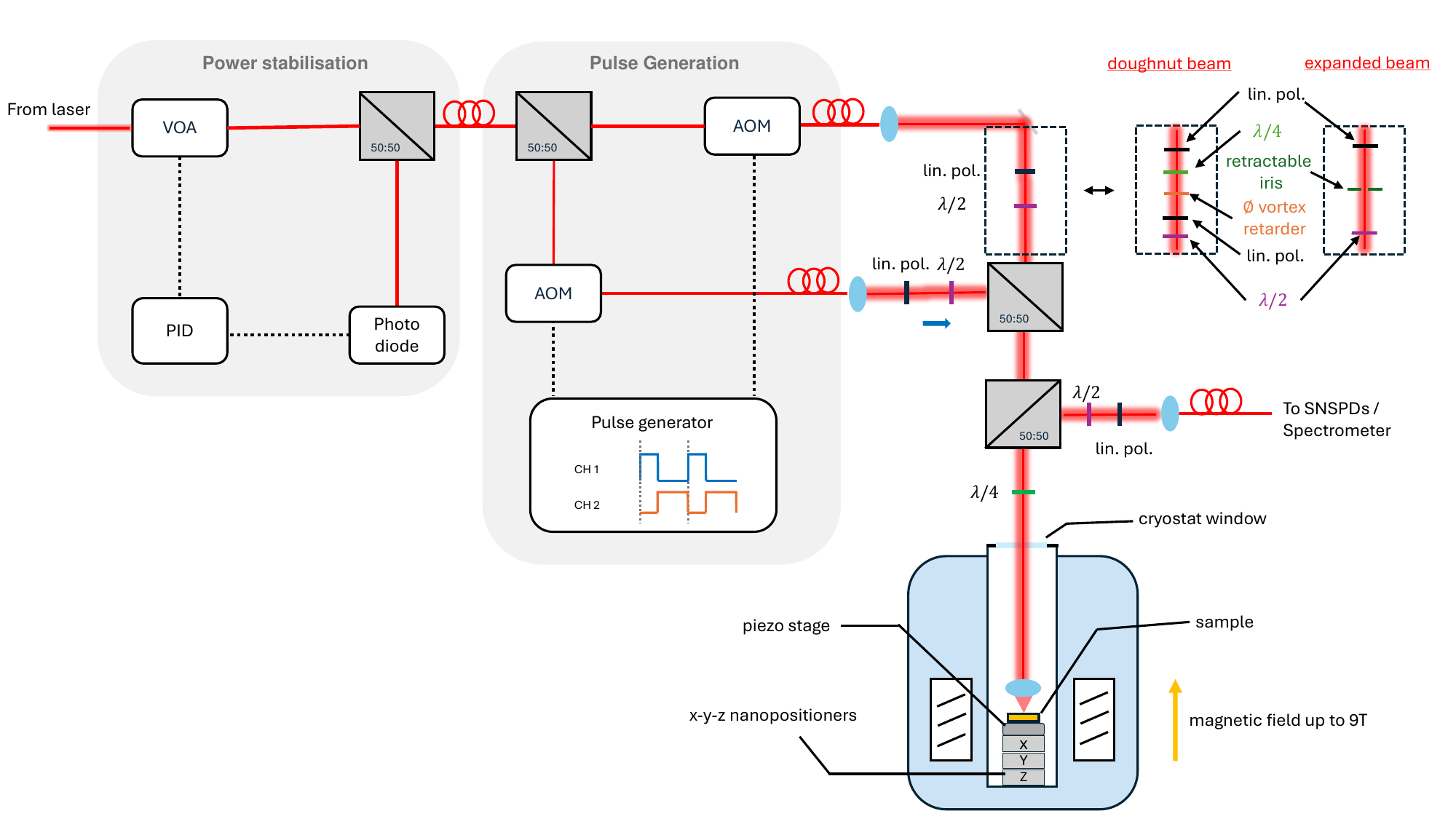}
	\caption{{\bf Experimental setup.} Schematic illustration of the measurement setup employed in this work. After passing through a home-built PID-based power-stabilization system and two fiber-based AOMs, the excitation light is delivered to two separate arms of the confocal microscope setup. Polarization optics, together with retractable irises and zero-order vortex retarders, enable independent control over the polarization and spatial profile of the two excitation beams. The light is focused onto a sample mounted on $x$-$y$-$z$ nanopositioners and a three-axis piezoelectric scanner inside a magneto-optical cryostat. Light reflected of the sample is directed to the collection path, passes through polarization optics, gets coupled into a single-mode fiber, and is finally detected using either an infrared SNSPD or an InGaAs CCD camera integrated with a 0.5-m spectrometer.}\label{fig:set_up}
\end{figure*}

Our experiments were carried out in a closed-cycle cryostat equipped with a variable-temperature insert (VTI) and a superconducting magnet, allowing the sample to be cooled down to 1.6~K and subjected to magnetic fields of up to 9~T perpendicular to its surface. The VTI was filled with He exchange gas to ensure efficient thermalization of the sample. It was further integrated with a confocal microscope setup, schematically illustrated in Extended Data Fig.~\ref{fig:set_up}.

The sample was excited using two different light sources. For white-light experiments, we employed a single-mode-fiber-coupled superluminescent light-emitting diode (SLED) generating continuous-wave (CW) light centered at $\sim1090$~nm with a bandwidth of $\sim100$~nm. The bandwidth was further reduced using a set of suitably rotated razor-edge short- and long-pass filters. Measurements involving resonant AP excitation were performed with a single-frequency CW laser, whose wavelength was tunable between 1080 and 1125~nm. The power of this laser was stabilized using a home-built PID feedback loop consisting of an InGaAs photodiode for monitoring the power and a variable optical attenuator (VOA) for adjusting it. The light, after passing through a wedged VTI window, was focused onto the sample surface to a diffraction-limited spot using an achromatic microscope objective with a numerical aperture of 0.7. The sample was placed on a set of $x$-$y$-$z$ nanopositioners and a three-axis voltage-controlled piezoelectric scanner. This allowed us to bring the sample to the focal plane of the objective, select a measurement spot, and reproducibly displace it with sub-50-nm precision, thus enabling spatially resolved mapping experiments from Fig.~\ref{fig:Fig_domains}. The light reflected off the sample, after exiting the VTI, was directed to the detection path with a beam splitter, coupled into a single-mode fiber, and sent to one of the two detectors: a 0.5-m spectrometer equipped with a liquid-nitrogen-cooled InGaAs camera for spectrally resolved measurements, or a fiber-coupled SNSPD operated at $\sim2$~K in a separate cryostat for sensitive time-resolved experiments.

The polarization of both the excitation and detected beams was controlled with a combination of linear polarizers as well as achromatic quarter- and half-waveplates (see Extended Data Fig.~\ref{fig:set_up}). In order to expand the excitation spot in the spatially resolved experiments from Fig.~\ref{fig:Fig_spatial}, the pump-beam diameter was controllably reduced using a retractable manual iris. Furthermore, to obtain a doughnut-shaped excitation spot used in domain writing experiments (Fig.~\ref{fig:Fig_domains}{\bf c}), such an expanded pump beam was passed through a zero-order vortex retarder plate placed in between two co-linear polarizers and a quarter-wave plate, as illustrated in Extended Data Fig.~\ref{fig:set_up}.

\subsection{Device fabrication \label{sec:device_fabrication}}

The devices were assembled using the flakes that were mechanically exfoliated from the bulk crystals (HQ Graphene 2H-MoTe$_2$, NIMS hBN, and natural graphite) onto SiO$_2$/Si substrates. Exfoliation of hBN and graphite was carried out under ambient conditions, whereas MoTe$_2$ monolayers were obtained inside an N$_2$-filled glovebox to preserve their high quality. All flakes were selected based on their thickness and uniformity, as assessed by optical microscopy. To ensure precise control of the twist angle, $t$-MoTe$_2$ bilayers were prepared from a single MoTe$_2$ monolayer, which was cut into two nearly equal pieces using a standard atomic-force-microscope (AFM) tip~\cite{cao2018correlated, cao2018unconventional}.

All flakes were then stacked inside the same glovebox using a standard dry-transfer method~\cite{Zomer2014}. At each step, a target flake was first aligned with submicrometer precision and subsequently picked up using a thin polycarbonate (PC) film mounted on a dome-shaped PDMS stamp. The stacking was performed at a temperature of around $100^\circ$C, which was lowered during pickup of the MoTe$_2$ layers to preserve their quality and reduce twist-angle disorder. The completed heterostructure consisted of an hBN-encapsulated $t$-MoTe$_2$ bilayer sandwiched between top and bottom few-layer-graphene (FLG) gates and electrically contacted by a third FLG flake. It was dropped onto a SiO$_2$/Si substrate with prepatterned gold electrodes that were aligned with the FLG flakes, enabling electrical contacting. For the release step, the PC film was melted at $180^\circ$C, after which its residues were removed by immersing the sample in chloroform at $50^\circ$C for approximately one hour.

In both devices, the thicknesses of both top and bottom hBN flakes were selected to be nearly identical, yielding around $34$~nm for device A and $23$~nm for device B. For the selected Si substrate with a 90-nm-thick SiO$_2$ layer, this choice ensured destructive interference between the light reflected off various interfaces of the heterostructure away from the $t$-MoTe$_2$ region. This resulted in nearly Lorentzian line shapes and high contrast of the exciton-polaron resonances in the $t$-MoTe$_2$ reflectance spectra.

\subsection{Twist angle homogeneity in device A \label{sec:homogeneity}}

\begin{figure*}[]
	\includegraphics[width = 70mm]{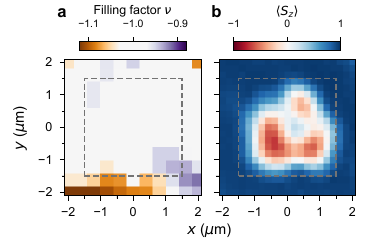}
	\caption{{\bf Spatial homogeneity of device A.} ({\bf a}) Spatial map of a local filling factor measured with top and bottom gate voltages set to the values nominally corresponding to $\nu=-1$ in the center of the probed area. The color scale is selected such that the $-1.03<\nu<-0.97$ range appears in white. ({\bf b}) Spatial profile of the zero-field magnetic domain created in initially spin-up-polarized ICI using $\sigma^-$-polarized doughnut-beam power of $P=1.2P_\mathrm{t}$ (data from Fig.~\ref{fig:Fig_domains}{\bf b}). The gray dashed line in both maps outlines a $3\times3~\mu$m$^2$ square. \label{fig:homogeneity}}
\end{figure*}

For the spatially resolved domain writing measurements shown in Fig.~\ref{fig:Fig_domains}, it was critical to ensure that both the twist angle and the corresponding filling factor are uniform across a sufficiently large region. Owing to its homogeneity, Device A uniquely fulfills this requirement. To quantify the twist angle disorder in the central region of this sample, we performed a spatially resolved experiment by measuring the reflectance contrast spectrum as a function of $n_\mathrm{e}$ at each spatial location $(x,y)$. This allowed us to extract the local filling factor calibration (using the procedure described in Methods Sec.~\ref{sec:gates_and_filling}) and thus retrieve the resulting spatial $\nu$ distribution at a fixed gate voltage setting, ensuring that the density in the central location $(0,0)$ corresponds to $\nu=-1$. As shown in Extended Data Fig.~\ref{fig:homogeneity}{\bf a}, the local filling factor deviates by no more than $\pm6\%$ from $-1$ within the $\sim3\times3~\mu$m$^2$ region in which the magnetic domains are imprinted, thus providing ideal conditions for the experimental comparison of optically written domains between insulating and metallic phases. The filling factor uncertainty can equivalently be expressed in terms of twist angle $\theta$ disorder: at fixed electron density $\nu\propto a_\mathrm{M}^2\propto 1/\theta^2$, where $a_\mathrm{M}$ is the moir\'e lattice period, giving $\delta\theta/\theta \approx 1/2 \cdot \delta\nu/|\nu|$. This translates a $\pm6\%$ uncertainty in $\nu$ into twist angle disorder within the region of interest of under $\pm3\%$.

\subsection{Gate-tunability and filling factor calibration \label{sec:gates_and_filling}}

As stated in the main text, the charge state of our devices is controlled by applying two voltages $V_\mathrm{TG}$ and $V_\mathrm{BG}$ to the top and bottom FLG gates, while keeping the $t$-MoTe$_2$ bilayer at ground potential. This enables independent control of the doping density and displacement field $D$ that, according to the parallel-plate capacitor model, are given by $n_\mathrm{e}=(\epsilon_0\epsilon_\mathrm{hBN}/t_\mathrm{t}e)\cdot(V_\mathrm{TG}+V_\mathrm{BG}t_\mathrm{t}/t_\mathrm{b}-V_0)$ and $D/\epsilon_0=(\epsilon_\mathrm{hBN}/2t_\mathrm{t})\cdot(V_\mathrm{TG}-V_\mathrm{BG}t_\mathrm{t}/t_\mathrm{b})-D_0/\epsilon_0$, where $t_\mathrm{t}$ ($t_\mathrm{b}$) are top (bottom) hBN thicknesses, while $\epsilon_\mathrm{hBN}\approx3.1$ is an out-of-plane hBN dielectric constant~\cite{xu2021creation, Popert2022, smolenski2019interaction}. The parameter $V_0$, yielding approximately $-1.0$~V for device~A, represents a finite voltage that needs to be applied in order to start injecting holes into our device. Similarly, $D_0/\epsilon_0$ ($\approx 6$~mV/nm for device A) is introduced to account for the residual asymmetry of our device, ensuring that $D$-induced phase transitions between ferromagnetic and paramagnetic phases occur symmetrically with respect to $D=0$. This is illustrated in Extended Data Fig.~\ref{fig:muE}, which shows the map of hole spin polarization degree obtained at $B=0.2$~T as a function of the above-defined $n_e$ and $D$, along with the corresponding circular-polarization-resolved $n_\mathrm{e}$ or $D$ evolutions of differentiated reflectance contrast spectra at $D=0$ and $\nu=-1$, respectively.

The moir\'e filling factor, $\nu=n_\mathrm{e}/|n_\mathrm{ICI}|$, was calibrated using the doping density $n_\mathrm{ICI}$ corresponding to the local maximum of the AP intensity observed upon formation of the ICI at $\nu=-1$. As shown in Fig.~\ref{fig:Fig1}{\bf d}, this calibration accurately reproduces the positions of the weaker AP maxima at $\nu=-2/3$ and $\nu=-3/5$, where the two most robust FCIs emerge. It also correctly captures the magnetic-field-induced density shifts of all these phases, governed by the St\v{r}eda formula $\Delta n_\mathrm{e}=(C/\phi_0)\Delta B$ (where $\phi_0$ is the magnetic flux quantum)~\cite{streda1982theory}, with the many-body Chern numbers $C$ determined by the corresponding filling factors (see Extended Data Fig.~\ref{fig:streda}).

\subsection{Analysis of reflectance spectra \label{sec:rc_analysis}}

\begin{figure*}[]
	\includegraphics[width = 120mm]{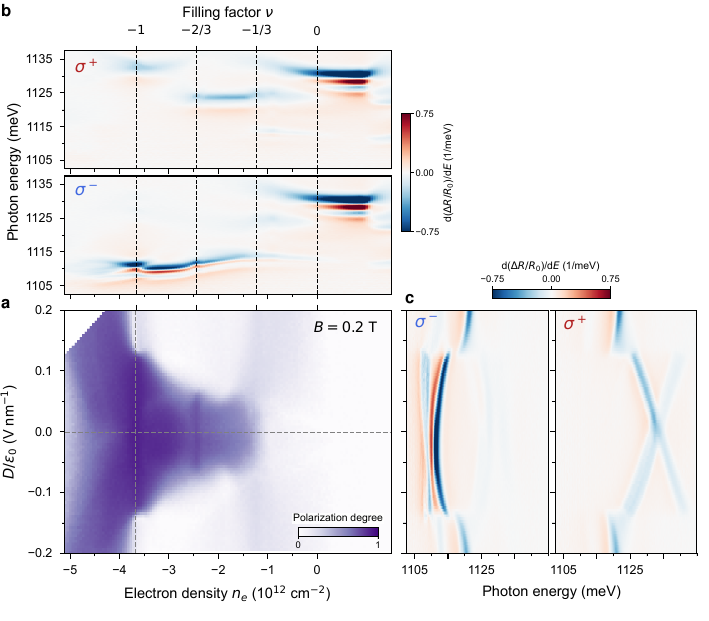}
	\caption{{\bf Magnetic phase diagram of device A.} ({\bf a}) Degree of circular polarization of the AP resonance, $\rho_\mathrm{AP}=[f_--f_+]/[f_-+f_+]$, measured at $B=0.2$~T as a function of doping density $n_\mathrm{e}$ and displacement field $D$. The AP spectral weights $f_\pm$ in $\sigma^\pm$ polarization are obtained by integrating the absolute value of the differentiated reflectance contrast, $d(\Delta R/R_0)/dE$, within a narrow spectral window around the AP resonance. The dark region marks the part of the $(n_\mathrm{e},D)$ phase space in which the hole system is ferromagnetic. In this region, the AP resonance is fully $\sigma^-$-polarized, but a finite noise floor in $f_+$  leads to internal structure in $\rho_\mathrm{AP} \approx 1 - 2f_+/f_- < 1$ that reflects variations in the $f_-$ spectral weight. Horizontal and vertical dashed lines mark the cuts in the $(n_\mathrm{e}, D)$ plane along which the spectral evolution is shown in panels {\bf b} and {\bf c}, respectively. ({\bf b,c})~Representative evolution of the circular-polarization-resolved differentiated reflectance-contrast spectra, measured under the same conditions as in {\bf a}, as a function of $n_\mathrm{e}$ at fixed $D\approx0$ ({\bf b}) and as a function of $D$ at fixed $\nu=-1$ ({\bf c}). \label{fig:muE}}
\end{figure*}

\begin{figure*}[]
	\includegraphics[width = \textwidth]{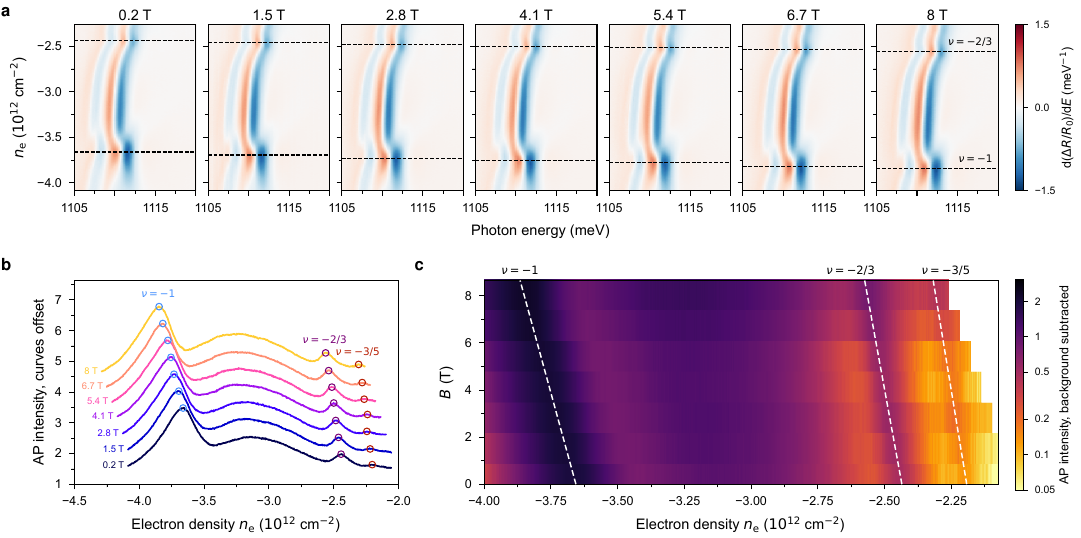}
	\caption{{\bf St\v{r}eda formula dispersion of Chern insulating states in device A.} ({\bf a}) Evolutions of differentiated reflectance contrast spectra with $n_\mathrm{e}$ acquired in the AP spectral range for device A in $\sigma^-$ polarization at $D=0$ and different magnetic fields (as indicated). Dashed lines mark the densities corresponding to the formation of Chern insulators at $\nu=-1$ and $\nu=-2/3$. ({\bf b})~Density-dependent AP intensities obtained by spectrally integrating the absolute value of data in panel {\bf a}. The curves for different $B$-fields are vertically offset for clarity. The local maxima (marked with circles) are due to the formation of the ICI at $\nu=-1$ and FCIs at $\nu=-2/3$ and $\nu=-3/5$. ({\bf c}) Color-scale map showing the background-corrected AP intensities from panel {\bf b} as a function of both $B$ and $n_\mathrm{e}$. Dashed lines represent the shifts in the density of all three insulating phases determined based on the St\v{r}eda formula with Chern numbers $C$ corresponding to the filling factor of each phase. Their agreement with the $B$-induced shift in the AP intensity maxima directly confirms the topological origin of all three insulating states.\label{fig:streda}}
\end{figure*}

The white-light reflectance spectrum $R$ measured in a given circular polarization was normalized to a co-polarized background spectrum, $R_0(E)$, acquired away from the $t$-MoTe$_2$ region of the sample. The resulting reflectance contrast, $R_\mathrm{c}\equiv\Delta R/R_0=(R-R_0)/R_0$, was then differentiated with respect to photon energy $E$. To reduce the noise and enhance the contrast of the differentiated signal, we evaluated the numerical derivative using a symmetric difference quotient over third-nearest-neighbor points, $R_\mathrm{c}'(E)=(R_{\mathrm{c},n+3}-R_{\mathrm{c},n-3})/(E_{n+3}-E_{n-3})$. The resulting $d(\Delta R/R_0)/dE$ signal is shown in Fig.~\ref{fig:Fig1}{\bf c} in the main text, as well as in Extended Data Figs.~\ref{fig:muE},~\ref{fig:streda}, and~\ref{fig:relaxation 2nd device}{\bf b}.

\subsection{Determination of the coercive field \label{sec:coercive_field_determination}}

To precisely determine the coercive field of the hole system at a given filling factor, we perform magnetic hysteresis measurements. To avoid perturbing the spins with the probe light, these measurements are carried out using a circularly polarized, low-power ($\sim4$~pW) single-frequency laser resonant with the AP transition. The time-integrated intensity $I_\mathrm{probe}(B)$ of the reflected laser light is detected with an SNSPD while the $B$-field is swept in a loop, first from negative to positive values and subsequently back to negative fields. Whenever a positive (negative) $B$ appreciably exceeds the coercive threshold, the hole spins are oriented upwards (downwards), resulting in a loss of AP resonance in $\sigma^+$ ($\sigma^-$) polarization. Hence, for $\sigma^+$-polarized probe laser, $I_\mathrm{probe}(B)$ changes from the background level $I_\mathrm{back}$ at $B\gg B_\mathrm{c}$ to that corresponding to maximal AP reflectance $I_\mathrm{max}$ at $B\ll -B_\mathrm{c}$, thus allowing us to directly determine the hole spin polarization degree as $\langle S_z\rangle(B)= 1-2[I_\mathrm{probe}(B)-I_\mathrm{back}]/[I_\mathrm{max}- I_\mathrm{back}]$ (and analogously for the $\sigma^-$ probe). As expected, the resulting $\langle S_z\rangle(B)$ exhibits a hysteresis loop, for example, in Fig.~\ref{fig:Fig1}{\bf e}, that is independent of the probe polarization. The coercive field $B_\mathrm{c}$ is extracted as the average, $(B_+ + |B_-|)/2$, of the positive $B_+$ and negative $B_-$ fields at which $\langle S_z\rangle$ crosses zero during the up- and down-sweeps of the magnetic field, respectively.

As is typical for ferromagnetic systems~\cite{bruno1990}, the measured hysteresis loop---and hence the extracted value of $B_\mathrm{c}$---depends on the magnetic-field sweep rate $\Gamma_\mathrm{sweep}$. In particular, because spin relaxation in the hole system is very slow for $|B|\sim B_\mathrm{c}$, the measured $B_\mathrm{c}$ decreases slightly as $\Gamma_\mathrm{sweep}$ is reduced. Likewise, $B_\mathrm{c}$ decreases with increasing temperature~\cite{anderson2023}. To minimize the former effect, our hysteresis experiments from Fig.~\ref{fig:Fig1}{\bf e} were carried out at low $\Gamma_\mathrm{sweep}\approx3$~mT/s. Nevertheless, owing to finite temperature, the value of $B_\mathrm{c}$ determined under such experimental conditions is still smaller than $B_\mathrm{c}^*$ at which the energy barrier separating the true- and false-vacuum states vanishes, as discussed further in Methods Sec.~\ref{sec:theory_coercive_field}.

\subsection{Efficiency and dynamics of optical spin orientation \label{sec:optical_orientation}}

In contrast to the spin relaxation dynamics, the characteristic timescale $\tau_\mathrm{orient}$ of light-induced spin orientation is only weakly dependent on the filling factor, always remaining inversely proportional to the excitation power $P$ (Fig.~\ref{fig:Fig_relax_dynamics}{\bf c}). In addition, the orientation rate $\tau_\mathrm{orient}^{-1}$ at a given $P$ exhibits a slight increase (decrease) with external $B$-field that aligns the spins along (against) the light-induced orientation direction, as shown in Extended Data Fig.~\ref{fig:ext_methods_orient_dynamics}. These changes are most appreciable in the vicinity of $\nu=-1$, yet even there $\tau_\mathrm{orient}$ varies by less than a factor of 2 when $|B|$ is increased to $\sim2B_\mathrm{c}$. 

In addition to modifying its rate, the external magnetic field also affects the efficiency of optical spin orientation, as shown in Fig.~\ref{fig:Fig1}{\bf f}. Specifically, although strong $\sigma^-$ ($\sigma^+$)-polarized light can orient the hole spins against a positive (negative) magnetic field $B$, the resulting steady-state light-induced spin polarization $\langle S_z\rangle_\mp(B)$ typically deviates from $-1$ ($+1$) once $|B|$ substantially exceeds the coercive field. These changes can be monitored by measuring the intensity $I_\mp(B)$ of $\sigma^-$ ($\sigma^+$) CW light reflected from the AP resonance. This intensity remains at the background level $I_\mathrm{back}$ for $B\lesssim B_\mathrm{c}$ ($B\gtrsim-B_\mathrm{c}$), but increases once the field sizably exceeds the coercive threshold, indicating a growth of $|\langle S_z\rangle_\mp(B)\pm1|\propto \Delta I_\mp(B)=I_\mp(B)-I_\mathrm{back}$. In order to normalize this intensity change, we perform time-resolved optical spin reorientation measurements in which we track the time-dependent count levels following a switch of the excitation helicity. For example, if the light polarization is changed from $\sigma^+$ to $\sigma^-$, the intensity just after the switch for $B\gtrsim B_\mathrm{c}$ corresponds to the maximal AP intensity $I_\mathrm{max}$, as in this setting the spins are fully oriented upwards prior to the switch by the combined effect of a positive $B$-field and $\sigma^+$-polarized excitation. This value of $I_\mathrm{max}$ can then be used to normalize the $\Delta I_\mp(B)$ acquired under CW $\sigma^\mp$ excitation, and hence to extract the $B$-field-dependent steady-state values of $\langle S_z\rangle_\mp(B)$ at different CW powers. In parallel, $\langle S_z\rangle_\mp(B)$ can be independently extracted based on the decay amplitude of the time-resolved optical spin reorientation signal itself. This amplitude is proportional to $1-\langle S_z\rangle_-(B)$ for $B\gtrsim B_\mathrm{c}$ and to $1+\langle S_z\rangle_+(B)$ for $B\lesssim-B_\mathrm{c}$, with each factor reaching 2 within the hysteresis loop at sufficiently high excitation powers ensuring complete optical spin orientation. As shown in Extended Data Fig.~\ref{fig:dynamical_sz_normalization}, these procedures yield consistent results, demonstrating the validity of our approach.

Extended Data Figs.~\ref{fig:hyst_power_series}{\bf a,b} show $\langle S_z\rangle(B)$ obtained in this way for both the ICI and metallic phases. As expected, the efficiency of optical spin orientation against the magnetic field for $|B|>B_\mathrm{c}$ decreases with increasing $|B|$ and improves at higher excitation powers, as shown in Extended Data Figs.~\ref{fig:hyst_power_series}{\bf c,d}. Nevertheless, for fields slightly exceeding the coercive threshold, optical spin reversal remains nearly complete at the excitation powers of a few hundred nW, enabling efficient initialization of magnetic domains in our time-resolved experiments.

\subsection{Time-resolved pump-probe experiments \label{sec:time_res_experiments}}

\begin{figure*}[]
	\includegraphics[width= \textwidth]{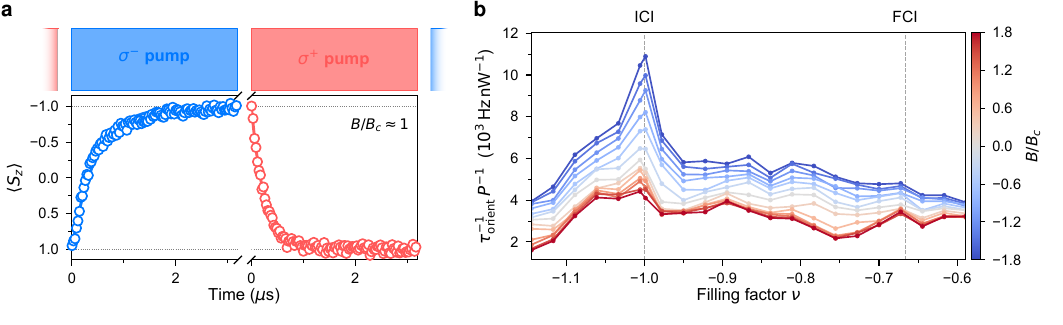}
	\caption{{\bf Optical spin orientation dynamics in the presence of the magnetic field.} ({\bf a}) Example measurement of spin orientation dynamics at $\nu\approx-1$ and positive $B\approx B_\mathrm{c}$ using a train of equally intense light pulses of alternating circular polarizations and the same power $P=500$~nW (as sketched in the top panel). The data points show time-dependent $\langle S_z\rangle(t)$ acquired for both $\sigma^\pm$ pump polarizations, while the solid lines mark exponential fits $\pm[1-2\exp(-t/\tau_\mathrm{orient})]$. The orientation time $\tau_\mathrm{orient}$ is shorter for $\sigma^+$ and longer for $\sigma^-$ polarization, corresponding to spins being oriented along and against the external magnetic field, respectively. ({\bf b}) Optical spin orientation rate, normalized by the power of the $\sigma^-$-polarized pump light, extracted as a function of $\nu$ and $B/B_\mathrm{c}$. While $\tau_\mathrm{orient}^{-1}P^{-1}$ increases (decreases) for negative (positive) $B$-fields that align the spins parallel (antiparallel) to the light-induced spin orientation, these variations remain comparatively small. As a consequence, $\tau_\mathrm{orient}$ at $P$ of a few hundred nW remains always much shorter than the spin relaxation timescales throughout the explored $B/B_\mathrm{c}$ range.\label{fig:ext_methods_orient_dynamics}}
\end{figure*}

\begin{figure*}[]
	\includegraphics[width = \textwidth]{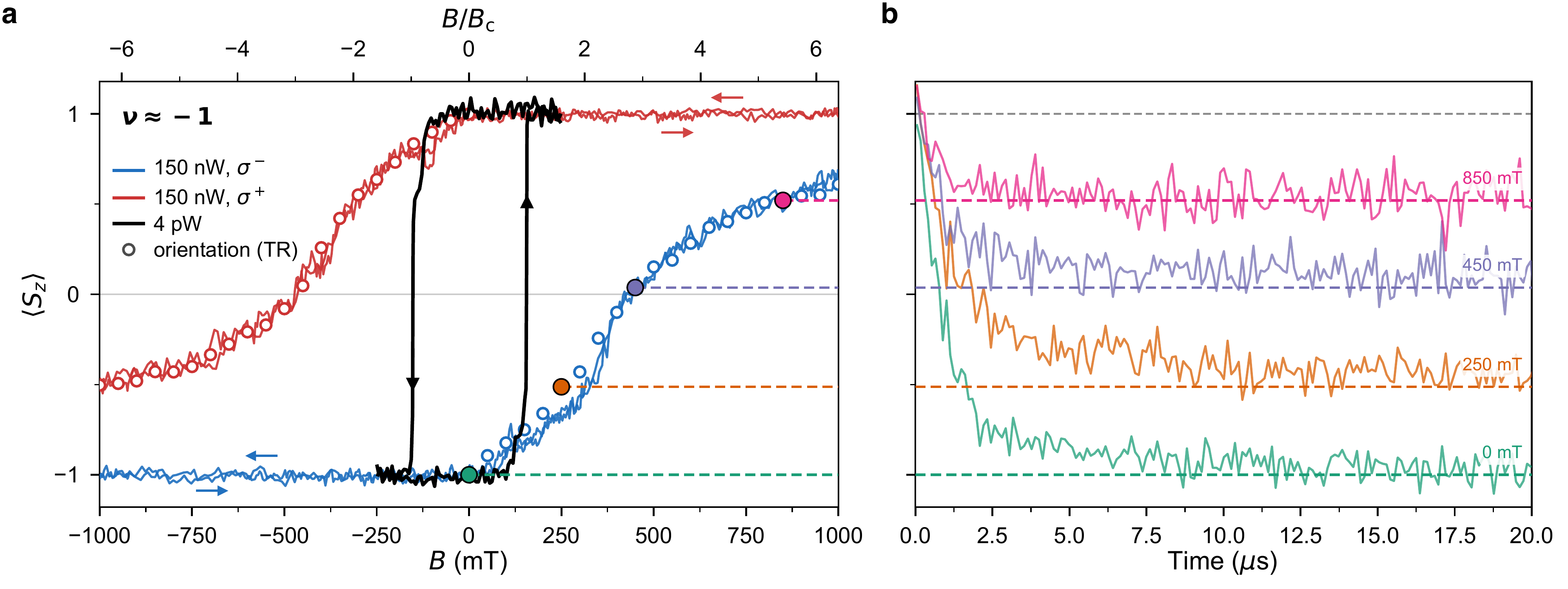}
	\caption{{\bf Determination of the steady-state light-induced spin polarization.} ({\bf a}) Steady-state hole spin polarization degree $\langle S_z\rangle (B)$ measured as a function of the external magnetic field at $\nu=-1$ under $\sigma^+$ (red) and $\sigma^-$-polarized CW excitation at a power of 150~nW. The data are obtained based on the normalized time-integrated intensity $I_\pm(B)$ of the reflected pump laser resonant with the AP transition, whose change $\Delta I_\pm(B)$ relative to the background level is proportional to $|\langle S_z\rangle(B)\mp1|$ for $\sigma^\pm$ polarization. The black curve depicts the magnetic hysteresis loop extracted at the same filling factor using a non-destructive low-power (4~pW) probe. The data points represent the corresponding $\langle S_z \rangle(B)$ values extracted from the optical reorientation measurements in panel {\bf b}. The agreement between these two datasets confirms the validity of our $\langle S_z\rangle (B)$ determination. ({\bf b}) Time-resolved optical spin reorientation dynamics measured after switching the pump helicity from $\sigma^+$ to $\sigma^-$ at selected magnetic fields (as indicated). The relative decay amplitudes of the time traces are proportional to $1-\langle S_z\rangle(B)$, where $\langle S_z\rangle(B)$ is the steady-state spin polarization under $\sigma^-$ excitation at $B\gtrsim B_\mathrm{c}$.
} \label{fig:dynamical_sz_normalization}
\end{figure*}

\begin{figure*}[]
	\includegraphics[width= \textwidth]{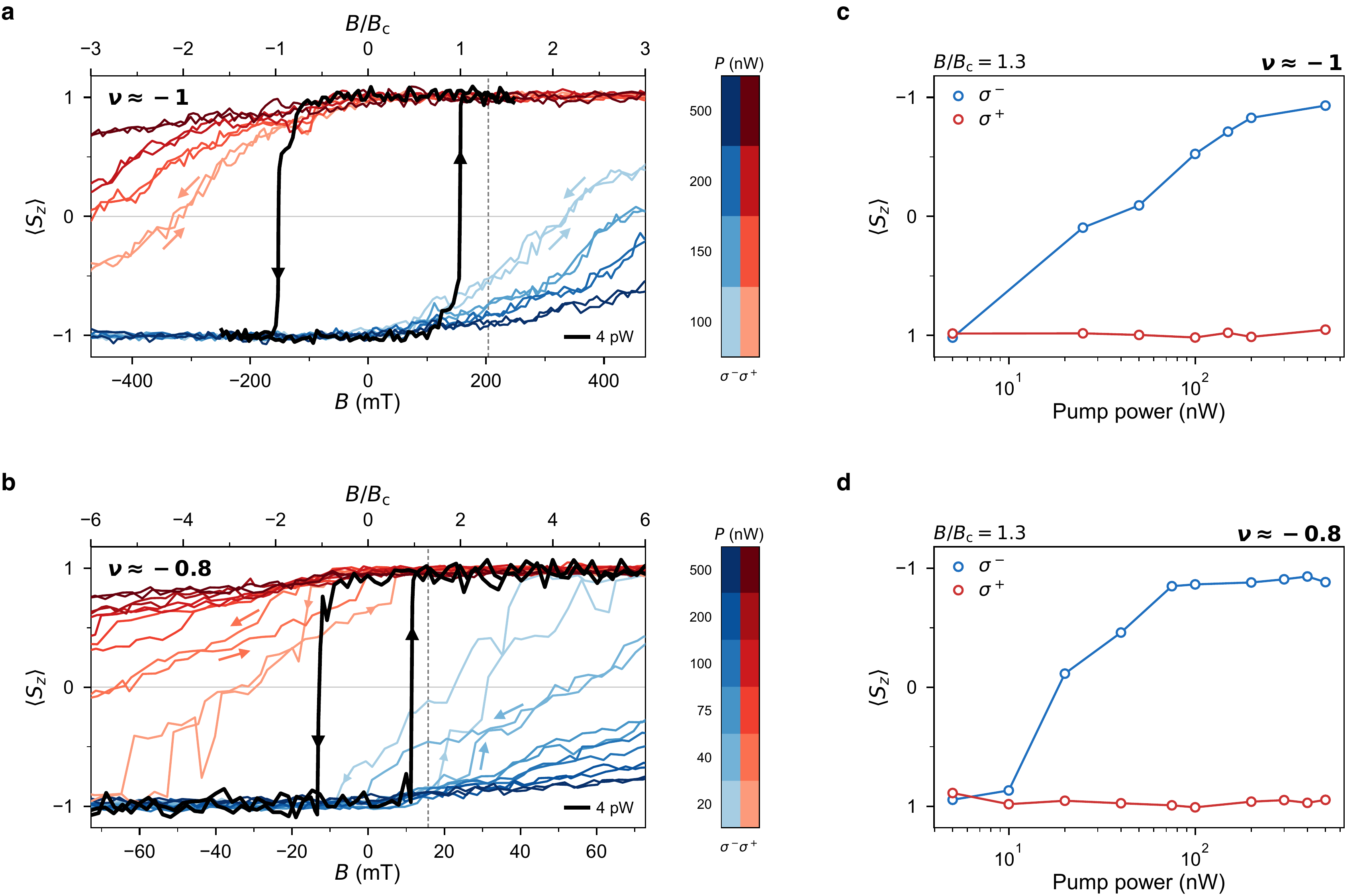}
	\caption{{\bf Dependence of light-induced spin orientation efficiency on power and magnetic field.} ({\bf a,b}) Steady-state hole spin polarization, $\langle S_z\rangle$, measured as a function of the magnetic field for the ICI ({\bf a}) and ferromagnetic metal ({\bf b}) under resonant CW excitation of the AP transition with $\sigma^\pm$-polarized light. The pump power is indicated by the color scales. The presented $\langle S_z\rangle$ are determined using the procedure described in Methods Sec.~\ref{sec:optical_orientation} and illustrated in Extended Data Fig.~\ref{fig:dynamical_sz_normalization}. At higher powers, the hysteretic behavior becomes indistinguishable, with $\langle S_z\rangle$ determined solely by the excitation power and independent of the $B$-field sweep direction (indicated by arrows). ({\bf c,d}) Linecuts through the $B$-down-sweep data in panels {\bf a,b}, showing $\langle S_z\rangle$ at a fixed $B=1.3B_\mathrm{c}$ (marked with vertical dashed lines) as a function of the excitation power for ICI ({\bf c}) and ferromagnetic metal ({\bf d}). In both cases, $\langle S_z\rangle$ remains at the field-induced value of $+1$ under $\sigma^+$ excitation, whereas increasing the power of the $\sigma^-$ pump progressively drives it towards $-1$. Nearly complete spin orientation against the applied positive field is reached at pump powers of a few hundred nW.\label{fig:hyst_power_series}}
\end{figure*}

\begin{figure*}[]
	\includegraphics[width = \textwidth]{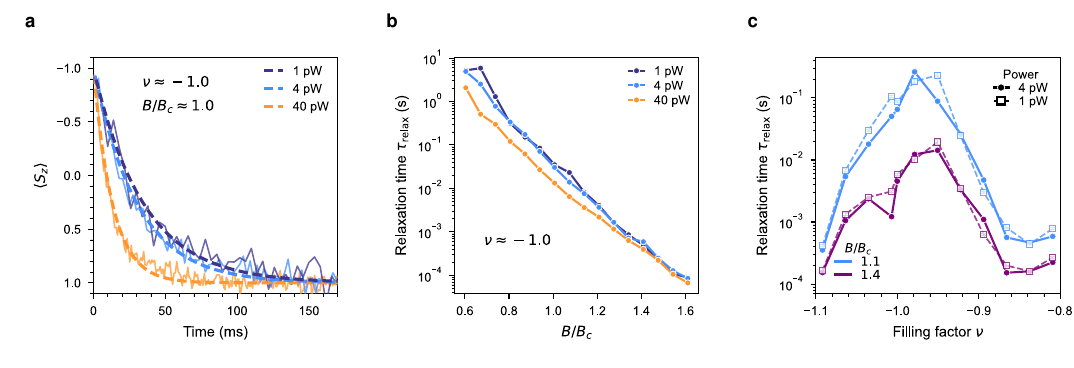}
	\caption{{\bf Non-destructive optical sensing of the hole spin relaxation.} ({\bf a}) Comparison of spin relaxation traces measured at $\nu\approx-1.0$ and $B/B_\mathrm{c}\approx1.0$ for three probe powers. Although relaxation is faster at $40$~pW, no discernible difference is observed between $4$~pW and $1$~pW, demonstrating that these lower powers enable non-destructive spin sensing. ({\bf b,c}) Spin relaxation timescales $\tau_\mathrm{relax}$ extracted by fitting relaxation profiles with exponential decays (marked by dashed lines in~{\bf a}) at fixed $\nu\approx-1.0$ as a function of $B/B_\mathrm{c}$ ({\bf b}) and at two fixed $B/B_\mathrm{c}$ values (as indicated) as a function of $\nu$ ({\bf c}). Notably, the probe-induced reduction of $\tau_\mathrm{relax}$ at the highest power, $40$~pW, is appreciable only at low $B/B_\mathrm{c}$, where relaxation is slowest and the spin system is therefore most susceptible to external perturbations. In contrast, the times obtained at $P\lesssim4$~pW agree within experimental uncertainty throughout the explored ranges of $\nu$ and $B/B_\mathrm{c}$, confirming non-destructive spin sensing under these conditions.}
\label{fig:non_destructive_readout}
\end{figure*}

To generate pump and probe pulses, a single-frequency CW laser resonant with the AP transition was first split into two beams using a fiber beam splitter. These beams were subsequently passed through two fast fiber-based acousto-optic modulators (AOMs) with rise times below 10~ns. Both AOMs were driven by synchronized channels of an arbitrary waveform generator (AWG), allowing independent control of their duration and delay. This enabled us to ensure that the probe pulse arrived at the sample precisely when the pump pulse ended, and vice versa. The resulting light pulses were coupled into two separate excitation arms of the confocal microscope setup (see Extended Data Fig.~\ref{fig:set_up}) and passed through independent polarization optics, ensuring they were cross-circularly polarized. For measurements of the optical spin orientation dynamics (e.g., shown in Extended Data Fig.~\ref{fig:ext_methods_orient_dynamics}), both pulses were sufficiently intense and long to establish a steady-state spin polarization. By contrast, for spin-relaxation experiments (Figs.~\ref{fig:Fig_relax_dynamics}{\bf e-g} and~\ref{fig:Fig_spatial}), the probe pulse power was reduced by orders of magnitude to approximately $4$~pW. As shown in Extended Data Fig.~\ref{fig:non_destructive_readout}, the spin relaxation dynamics measured under such conditions remains unchanged upon further reducing the probe power, demonstrating that this power enables non-perturbative optical spin readout.

To monitor the time-dependent spin polarization degree $\langle S_z\rangle(t)$ of the hole system, the reflected probe light was sent to an SNSPD. Importantly, this light was collected in the same circular polarization as that of the probe itself, suppressing the cross-circularly polarized pump. As a consequence, the count rate of photons detected by the SNSPD remained well below the saturation threshold at all times, thus ensuring a linear response of the detector. Individual photon arrival events were recorded using synchronized time-tagging electronics triggered by the leading edge of the voltage signal driving the AOM in the pump beam. The resulting signal was accumulated over a large number of pump-probe cycles, allowing us to retrieve time-dependent probe light intensity $I_\mathrm{probe}(t)=I_0R_0[1+f_\mathrm{+}(t)]$, where $I_0$ is the probe laser intensity, $f_\mathrm{+}(t)$ is the time-dependent AP spectral weight in $\sigma^+$ polarization of the probe, while $R_0$ denotes a background reflectance of the sample at AP energy. Given that the probe pulse was always sufficiently long for the hole system to reach its steady state with $f_\mathrm{+}=0$, at long times $I_\mathrm{probe}(t\rightarrow\infty)$ saturates at the level of $\approx I_0R_0$, thus allowing us to extract $f_\mathrm{+}(t)=I_\mathrm{probe}(t)/I_\mathrm{probe}(t\rightarrow\infty)-1$. Since the total AP intensity in both circular polarizations $f_\mathrm{+}(t)+f_\mathrm{-}(t)=f_0$ remains constant, we obtain $\langle S_z\rangle(t)=[f_\mathrm{-}(t)-f_\mathrm{+}(t)]/[f_\mathrm{-}(t)+f_\mathrm{+}(t)]=1-2f_\mathrm{+}(t)/f_0$. By determining $f_0$ based on $f_\mathrm{+}(t=0)$ when the pump is sufficiently strong to ensure complete spin flip in the detected sample area, we can thus directly determine $\langle S_z\rangle(t)$, as plotted, e.g., in Fig.~\ref{fig:Fig_spatial}. An exactly the same procedure can be applied to obtain $\langle S_z\rangle(t)$ during the pump pulse if the light is collected in $\sigma^-$ polarization instead of $\sigma^+$, as in Figs.~\ref{fig:Fig_relax_dynamics}{\bf c,d}. 

In all cases, the characteristic timescale $\tau$ of either relaxation or light-induced spin orientation was extracted by fitting the $f_\pm(t)$ data with a phenomenological exponential function, $\propto\exp(-t/\tau)$. In certain regimes, these fits are subject to small systematic uncertainties due to deviations of the spin-relaxation traces from purely exponential behavior (e.g., in metallic states at low $B/B_\mathrm{c}$). Nevertheless, this procedure provides a robust measure of the characteristic timescales of the underlying relaxation processes.

\subsection{Dependence of the shape of spin relaxation profiles on the area and power of the pump beam} \label{sec:profile_dependence_on_area_and_power}

\begin{figure*}[]
	\includegraphics[width = 1.0\textwidth]{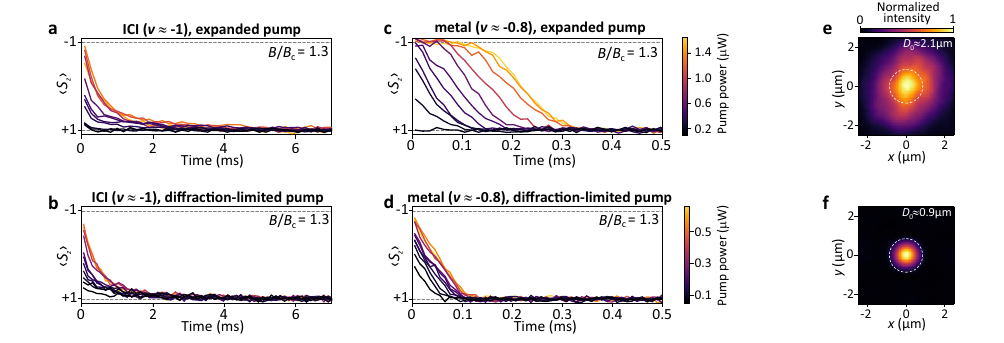}
	\caption{{\bf Relaxation dynamics of ICI and ferromagnetic metal vs. area and power of the pump beam.} ({\bf a-d}) Spin relaxation time traces measured at a fixed $B/B_\mathrm{c}=1.3$ for ICI ({\bf a,b}) and ferromagnetic metal ({\bf c,d}) as a function of the pump power using expanded ({\bf a,c}) or diffraction-limited ({\bf b,d}) pump beam with $D_0\approx2.1~\mu$m and $D_0\approx0.9~\mu$m, respectively. In all measurements, the probe beam was diffraction-limited and co-centered, while the pump pulse duration was fixed at $0.5$~ms. ({\bf e,f}) Color-scale plots showing spatial intensity profiles of an expanded ({\bf e}) and narrow ({\bf f}) pump beam used in measurements from panels {\bf a-d}. \label{fig:method_expanded_vs_normal_spot}}
\end{figure*}

A key signature of spin relaxation via domain shrinking is the initial plateau in the temporal relaxation profiles. This plateau appears exclusively in the metallic phases when the initial magnetic domain diameter $D_0$ exceeds the probe spot diameter $D_\mathrm{meas}$, as demonstrated in Fig.~\ref{fig:Fig_spatial} by comparing $\langle S_z\rangle(t)$ traces measured with different pump-beam diameters. This conclusion is further supported by Extended Data Fig.~\ref {fig:method_expanded_vs_normal_spot}, which shows temporal profiles measured at fixed $B=1.3B_\mathrm{c}$ as a function of pump-pulse power $P$ for both expanded and diffraction-limited pump beams. Strikingly, upon reducing the power of the expanded beam, the plateau at $\nu\approx-0.8$ becomes shorter and eventually disappears in the low-power regime (Extended Data Fig.~\ref {fig:method_expanded_vs_normal_spot}{\bf c}). This behavior is fully consistent with the expected reduction of the initial domain size at lower pump powers, as the sample area in which spins are effectively optically oriented by the pump pulse is reduced. In contrast, no plateau is observed for excitation with a diffraction-limited spot (Extended Data Fig.~\ref {fig:method_expanded_vs_normal_spot}{\bf d}), even when the relaxation trace approaches its saturation in the high power limit. This finding remains in perfect agreement with the fact that the maximum initial domain size under such conditions remains comparable to $D_\mathrm{meas}$.

Finally, no plateau is observed in spin relaxation profiles acquired near the Chern insulating phase at $\nu\approx-1$, regardless of the diameter or power of the pump beam (Extended Data Fig.~\ref{fig:method_expanded_vs_normal_spot}{\bf a,b}). This finding is likewise consistent with the expected dominance of relaxation through the creation of local internal domains, which explains why the measured $\langle S_z\rangle(t)$ in this regime retains a nearly exponential shape independently of pump power, with the power effect limited to reducing the initial amplitude.

\subsection{Dependence of the shape of spin relaxation profiles on the magnetic field \label{sec:profile_dependence_on_B}}

\begin{figure*}[]
	\includegraphics[width = 1.0\textwidth]{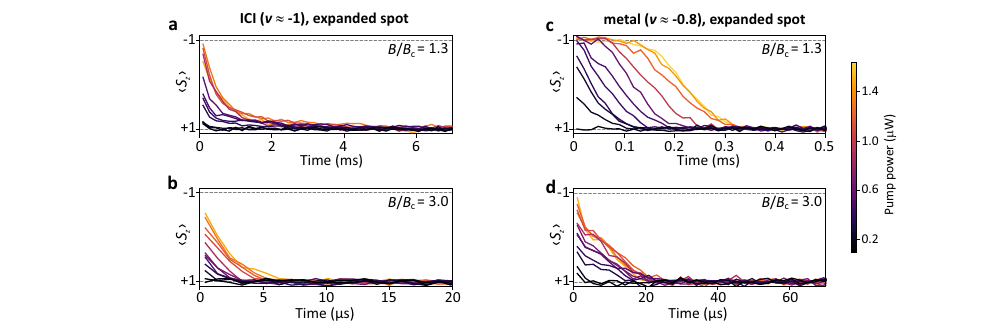}
	\caption{{\bf Relaxation dynamics of large magnetic domains at different magnetic fields.} ({\bf a-d}) Excitation-power-dependent spin relaxation time traces measured at ICI ({\bf a,b}) and ferromagnetic metal ({\bf c,d}) using an expanded pump beam with $D_0\approx2.1~\mu$m at two different $B/B_\mathrm{c}=1.3$ ({\bf a,c}) and 3.0 ({\bf b,d}). In all measurements, the pump pulse duration was fixed at 0.5~ms, while the probe beam was co-centered and diffraction-limited. As expected, the initial plateau indicating spin relaxation via domain shrinking is visible exclusively for the metallic phase at low $B/B_\mathrm{c}$ and for sufficiently large pump powers ensuring a large initial magnetic domain diameter.\label{fig:method_power_dependance_expanded_spot}}
\end{figure*}

As predicted by our theoretical model, the domain-shrinking mechanism becomes inefficient even for ferromagnetic metals once the energy barrier between the false- and true-vacuum states vanishes at high magnetic fields. This prediction is fully supported in Extended Data Fig.~\ref {fig:method_power_dependance_expanded_spot}, which compares pump-power-dependent $\langle S_z\rangle(t)$ profiles measured with an expanded pump beam for the ICI and ferromagnetic metal at $B=1.3B_\mathrm{c}$ and $B=3.0B_\mathrm{c}$. In the metallic phase, the initial plateau observed at the lower field clearly disappears at $B=3.0B_\mathrm{c}$. By contrast, despite their markedly different relaxation dynamics, the ICI profiles retain essentially the same shape in both explored $B/B_\mathrm{c}$ regimes. The disappearance of the initial plateau for the metallic phase at high field thus provides further evidence that this feature originates from collective magnetic-domain shrinking.

\subsection{Reproducibility of spin relaxation profiles on a second spot \label{sec:relaxation_reproducibility}}

\begin{figure*}[]
	\includegraphics[width = 140mm]{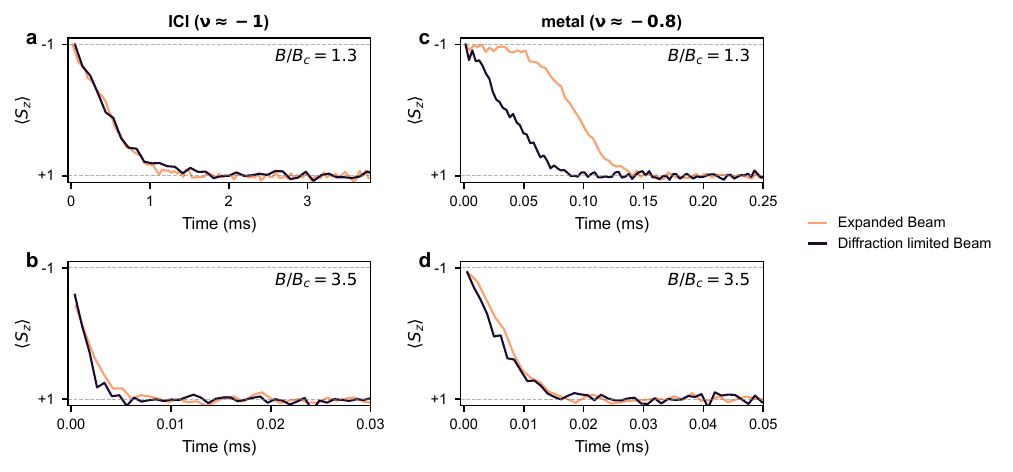}
	\caption{{\bf Relaxation dynamics vs area of magnetic domain on a second spot on device A}. ({\bf a-d}) Spin relaxation time traces measured on a second spot on device A for ICI ({\bf a,b}) and ferromagnetic metal ({\bf c,d}) at a fixed $B/B_{c}$ = 1.3 ({\bf a,c}) and 3.5 ({\bf b,d}). Each panel compares traces acquired using an expanded pump beam with $D_{0} \approx2.1\ \mu$m and a diffraction-limited pump beam with $D_{0} \approx0.9\ \mu$m, respectively. All traces were measured with a diffraction-limited, co-centered probe beam and with pump pulses of sufficiently high power and duration to ensure complete preparation of the initial magnetic domain.\label{fig:relaxation_2nd_spot}}
\end{figure*}

The key findings of our work were reproduced on multiple spots in device A. This, in particular, concerns the shape of the temporal spin relaxation profiles for metallic and Chern insulating phases. Extended Data Fig.~\ref{fig:relaxation_2nd_spot} shows the comparison of such profiles acquired in the two cases with expanded and diffraction-limited pump beams at both low and high-$B$-field. As expected, the initial plateau is observed exclusively for ferromagnetic metal excited with a sufficiently strong expanded pump beam at low $B/B_\mathrm{c}$, in perfect agreement with theoretical model and with the results from the main text.

\subsection{Analysis of spatially resolved zero-field domain writing experiments \label{sec:domain_writing}}

\begin{figure*}[]
	\includegraphics[width = \textwidth]{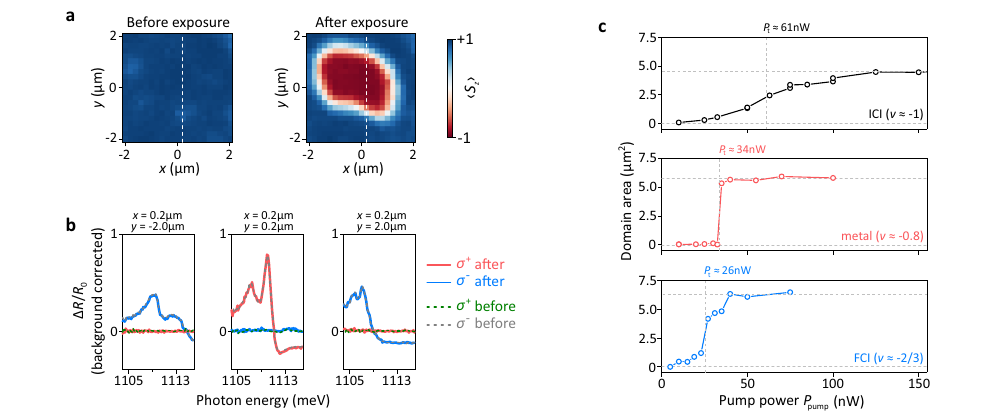}
	\caption{{\bf Determination of magnetic domain size in spatially resolved zero-field measurements.} ({\bf a}) Spatial maps illustrating the spin polarization degree $\langle S_z\rangle({\bf r})$ of the ferromagnetic metal ($\nu\approx-0.8$) measured before (left) and after (right) writing a magnetic domain with an expanded doughnut-beam with excitation power of $P=55$~nW. Before the experiment, the hole spins were oriented upwards with $B=0.2$~T, which was subsequently ramped down to zero. The value of $\langle S_z\rangle({\bf r})$ is determined based on the spectrally integrated, background-corrected reflectance contrast spectra $R_\mathrm{c}^\pm$ measured in $\sigma^-$ and $\sigma^+$ circular polarizations. ({\bf b}) Examples of such $R_\mathrm{c}^\pm$ spectra measured before and after domain writing at various spatial locations along the white dashed line in {\bf a}. ({\bf c}) The areas of magnetic domains created in ICI (top), ferromagnetic metal (middle), and FCI (bottom) as a function of the power of the doughnut-beam. The horizontal dashed lines label saturated domain areas, while the vertical dashed line marks the threshold power $P_\mathrm{t}$ at which the domain size reaches half of its saturated value. \label{fig: Analysis_domain_writing}}
\end{figure*}

To visualize the shape of optically written magnetic domains in Fig.~\ref{fig:Fig_domains}{\bf b}, we first prepare the device at a given filling factor and orient the hole spins upwards using a positive magnetic field $B=0.2$~T. After ramping the field back to zero, we map the resulting $\langle S_z\rangle({\bf r})$ using a weak, a-few-nm-bandwidth white light filtered around the AP energy. To this end, we measure two reflectance spectra $R^\pm({\bf r})$ in the two circular polarizations at each spatial location ${\bf r}=(x,y)$, and then normalize them by the background spectra $R^\pm_{0}({\bf r})$  acquired beforehand at the same locations when the device was charge neutral and therefore did not exhibit an AP resonance in the examined spectral range. To account for the residual influence of long-wavelength tails of strong exciton resonances on $R^\pm_{0}(\bf{r})$, the resulting spatially dependent reflectance contrast $R_\mathrm{c}^\pm({\bf r})= R^\pm({\bf r})/R^\pm_{0}({\bf r}) - 1$ is further corrected by subtracting a smooth background term $W({\bf r})$ obtained by fitting a polynomial to $R_\mathrm{c}^+({\bf r})$, which does not exhibit any AP resonance. The spectrally integrated absolute value of the resulting background-corrected reflectance contrast spectra---termed $I^\pm({\bf r})$---is directly proportional to the AP spectral weight in $\sigma^\pm$ polarization. This allows us to determine the local spin polarization degree as $\langle S_z\rangle_\mathrm{before}({\bf r})=[I^-({\bf r})-I^+({\bf r})]/[I^-({\bf r})+I^+({\bf r})]$, as shown in the left panel of Extended Data Fig.~\ref{fig: Analysis_domain_writing}{\bf a} for an example case of a ferromagnetic metal.

After this initialization, the sample is illuminated for a few tens of seconds with an expanded $\sigma^-$-polarized doughnut beam of controlled power $P$ to orient the spins opposite to their original direction. The spectral profile of this beam is identical to that used for probing the sample, while the beam position is aligned with the center of the examined spatial region. Finally, the resulting $\langle S_z\rangle_\mathrm{after}({\bf r})$ is determined similarly as during the initialization, by mapping $\sigma^\pm$-polarized reflectance contrast spectra (background corrected with the original $W({\bf r})$ polynomial) using a diffraction-limited weak probe beam and extracting the corresponding spectrally integrated polarization degree, as shown in the right panel of Extended Data Fig.~\ref{fig: Analysis_domain_writing}{\bf a}. Example $R_\mathrm{c}^\pm$ spectra measured at various spatial locations before and after domain writing are shown in Extended Data Fig.~\ref{fig: Analysis_domain_writing}{\bf b}.

The above procedure was employed to obtain all maps in Fig.~\ref{fig:Fig_domains}{\bf b}. To determine the area of the optically created magnetic domains in each map, we integrate out the obtained spin-polarization degree over the probed spatial area, $A(P)=\int\frac{1}{2}[1-\langle S_z\rangle_\mathrm{after}({\bf r})]d^2{\bf r}$, and subtract the analogous integral obtained based on $\langle S_z\rangle_\mathrm{before}({\bf r})$. Extended Data Fig.~\ref{fig: Analysis_domain_writing}{\bf c} shows the resulting areas determined as a function of the pump power $P$ for both Chern insulators and the ferromagnetic metal. In each case, the area clearly saturates at $A_\mathrm{sat}(\nu)$ in the high-power regime, with a $P$-induced increase that is abrupt in the metal and gradual for both the FCI and ICI phases. To facilitate the comparison between these curves, we determine the threshold power $P_\mathrm{t}(\nu)$ at which the domain area reaches half of its saturated value, and then plot the normalized area $A/A_\mathrm{sat}$ as a function of normalized power $P/P_\mathrm{t}$, as shown in Fig.~\ref{fig:Fig_domains}{\bf e} in the main text.

\subsection{Data for the device B \label{sec:2nd_device_data}}
\begin{figure*}[]
	\includegraphics[width = \textwidth]{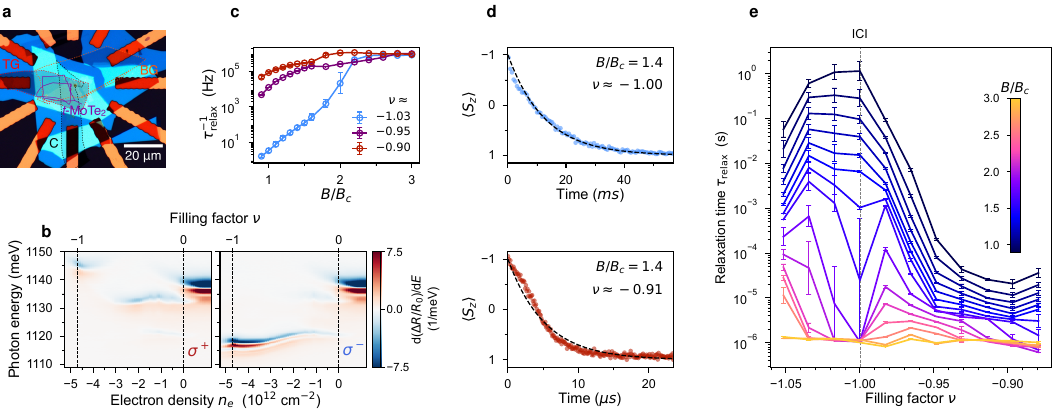}
	\caption{{\bf Summary of the results obtained on device B.}
    ({\bf a}) Optical micrograph of the dual-gated twisted $\text{MoTe}_2$ bilayer in device B with $\sim4.1^\circ$ twist angle (TG: top gate, BG: bottom gate, C: contact). ({\bf b}) Filling-factor evolution of reflectance-contrast spectra differentiated with respect to photon energy. The data were acquired at $B=0.2$~T in $\sigma^+$ (left) and $\sigma^-$ (right) circular polarizations. The cusp in the energy and intensity of the AP transition is due to the ICI formation around $\nu = -1$. ({\bf c}) Spin relaxation rate ($\tau_{\text{relax}}^{-1}$) as a function of $B/B_\mathrm{c}$ measured for selected filling factors (as indicated). ({\bf d})  Normalized spin relaxation time traces measured at $B/B_\mathrm{c} = 1.4$ for the ICI ($\nu \approx -1.00$; top) and the ferromagnetic metal ($\nu \approx -0.91$; bottom). Dashed lines in both plots represent exponential fits. ({\bf e}) Spin relaxation times $\tau_{\text{relax}}$ as a function of the filling factor and $B/B_\mathrm{c}$. Similarly as in Fig.~\ref{fig:Fig_relax_dynamics}{\bf f} for device A, at low fields ($B/B_\mathrm{c} \sim 1.0$), relaxation is sensitively dependent on the underlying many-body ground state, peaking sharply in the vicinity of the ICI state. At elevated fields ($B/B_\mathrm{c} \sim 2-3$), this filling-factor dependence gradually washes out. In these experiments, the sampled $B/B_\mathrm{c}$ values differed between filling factors; therefore, to obtain a common $B/B_\mathrm{c}$ grid, the $\log[\tau_\mathrm{relax}(B/B_\mathrm{c})]$ dependencies were linearly interpolated for each $\nu$ value. The error bars indicate the interpolation uncertainty, defined as the time range separating each interpolated $\tau_\mathrm{relax}$ from the nearest measured value at the same $\nu$.
    \label{fig:relaxation 2nd device}}
\end{figure*}

The experimental findings detailed in the main text were reproduced and verified on a second device (device B) with a twist angle of $\sim4.1^\circ$ (see Extended Data Fig.~\ref{fig:relaxation 2nd device}{\bf a}). Notably, due to smaller hBN thicknesses, the optical resonances in device B exhibited almost perfectly Lorentzian lineshape and larger contrast compared to device A. To track the evolution of the AP resonance with doping density, we examined the derivative of the reflectance-contrast spectra with respect to photon energy $d(\Delta R/R_0)/dE$ as a function of the filling factor $\nu$ (Extended Data Fig.~\ref{fig:relaxation 2nd device}{\bf b}), which revealed a similar cusp in the energy and intensity around the ICI state. Owing to a larger twist angle, this device showed no robust signature of the FCI~\cite{li2025}. Furthermore, it was considerably less homogeneous than device A, making it less suitable for spatially resolved domain writing measurements or experiments requiring an expanded excitation beam. Nevertheless, the characteristic strong dependence of the spin relaxation rates on the underlying many-body ground state remained clearly observable in time-resolved experiments carried out using diffraction-limited pump and probe beams, as shown in Extended Data Figs.~\ref{fig:relaxation 2nd device}{\bf c,e}. At low $B/B_\mathrm{c}$ ratios, the relaxation times are several orders of magnitude longer in the vicinity of ICI than for the metallic phases. Correspondingly, at elevated fields ($B/B_\mathrm{c}>2$), this difference gradually disappears, showing behavior very similar to that observed in device A. Finally, despite being obtained using pump and probe beams of the same diameter, the spin-relaxation profiles for the ICI and ferromagnetic metals exhibit distinct differences (Extended Data Fig.~\ref{fig:relaxation 2nd device}{\bf d}): the former is nearly exponential, whereas the latter initially decays more slowly before accelerating, consistent with domain shrinking dominating relaxation in the metallic phase.

\subsection{Theory: Effective model \label{sec:theory_model}}
To capture the out-of-equilibrium magnetization dynamics in twisted MoTe$_2$ bilayers, we use \emph{model-A} dynamics~\cite{HohenbergHalperin77}, describing a non-conserved order-parameter $m(\mathbf{r}, t)$:
\begin{equation}
    \partial_t m(\mathbf{r},t) = -\Gamma\frac{\delta f[m]}{\delta m(\mathbf{r},t)} + \xi(\mathbf{r},t),
    \label{eq:model_a}
\end{equation}
where $t$ and $\mathbf{r}=(x,y)$ denote, respectively, time and spatial coordinates, $f[m]$ is the free energy density, $\Gamma$ is an intrinsic equilibration rate fixing the time scale of our system, and $\xi(\mathbf{r}, t)$ are temperature-dependent fluctuations of a thermal bath. As argued in Ref.~\cite{pichler2026}, the presence of topological bands with opposite chirality for the two electronic spin states leads to an Ising anisotropy~\cite{Qiu2025} suppressing spin-wave fluctuations and justifying the description in terms of an Ising-type order parameter.
We write the free energy as
\begin{equation}
    f[m] = V[m] + \kappa_2 |\nabla m|^2 + \cdots \label{eq:freeEnergy_generic}
\end{equation}
where $V[m]$ is the free energy potential, which we compute from a microscopic model for $t$-MoTe$_2$ on a mean-field level, and $\kappa_2$ is the spin stiffness. In principle, higher-order gradient terms can be added to the free energy, as we discuss below. We find that the spin stiffness $\kappa_2$, which determines the shape of magnetic domain walls, strongly affects the dynamics.

To provide an estimate for the spin stiffness and the potential $V[m]$ of $t$-MoTe$_2$ as a function of the filling factor, we use a microscopic lattice model describing the two topmost hole bands, closely following Ref.~\cite{pichler2026}. Concretely, we consider the extended Haldane model on a honeycomb lattice with up to third-nearest-neighbor hopping, adapted from Ref.~\cite{chen2025fractionalcherninsulatorquantum}
\begin{equation}
    H_0^\sigma = \sum_{ij} t_{ij}^\sigma c_{\sigma i}^\dagger c_{\sigma j}.
\end{equation}
We work with the following hopping parameters: $t_1=3.225$ meV, $t_2=-2.210$ meV, $t_3=-0.947$ meV. Nearest- and third-nearest-neighbor hoppings are assumed to be purely real, while the next-nearest neighbor hopping has a complex phase of $e^{i\sigma\frac{2\pi}{3}}$, with $\sigma=\pm1$ for spin $\uparrow$ and $\downarrow$, respectively. These parameters have been shown to approximate the particle-hole transformed two topmost valence bands of $t$-MoTe$_2$ at a twist angle around $\theta\approx4^\circ$~\cite{chen2025fractionalcherninsulatorquantum}. Upon increasing repulsive interactions, the model transitions to a ferromagnetic phase that breaks discrete time-reversal symmetry. We consider repulsive on-site interactions $U$:
\begin{equation}
    H_{\mathrm{int}} = U \sum_i c^\dagger_{\uparrow i} c^\dagger_{\downarrow i} c_{\downarrow i} c_{\uparrow i}.
\end{equation}
In a more accurate description of $t$-MoTe$_2$, longer-range interactions are expected to be relevant. However, we have verified that including interaction terms beyond the on-site term does not qualitatively modify our results. 

To compute the potential $V[m]$, we perform a mean-field decoupling of the interaction Hamiltonian $H_{\mathrm{int}}$, assuming a finite and translationally uniform magnetization  $m = \langle S_z\rangle = (c^\dagger_{\uparrow i} c_{\uparrow i} -c^\dagger_{\downarrow i} c_{\downarrow i})/2$. The resulting mean-field Hamiltonian is
\begin{equation}
  H_\mathrm{MF}[m] =  \begin{pmatrix}
        H_0^\uparrow & 0 \\
        0            &  H_0^\downarrow
    \end{pmatrix} - \frac{1}{2}\begin{pmatrix}
        U m +B & 0 \\
        0            &  - Um -B
    \end{pmatrix}, \label{eq:MF_magnHamiltonian}
\end{equation}
where we have also introduced a Zeeman term coupling to an external magnetic field $B$.
The potential $V[m]$ is then obtained by solving for the eigenenergies $E_n$ of Eq.~\eqref{eq:MF_magnHamiltonian} as a function of the magnetization $m$ and summing over all filled states:
\begin{equation}
    V[m] = \sum_{n \; \mathrm{filled}} E_n[m] + \frac{U m^2}{2}. \label{eq:FreeEnergyPotential}
\end{equation}
This fixes the potential part of the free energy in Eq.~\eqref{eq:freeEnergy_generic}. To extract the spin stiffness $\kappa_2$, we use the same microscopic model, now assuming translational invariance only along the $y$ direction while introducing two domain walls along the $x$ direction. We impose periodic boundary conditions in both directions and consider a system with $L_x\times L_y$ unit cells. The stiffness can then be computed from the structure of the magnetic domain walls, which we obtain through self-consistent Hartree-Fock calculations~\cite{pichler2026, wang2022structuredomainwallschiral}. We rewrite the full Hamiltonian as
\begin{equation}
    H = H_0 + U \displaystyle\sum_{\substack{x, y \\ a}} c^\dagger_{a\uparrow}(x, y) c^\dagger_{a\downarrow}(x, y) c_{a\downarrow}(x, y) c_{a\uparrow}(x, y), \label{eq:generalHamiltonian}
\end{equation}
where $H_0 = H_0^\uparrow + H_{0}^\downarrow$ is the kinetic term and $a$ is the sublattice degree of freedom of the honeycomb lattice. Upon mean-field decoupling of the interaction term, we obtain the following Hamiltonian
\begin{equation}
    \begin{aligned}
        H_\mathrm{MF} = H_0 +  U& \displaystyle\sum_{k = 1}^{L_y} \displaystyle\sum_{x, a, \sigma} \big[ \chi_{aa, \sigma\sigma}(x) c^\dagger_{a \bar{\sigma} k}(x) c_{a \bar{\sigma} k}(x) \\ &\quad- \chi_{aa, \sigma \bar{\sigma}}(x) c^\dagger_{a\bar{\sigma}k}(x) c_{a {\sigma} k}(x)\big] \\
        - U L_y &\displaystyle\sum_{x, a} \big[\chi_{aa, \uparrow\uparrow}(x) \chi_{aa, \downarrow\downarrow}(x) - |\chi_{aa, \uparrow \downarrow}(x)|^2 \big]&
    \end{aligned}
    \label{eq:MF_Hamiltonian}
\end{equation} 
with 
\begin{equation}
    \chi_{ab, \sigma \sigma'}(x) = \frac{1}{L_y} \displaystyle\sum_{k=1}^{L_y} \displaystyle\sum_{\lambda} \psi^*_{\lambda, a\sigma k}(x) \psi_{\lambda, b\sigma' k}(x) n_{k}^\lambda.
\end{equation}
We sum over all eigenstates $\lambda$ of the mean-field Hamiltonian, with eigenvalues $\varepsilon_{\lambda k}$ and eigenvectors $\psi_{\lambda, a\sigma k}(x)$. Here, $n_k^\lambda = 1/(e^{\varepsilon_{\lambda k}/k_B T} + 1)$ is the Fermi-Dirac distribution at a temperature $T$. We use the following workflow adapted from Ref.~\cite{pichler2026} to obtain the domain-wall structures, from which the stiffness can be extracted: First, we use an initial state $\chi_{aa, \sigma \sigma}(x)$ that is uniform in the $x$ direction and solve Eq.~\eqref{eq:MF_Hamiltonian} iteratively until both energy and $\chi$ converge. We then use this converged uniform state to construct an initial state with two domain walls. Since we are working with periodic boundary conditions, the number of domain walls must be even. We then again solve Eq.~\eqref{eq:MF_Hamiltonian} iteratively, using the two-domain-wall initial state. Convergence is verified by introducing small random fluctuations to every initial state, which we find does not affect the converged result. We treat the chemical potential as a Lagrange multiplier, adjusting it at each iteration in order to ensure that the total particle number remains constant, thus fixing the average filling factor.

As discussed in the main text, we find qualitatively different domain-wall structures for the metal and the Chern insulator at $\nu=-1$, consistent with what was reported in Ref.~\cite{pichler2026}. In the metal, the domain walls exhibit standard functional form $m(x) = m_0 \tanh(x/\ell_\mathrm{DW})$, where the width $\ell_{\mathrm{DW}} \propto \sqrt{\kappa_2}$ scales with the square root of the stiffness. In stark contrast, the domain walls in the Chern insulating phase are much sharper and do not have the usual $\tanh$-like form. Example domain walls obtained from solving Eq.~\eqref{eq:diffEq_domainWall} are presented in Fig.~\ref{fig:Fig_spatial}\textbf{h}.

We find that in the vicinity of the Chern insulating phase, the shape of the domain wall can be reproduced by introducing a higher-order gradient term $\kappa_4 |\nabla m|^2$ to the free energy. Concretely, for a given $\kappa_2$ and $\kappa_4$, the domain-wall structure along the $x$ direction is obtained by solving the following differential equation
\begin{equation}
    \kappa_4 \partial_x^4 m(x) - \kappa_2 \partial_x^2 m(x) + \frac{\delta V[m]}{\delta m(x)} = 0 \label{eq:diffEq_domainWall}
\end{equation}
with boundary conditions $m(x\rightarrow\pm \infty) = \pm1$ and $\partial^2_x m(x\rightarrow \pm\infty) = 0$. By fitting $\kappa_2$ and $\kappa_4$ to the self-consistently determined domain walls in the ICI phase, we find a negative stiffness $\kappa_2 < 0$, and $\kappa_4>0$, which is required for stability against short-range fluctuations. 
If we introduce the higher-order term $\kappa_4$ also in the metallic phase, and fit it to the domain wall profiles, we find $\kappa_4=0$, justifying neglecting this term in case of ferromagnetic metals. Physically, we attribute the negative sign of the lowest-order stiffness parameter in the ICI phase to the presence of chiral edge modes at domain-wall boundaries in this case. These topologically protected modes affect the domain walls, allowing for the formation of very sharp boundaries between regions with different spin orientations. We find that the resulting sharp domain-wall shape is already present in the vicinity of $\nu=-1$, with its onset coinciding with the filling factor at which the chiral edge modes start to populate on a mean-field level. As argued in the main text, these qualitative differences in the domain-wall structure are the underlying reason behind striking differences in the relaxation dynamics we observe in our experiments. At the same time, we note that the spatial resolution of our domain-mapping experiments in Fig.~\ref{fig:Fig_domains}{\bf b}, set by the diffraction-limited probe-beam size, is not sufficient to directly resolve the domain walls and thus visualize their distinct structures in the metal and Chern insulators.

While our mean-field treatment cannot capture fractional Chern insulators directly, we expect the domain-wall structure to be similar to the case of ICI, in agreement with our experimental observations. For future work, it will be interesting to further explore the role of fractionalization on the domain-wall formation and the resulting magnetization dynamics.

\subsection{Theory: Coercive field \label{sec:theory_coercive_field}}

\begin{figure*}[]
	\includegraphics[width = \textwidth]{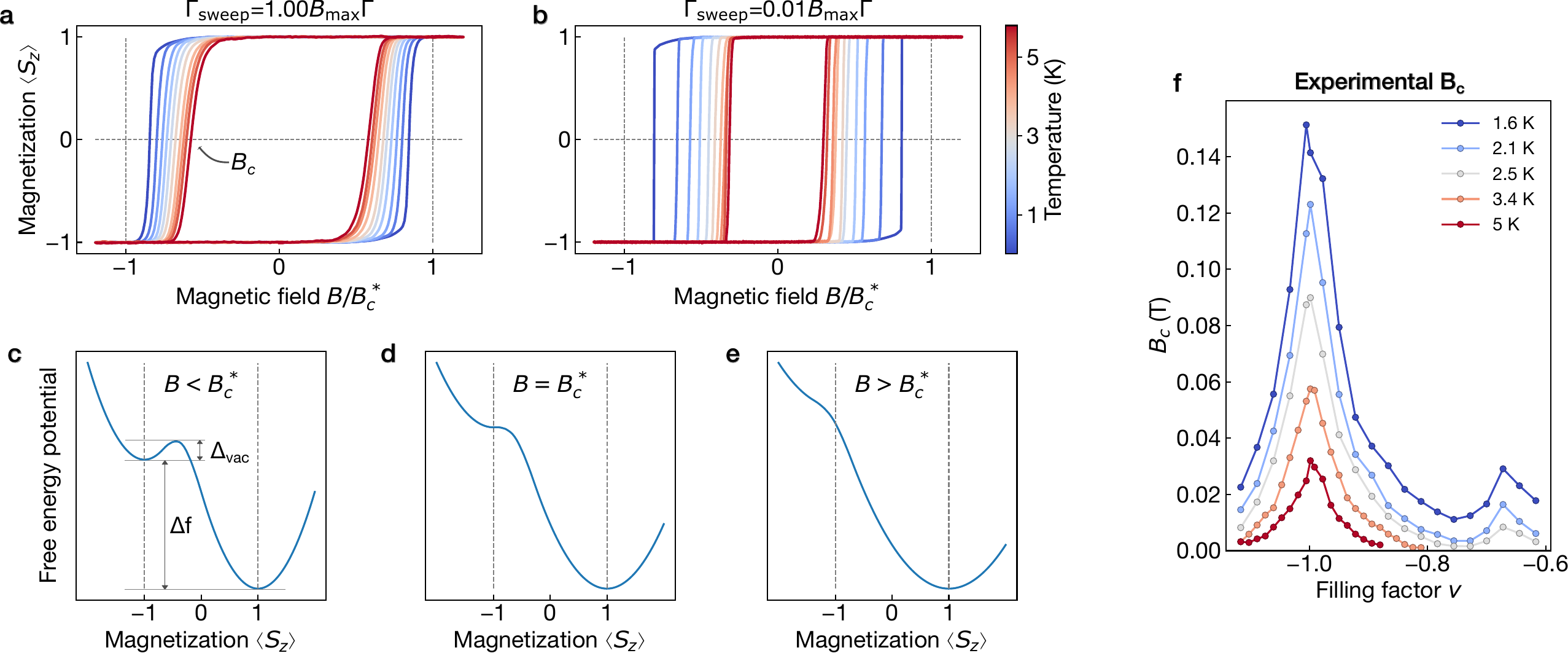}
	\caption{{\bf Coercive field dependence on sweep rate and temperature.} (\textbf{a-b}) Theoretically simulated hysteresis loops for a fast ({\bf a}) and slow {(\bf b)} constant $B$-field sweep rate, and different temperatures at $\nu=-0.8$. Following the experimental procedure, the temperature- and sweep-rate-dependent coercive field $B_\mathrm{c}$ is defined as the $B$-value where the magnetization crosses zero. ({\bf c-e}) Free energy potential $V[m]$, calculated with Eq.~\eqref{eq:FreeEnergyPotential}, for three different magnetic field values. For small magnetic fields $B<B_\mathrm{c}^*$, there are two energy minima corresponding to true- and false-vacuum states with free-energy difference $\Delta f$. The minima are separated by an energy barrier $\Delta_\mathrm{vac}$ ({\bf c}). The second minimum of the false-vacuum state disappears at $B=B_\mathrm{c}^*$ ({\bf d}), leaving only a single global minimum of a true-vacuum state ({\bf e}) for $B>B_\mathrm{c}^*$. ({\bf f}) Experimentally determined values of $B_\mathrm{c}$ for different temperatures as a function of the filling factor.
    \label{fig:method_coercivefield_theory}}
\end{figure*}

Before discussing the relaxation dynamics, we clarify our definition of the coercive field, which plays a crucial role in our discussion of both the experimental results and theoretical simulations. 
At zero magnetic field, the free energy potential has two equivalent minima at $\langle S_z \rangle = \pm 1$. A finite magnetic field $B$ introduces an energy shift between them. The spin orientation favored by the magnetic field becomes the true-vacuum, while the disfavored orientation is a metastable false-vacuum state. If the external magnetic field is increased sufficiently far, the local minimum corresponding to the false-vacuum vanishes, leaving only a single, global minimum in the free energy potential. We denote the critical $B$-field strength at which the second minimum disappears as $B_\mathrm{c}^*$; see Extended Data Fig.~\ref{fig:method_coercivefield_theory}\textbf{c}-\textbf{e}. In general, $B_\mathrm{c}^*$ is different from the experimentally extracted coercive field $B_\mathrm{c}$, which depends both on the temperature and the sweep rate, as discussed in Methods Sec.~\ref{sec:coercive_field_determination}. To gain a better theoretical understanding of how $B_\mathrm{c}$ and $B_\mathrm{c}^*$ are related in our system, we simulate a magnetic hysteresis experiment with the model described by Eq.~\eqref{eq:model_a} as follows: We initialize the state at $t=0$ in a uniformly fully spin-down-polarized state and then increase the magnetic field $B=B(t)$ with a constant rate $\Gamma_\mathrm{sweep}$ up to a maximum field $B_\mathrm{max} = 1.2 B_\mathrm{c}^*$, updating the potential $V[m]$ at each time step. Then, we decrease the magnetic field at the same rate until we reach $-B_\mathrm{max}$. Similarly as in our experiments, the coercive field $B_\mathrm{c}$ is defined as the field value, at which the coercive loop crosses zero magnetization; see Extended Data Fig.~\ref{fig:method_coercivefield_theory}\textbf{a},\textbf{b}. We generally find $B_\mathrm{c} < B_\mathrm{c}^*$, with a strong temperature and sweep rate dependence, qualitatively consistent with the experimentally extracted temperature dependence of $B_\mathrm{c}$ (see Extended Data Fig.~\ref{fig:method_coercivefield_theory}\textbf{f} and Methods Sec.~\ref{sec:coercive_field_determination}). A higher temperature and a slower sweep rate lead to a smaller $B_\mathrm{c}$, for fixed $B_\mathrm{c}^*$. Physically, this is because a slower sweep rate leaves more time for thermal fluctuations to flip the system into the true-vacuum. And higher temperatures increase thermal fluctuations, making such flips more likely to begin with. 

Given that the experimental hysteresis loops are measured at a comparatively slow sweep rate, the above analysis indicates that the experimental $B_\mathrm{c}$ values obtained at $T=1.6$~K are smaller than $B_\mathrm{c}^*$. As a result, $B>B_\mathrm{c}$ does not imply $B>B_\mathrm{c}^*$; thus, for $B$-fields only slightly exceeding $B_\mathrm{c}$, a metastable local minimum remains in the free-energy potential.

\vspace{-0.3cm}
\subsection{Theory: Out-of-equilibrium dynamics \label{sec:theory_dynamics}}\vspace{-0.1cm}
We simulate the spin-relaxation experiment by initializing at time $t=0$ a magnetic domain of diameter $D_0$ and magnetization $m=-1$ opposite to the background magnetization, which is energetically favored by the magnetic field $B>0$. We normalize the magnetization such that $m=\pm1$ corresponds to the fully polarized state. We then use Eq.~\eqref{eq:model_a} to calculate the temporal spin evolution of the system and compute the average magnetization within a central disc of diameter $D_{\mathrm{meas}}$ as a function of time. In general, $D_0$ and $D_{\mathrm{meas}}$ are different, reflecting the difference in the optical spot used to prepare the initial state and the spot used to detect the magnetization (see Fig.~\ref{fig:Fig_spatial}\textbf{c} and Methods Sec.~\ref{sec:profile_dependence_on_area_and_power}). In our non-equilibrium simulations, we include $N_x\times N_y = 200 \times 200$ unit cells and fix $D_\mathrm{meas} = 80$ unit cells. First, we consider the case close to the coercive field $B \lesssim B^*_\mathrm{c}$, where there still exists a second local minimum in the free energy potential (false-vacuum), separated from the global minimum (true-vacuum) by an energy difference $\Delta f$. Note that while we assume $B < B_\mathrm{c}^*$, the magnetic field $B$ might already be above the experimentally measured coercive field $B_\mathrm{c}$; see Methods Sec.~\ref{sec:theory_coercive_field}. 

As discussed in the main text, there are two general mechanisms for the spin relaxation: thermal fluctuations and domain shrinking. Thermal fluctuations can lead to a local magnetization flip by overcoming the activation barrier $\Delta_\mathrm{vac}$ that separates the false vacuum from the true vacuum. They are exponentially suppressed at low temperatures, leading to a relaxation rate $\Gamma_\mathrm{thermal} \sim \Gamma e^{-\frac{\Delta_\mathrm{vac}}{k_\mathrm{B} T}}$. The thermal relaxation rate can be increased either by increasing the temperature or by increasing the external magnetic field, which reduces the activation gap $\Delta_\mathrm{vac}$. While relaxation mediated by thermal fluctuations is generic and applies equally to the metallic and the ICI phases, the second relaxation mechanism, namely domain shrinking, depends sensitively on the spin-stiffness of the underlying state: Close to the coercive field, the radius $R(t) = R_0 - v_B t$ of a magnetic domain in the false-vacuum will shrink with a velocity $v_B = \Gamma \Delta f \kappa_2/\sigma $, where $\sigma \propto \sqrt{\kappa_2}$ is the surface tension at magnetic domain walls, assuming $\kappa_2>0$~\cite{pichler2026}. Importantly, the shrinking velocity depends sensitively on the stiffness $v_B \propto B_z \sqrt{\kappa_2}$. 
At sufficiently low temperatures $k_B T \ll \Delta_\mathrm{vac}$, relaxation through shrinking dominates over thermal relaxation, which is exponentially suppressed with inverse temperature. This explains the much faster experimental relaxation rate of the metal compared to that of the Chern insulator, as the stiffness in the metal is much larger than in the Chern insulator. 


%
\vspace{-0.3cm}
\subsection{Theory: Stability of magnetic domains at $B=0$ \label{sec:theory_domain_stability}}\vspace{-0.1cm}
For our discussion of Fig.~\ref{fig:Fig_domains}, we perform simulations at zero magnetic field $B=0$. Here, the two states with opposite spin orientations have the same energy, and for the metal we expect to observe coarsening dynamics, governed by the Allen-Cahn equation~\cite{bray1994theory}:
\begin{equation}
    \dot{R}(t) = -\frac{\Gamma \kappa_2}{R(t)}. \label{eq:allen_cahn}
\end{equation}
In this regime, the shrinking velocity scales inversely with the domain radius. Consequently, small initial domains quickly disappear, while large initial domains are stable over very long times. 
To simulate the effect of domain initialization with increasing excitation power, we note that optical spin orientation in $t$-MoTe$_2$~\cite{holtzmann2026optical,huber2026optical,cai2026optical}, regardless of its precise microscopic mechanism, must proceed through local spin-flip events. The resulting randomly distributed holes with optically inverted spins then form small seed domains, which eventually merge into a continuous domain at sufficiently high power. To provide a theoretical proxy for this process, we use the following protocol: we initialize our state by creating a fixed number $N_\mathrm{bub}$ of small seed bubbles with random radius $R_0 \in [5, 10]$ unit cells, placed randomly (with uniform distribution) within a ring-shaped area, emulating excitation with a doughnut-shaped beam. We use $R_\mathrm{inner} = 20$ and $R_\mathrm{outer} = 60$ unit cells for the boundaries of the ring-shaped area. Increased excitation power is simulated by increasing $N_\mathrm{bub}$. We then compute the quasi-steady-state spatial magnetization profiles at a very long time $t=t_\mathrm{meas}=116\Gamma^{-1}$. As shown in Fig.~\ref{fig:Fig_domains}\textbf{a} and discussed in the main text, for the ICI, we find that the state remains frozen in its initial configuration, independent of $N_\mathrm{bub}$, and retains the fine structure of the small initial bubbles. Consequently, the average magnetization of the system, measured at $t_\mathrm{meas}$, increases roughly linearly with $N_\mathrm{bub}$, until it saturates at large $N_\mathrm{bub}$. In contrast, for the metal, coarsening dynamics leads to a quick decay of the small initial bubbles if their density is too low (small $N_\mathrm{bub}$), until a critical $N_\mathrm{bub}$ is reached, at which point the small bubbles merge into one big domain, which is stable over very long times. Hence, the average magnetization measured at $t_\mathrm{meas}$ is zero for small (but finite) $N_\mathrm{bub}$, until it suddenly jumps at a finite threshold value. 

We remark that in the ferromagnetic metal, even the very large domains will eventually disappear in our simple theoretical simulation, for $t\gg t_\mathrm{meas}$. We expect that in the experiment, disorder can provide a pinning potential, which stabilizes large domains even at very late times. Such a pinning potential must compensate for the curvature-induced pressure leading to the coarsening dynamics of Eq.~\eqref{eq:allen_cahn}. For large domains, with a correspondingly small curvature, small pinning is already sufficient for stabilization. 

For the theoretical magnetization maps plotted in Fig.~\ref{fig:Fig_spatial}\textbf{f},\textbf{g} and Fig.~\ref{fig:Fig_domains}\textbf{a}, a Gaussian filter with a standard deviation of $\sigma=2$ unit cells (corresponding to $\approx 10$ nm for the twist angle considered in the main text) was applied to the raw simulation data for visual clarity.

%


\begin{thebibliography}{63}%
\makeatletter
\providecommand \@ifxundefined [1]{%
 \@ifx{#1\undefined}
}%
\providecommand \@ifnum [1]{%
 \ifnum #1\expandafter \@firstoftwo
 \else \expandafter \@secondoftwo
 \fi
}%
\providecommand \@ifx [1]{%
 \ifx #1\expandafter \@firstoftwo
 \else \expandafter \@secondoftwo
 \fi
}%
\providecommand \natexlab [1]{#1}%
\providecommand \enquote  [1]{``#1''}%
\providecommand \bibnamefont  [1]{#1}%
\providecommand \bibfnamefont [1]{#1}%
\providecommand \citenamefont [1]{#1}%
\providecommand \href@noop [0]{\@secondoftwo}%
\providecommand \href [0]{\begingroup \@sanitize@url \@href}%
\providecommand \@href[1]{\@@startlink{#1}\@@href}%
\providecommand \@@href[1]{\endgroup#1\@@endlink}%
\providecommand \@sanitize@url [0]{\catcode `\\12\catcode `\$12\catcode
  `\&12\catcode `\#12\catcode `\^12\catcode `\_12\catcode `\%12\relax}%
\providecommand \@@startlink[1]{}%
\providecommand \@@endlink[0]{}%
\providecommand \url  [0]{\begingroup\@sanitize@url \@url }%
\providecommand \@url [1]{\endgroup\@href {#1}{\urlprefix }}%
\providecommand \urlprefix  [0]{URL }%
\providecommand \Eprint [0]{\href }%
\providecommand \doibase [0]{http://dx.doi.org/}%
\providecommand \selectlanguage [0]{\@gobble}%
\providecommand \bibinfo  [0]{\@secondoftwo}%
\providecommand \bibfield  [0]{\@secondoftwo}%
\providecommand \translation [1]{[#1]}%
\providecommand \BibitemOpen [0]{}%
\providecommand \bibitemStop [0]{}%
\providecommand \bibitemNoStop [0]{.\EOS\space}%
\providecommand \EOS [0]{\spacefactor3000\relax}%
\providecommand \BibitemShut  [1]{\csname bibitem#1\endcsname}%
\let\auto@bib@innerbib\@empty
\bibitem [{\citenamefont {Anderson}\ \emph {et~al.}(2023)\citenamefont
  {Anderson}, \citenamefont {Fan}, \citenamefont {Cai}, \citenamefont
  {Holtzmann}, \citenamefont {Taniguchi}, \citenamefont {Watanabe},
  \citenamefont {Xiao}, \citenamefont {Yao},\ and\ \citenamefont
  {Xu}}]{anderson2023}%
  \BibitemOpen
  \bibinfo {author} {\bibfnamefont {E.}~\bibnamefont {Anderson}}, \bibinfo
  {author} {\bibfnamefont {F.-R.}\ \bibnamefont {Fan}}, \bibinfo {author}
  {\bibfnamefont {J.}~\bibnamefont {Cai}}, \bibinfo {author} {\bibfnamefont
  {W.}~\bibnamefont {Holtzmann}}, \bibinfo {author} {\bibfnamefont
  {T.}~\bibnamefont {Taniguchi}}, \bibinfo {author} {\bibfnamefont
  {K.}~\bibnamefont {Watanabe}}, \bibinfo {author} {\bibfnamefont
  {D.}~\bibnamefont {Xiao}}, \bibinfo {author} {\bibfnamefont {W.}~\bibnamefont
  {Yao}},\ \bibnamefont {and}\ \bibinfo {author} {\bibfnamefont
  {X.}~\bibnamefont {Xu}}.\EOS\
\newblock \emph {\bibinfo {title} {{Programming correlated magnetic states with
  gate-controlled moir\'e geometry}}}.\EOS\
\newblock \href {\doibase 10.1126/science.adg4268} {\bibfield  {journal} {\emph
  {\bibinfo  {journal} {Science}}\ }\textbf {\bibinfo {volume} {381}}, \bibinfo
  {pages} {325--330} (\bibinfo {year} {2023})}\BibitemShut {NoStop}%
\bibitem [{\citenamefont {Cai}\ \emph {et~al.}(2023)\citenamefont {Cai},
  \citenamefont {Anderson}, \citenamefont {Wang}, \citenamefont {Zhang},
  \citenamefont {Liu}, \citenamefont {Holtzmann}, \citenamefont {Zhang},
  \citenamefont {Fan}, \citenamefont {Taniguchi}, \citenamefont {Watanabe},
  \citenamefont {Ran}, \citenamefont {Cao}, \citenamefont {Fu}, \citenamefont
  {Xiao}, \citenamefont {Yao},\ and\ \citenamefont {Xu}}]{cai2023}%
  \BibitemOpen
  \bibinfo {author} {\bibfnamefont {J.}~\bibnamefont {Cai}}, \bibinfo {author}
  {\bibfnamefont {E.}~\bibnamefont {Anderson}}, \bibinfo {author}
  {\bibfnamefont {C.}~\bibnamefont {Wang}}, \bibinfo {author} {\bibfnamefont
  {X.}~\bibnamefont {Zhang}}, \bibinfo {author} {\bibfnamefont
  {X.}~\bibnamefont {Liu}}, \bibinfo {author} {\bibfnamefont {W.}~\bibnamefont
  {Holtzmann}}, \bibinfo {author} {\bibfnamefont {Y.}~\bibnamefont {Zhang}},
  \bibinfo {author} {\bibfnamefont {F.}~\bibnamefont {Fan}}, \bibinfo {author}
  {\bibfnamefont {T.}~\bibnamefont {Taniguchi}}, \bibinfo {author}
  {\bibfnamefont {K.}~\bibnamefont {Watanabe}}, \bibinfo {author}
  {\bibfnamefont {Y.}~\bibnamefont {Ran}}, \bibinfo {author} {\bibfnamefont
  {T.}~\bibnamefont {Cao}}, \bibinfo {author} {\bibfnamefont {L.}~\bibnamefont
  {Fu}}, \bibinfo {author} {\bibfnamefont {D.}~\bibnamefont {Xiao}}, \bibinfo
  {author} {\bibfnamefont {W.}~\bibnamefont {Yao}},\ \bibnamefont {and}\
  \bibinfo {author} {\bibfnamefont {X.}~\bibnamefont {Xu}}.\EOS\
\newblock \emph {\bibinfo {title} {{Signatures of Fractional Quantum Anomalous
  Hall States in Twisted MoTe$_2$}}}.\EOS\
\newblock \href {\doibase 10.1038/s41586-023-06289-w} {\bibfield  {journal}
  {\emph {\bibinfo  {journal} {Nature}}\ }\textbf {\bibinfo {volume} {622}},
  \bibinfo {pages} {63--68} (\bibinfo {year} {2023})}\BibitemShut {NoStop}%
\bibitem [{\citenamefont {Zeng}\ \emph {et~al.}(2023)\citenamefont {Zeng},
  \citenamefont {Xia}, \citenamefont {Kang}, \citenamefont {Zhu}, \citenamefont
  {Kn{\"u}ppel}, \citenamefont {Vaswani}, \citenamefont {Watanabe},
  \citenamefont {Taniguchi}, \citenamefont {Mak},\ and\ \citenamefont
  {Shan}}]{zeng2023}%
  \BibitemOpen
  \bibinfo {author} {\bibfnamefont {Y.}~\bibnamefont {Zeng}}, \bibinfo {author}
  {\bibfnamefont {Z.}~\bibnamefont {Xia}}, \bibinfo {author} {\bibfnamefont
  {K.}~\bibnamefont {Kang}}, \bibinfo {author} {\bibfnamefont {J.}~\bibnamefont
  {Zhu}}, \bibinfo {author} {\bibfnamefont {P.}~\bibnamefont {Kn{\"u}ppel}},
  \bibinfo {author} {\bibfnamefont {C.}~\bibnamefont {Vaswani}}, \bibinfo
  {author} {\bibfnamefont {K.}~\bibnamefont {Watanabe}}, \bibinfo {author}
  {\bibfnamefont {T.}~\bibnamefont {Taniguchi}}, \bibinfo {author}
  {\bibfnamefont {K.~F.}\ \bibnamefont {Mak}},\ \bibnamefont {and}\ \bibinfo
  {author} {\bibfnamefont {J.}~\bibnamefont {Shan}}.\EOS\
\newblock \emph {\bibinfo {title} {{Thermodynamic Evidence of Fractional Chern
  Insulator in Moir\'e MoTe$_2$}}}.\EOS\
\newblock \href {\doibase 10.1038/s41586-023-06452-3} {\bibfield  {journal}
  {\emph {\bibinfo  {journal} {Nature}}\ }\textbf {\bibinfo {volume} {622}},
  \bibinfo {pages} {69--73} (\bibinfo {year} {2023})}\BibitemShut {NoStop}%
\bibitem [{\citenamefont {Xu}\ \emph {et~al.}(2023)\citenamefont {Xu},
  \citenamefont {Sun}, \citenamefont {Jia}, \citenamefont {Liu}, \citenamefont
  {Xu}, \citenamefont {Li}, \citenamefont {Gu}, \citenamefont {Watanabe},
  \citenamefont {Taniguchi}, \citenamefont {Tong}, \citenamefont {Jia},
  \citenamefont {Shi}, \citenamefont {Jiang}, \citenamefont {Zhang},
  \citenamefont {Liu},\ and\ \citenamefont {Li}}]{xu2023observation}%
  \BibitemOpen
  \bibinfo {author} {\bibfnamefont {F.}~\bibnamefont {Xu}}, \bibinfo {author}
  {\bibfnamefont {Z.}~\bibnamefont {Sun}}, \bibinfo {author} {\bibfnamefont
  {T.}~\bibnamefont {Jia}}, \bibinfo {author} {\bibfnamefont {C.}~\bibnamefont
  {Liu}}, \bibinfo {author} {\bibfnamefont {C.}~\bibnamefont {Xu}}, \bibinfo
  {author} {\bibfnamefont {C.}~\bibnamefont {Li}}, \bibinfo {author}
  {\bibfnamefont {Y.}~\bibnamefont {Gu}}, \bibinfo {author} {\bibfnamefont
  {K.}~\bibnamefont {Watanabe}}, \bibinfo {author} {\bibfnamefont
  {T.}~\bibnamefont {Taniguchi}}, \bibinfo {author} {\bibfnamefont
  {B.}~\bibnamefont {Tong}}, \bibinfo {author} {\bibfnamefont {J.}~\bibnamefont
  {Jia}}, \bibinfo {author} {\bibfnamefont {Z.}~\bibnamefont {Shi}}, \bibinfo
  {author} {\bibfnamefont {S.}~\bibnamefont {Jiang}}, \bibinfo {author}
  {\bibfnamefont {Y.}~\bibnamefont {Zhang}}, \bibinfo {author} {\bibfnamefont
  {X.}~\bibnamefont {Liu}},\ \bibnamefont {and}\ \bibinfo {author}
  {\bibfnamefont {T.}~\bibnamefont {Li}}.\EOS\
\newblock \emph {\bibinfo {title} {{Observation of Integer and Fractional
  Quantum Anomalous Hall Effects in Twisted Bilayer {MoTe}$_{2}$}}}.\EOS\
\newblock \href {\doibase 10.1103/PhysRevX.13.031037} {\bibfield  {journal}
  {\emph {\bibinfo  {journal} {Phys. Rev. X}}\ }\textbf {\bibinfo {volume}
  {13}}, \bibinfo {pages} {031037} (\bibinfo {year} {2023})}\BibitemShut
  {NoStop}%
\bibitem [{\citenamefont {Park}\ \emph {et~al.}(2023)\citenamefont {Park},
  \citenamefont {Cai}, \citenamefont {Anderson}, \citenamefont {Zhang},
  \citenamefont {Zhu}, \citenamefont {Liu}, \citenamefont {Wang}, \citenamefont
  {Holtzmann}, \citenamefont {Hu}, \citenamefont {Liu}, \citenamefont
  {Taniguchi}, \citenamefont {Watanabe}, \citenamefont {Chu}, \citenamefont
  {Cao}, \citenamefont {Fu}, \citenamefont {Yao}, \citenamefont {Chang},
  \citenamefont {Cobden}, \citenamefont {Xiao},\ and\ \citenamefont
  {Xu}}]{Park2023}%
  \BibitemOpen
  \bibinfo {author} {\bibfnamefont {H.}~\bibnamefont {Park}}, \bibinfo {author}
  {\bibfnamefont {J.}~\bibnamefont {Cai}}, \bibinfo {author} {\bibfnamefont
  {E.}~\bibnamefont {Anderson}}, \bibinfo {author} {\bibfnamefont
  {Y.}~\bibnamefont {Zhang}}, \bibinfo {author} {\bibfnamefont
  {J.}~\bibnamefont {Zhu}}, \bibinfo {author} {\bibfnamefont {X.}~\bibnamefont
  {Liu}}, \bibinfo {author} {\bibfnamefont {C.}~\bibnamefont {Wang}}, \bibinfo
  {author} {\bibfnamefont {W.}~\bibnamefont {Holtzmann}}, \bibinfo {author}
  {\bibfnamefont {C.}~\bibnamefont {Hu}}, \bibinfo {author} {\bibfnamefont
  {Z.}~\bibnamefont {Liu}}, \bibinfo {author} {\bibfnamefont {T.}~\bibnamefont
  {Taniguchi}}, \bibinfo {author} {\bibfnamefont {K.}~\bibnamefont {Watanabe}},
  \bibinfo {author} {\bibfnamefont {J.-H.}\ \bibnamefont {Chu}}, \bibinfo
  {author} {\bibfnamefont {T.}~\bibnamefont {Cao}}, \bibinfo {author}
  {\bibfnamefont {L.}~\bibnamefont {Fu}}, \bibinfo {author} {\bibfnamefont
  {W.}~\bibnamefont {Yao}}, \bibinfo {author} {\bibfnamefont {C.-Z.}\
  \bibnamefont {Chang}}, \bibinfo {author} {\bibfnamefont {D.}~\bibnamefont
  {Cobden}}, \bibinfo {author} {\bibfnamefont {D.}~\bibnamefont {Xiao}},\
  \bibnamefont {and}\ \bibinfo {author} {\bibfnamefont {X.}~\bibnamefont
  {Xu}}.\EOS\
\newblock \emph {\bibinfo {title} {{Observation of fractionally quantized
  anomalous Hall effect}}}.\EOS\
\newblock \href {\doibase 10.1038/s41586-023-06536-0} {\bibfield  {journal}
  {\emph {\bibinfo  {journal} {Nature}}\ }\textbf {\bibinfo {volume} {622}},
  \bibinfo {pages} {74--79} (\bibinfo {year} {2023})}\BibitemShut {NoStop}%
\bibitem [{\citenamefont {Li}\ \emph {et~al.}(2025)\citenamefont {Li},
  \citenamefont {Redekop}, \citenamefont {Wang~Beach}, \citenamefont {Zhang},
  \citenamefont {Zhang}, \citenamefont {Liu}, \citenamefont {Holtzmann},
  \citenamefont {Hu}, \citenamefont {Anderson}, \citenamefont {Park},
  \citenamefont {Taniguchi}, \citenamefont {Watanabe}, \citenamefont {Chu},
  \citenamefont {Fu}, \citenamefont {Cao}, \citenamefont {Xiao}, \citenamefont
  {Young},\ and\ \citenamefont {Xu}}]{li2025}%
  \BibitemOpen
  \bibinfo {author} {\bibfnamefont {W.}~\bibnamefont {Li}}, \bibinfo {author}
  {\bibfnamefont {E.}~\bibnamefont {Redekop}}, \bibinfo {author} {\bibfnamefont
  {C.}~\bibnamefont {Wang~Beach}}, \bibinfo {author} {\bibfnamefont
  {C.}~\bibnamefont {Zhang}}, \bibinfo {author} {\bibfnamefont
  {X.}~\bibnamefont {Zhang}}, \bibinfo {author} {\bibfnamefont
  {X.}~\bibnamefont {Liu}}, \bibinfo {author} {\bibfnamefont {W.}~\bibnamefont
  {Holtzmann}}, \bibinfo {author} {\bibfnamefont {C.}~\bibnamefont {Hu}},
  \bibinfo {author} {\bibfnamefont {E.}~\bibnamefont {Anderson}}, \bibinfo
  {author} {\bibfnamefont {H.}~\bibnamefont {Park}}, \bibinfo {author}
  {\bibfnamefont {T.}~\bibnamefont {Taniguchi}}, \bibinfo {author}
  {\bibfnamefont {K.}~\bibnamefont {Watanabe}}, \bibinfo {author}
  {\bibfnamefont {J.-h.}\ \bibnamefont {Chu}}, \bibinfo {author} {\bibfnamefont
  {L.}~\bibnamefont {Fu}}, \bibinfo {author} {\bibfnamefont {T.}~\bibnamefont
  {Cao}}, \bibinfo {author} {\bibfnamefont {D.}~\bibnamefont {Xiao}}, \bibinfo
  {author} {\bibfnamefont {A.~F.}\ \bibnamefont {Young}},\ \bibnamefont {and}\
  \bibinfo {author} {\bibfnamefont {X.}~\bibnamefont {Xu}}.\EOS\
\newblock \emph {\bibinfo {title} {{Universal Magnetic Phases in Twisted
  Bilayer MoTe$_2$}}}.\EOS\
\newblock \href {\doibase 10.1021/acs.nanolett.5c04751} {\bibfield  {journal}
  {\emph {\bibinfo  {journal} {Nano Lett.}}\ }\textbf {\bibinfo {volume} {25}},
  \bibinfo {pages} {18044--18050} (\bibinfo {year} {2025})}\BibitemShut
  {NoStop}%
\bibitem [{\citenamefont {Ji}\ \emph {et~al.}(2024)\citenamefont {Ji},
  \citenamefont {Park}, \citenamefont {Barber}, \citenamefont {Hu},
  \citenamefont {Watanabe}, \citenamefont {Taniguchi}, \citenamefont {Chu},
  \citenamefont {Xu},\ and\ \citenamefont {Shen}}]{ji2024}%
  \BibitemOpen
  \bibinfo {author} {\bibfnamefont {Z.}~\bibnamefont {Ji}}, \bibinfo {author}
  {\bibfnamefont {H.}~\bibnamefont {Park}}, \bibinfo {author} {\bibfnamefont
  {M.~E.}\ \bibnamefont {Barber}}, \bibinfo {author} {\bibfnamefont
  {C.}~\bibnamefont {Hu}}, \bibinfo {author} {\bibfnamefont {K.}~\bibnamefont
  {Watanabe}}, \bibinfo {author} {\bibfnamefont {T.}~\bibnamefont {Taniguchi}},
  \bibinfo {author} {\bibfnamefont {J.-H.}\ \bibnamefont {Chu}}, \bibinfo
  {author} {\bibfnamefont {X.}~\bibnamefont {Xu}},\ \bibnamefont {and}\
  \bibinfo {author} {\bibfnamefont {Z.-X.}\ \bibnamefont {Shen}}.\EOS\
\newblock \emph {\bibinfo {title} {{Local probe of bulk and edge states in a
  fractional Chern insulator}}}.\EOS\
\newblock \href {\doibase 10.1038/s41586-024-08092-7} {\bibfield  {journal}
  {\emph {\bibinfo  {journal} {Nature}}\ }\textbf {\bibinfo {volume} {635}},
  \bibinfo {pages} {578--583} (\bibinfo {year} {2024})}\BibitemShut {NoStop}%
\bibitem [{\citenamefont {Redekop}\ \emph {et~al.}(2024)\citenamefont
  {Redekop}, \citenamefont {Zhang}, \citenamefont {Park}, \citenamefont {Cai},
  \citenamefont {Anderson}, \citenamefont {Sheekey}, \citenamefont {Arp},
  \citenamefont {Babikyan}, \citenamefont {Salters}, \citenamefont {Watanabe},
  \citenamefont {Taniguchi}, \citenamefont {Huber}, \citenamefont {Xu},\ and\
  \citenamefont {Young}}]{redekop2024}%
  \BibitemOpen
  \bibinfo {author} {\bibfnamefont {E.}~\bibnamefont {Redekop}}, \bibinfo
  {author} {\bibfnamefont {C.}~\bibnamefont {Zhang}}, \bibinfo {author}
  {\bibfnamefont {H.}~\bibnamefont {Park}}, \bibinfo {author} {\bibfnamefont
  {J.}~\bibnamefont {Cai}}, \bibinfo {author} {\bibfnamefont {E.}~\bibnamefont
  {Anderson}}, \bibinfo {author} {\bibfnamefont {O.}~\bibnamefont {Sheekey}},
  \bibinfo {author} {\bibfnamefont {T.}~\bibnamefont {Arp}}, \bibinfo {author}
  {\bibfnamefont {G.}~\bibnamefont {Babikyan}}, \bibinfo {author}
  {\bibfnamefont {S.}~\bibnamefont {Salters}}, \bibinfo {author} {\bibfnamefont
  {K.}~\bibnamefont {Watanabe}}, \bibinfo {author} {\bibfnamefont
  {T.}~\bibnamefont {Taniguchi}}, \bibinfo {author} {\bibfnamefont {M.~E.}\
  \bibnamefont {Huber}}, \bibinfo {author} {\bibfnamefont {X.}~\bibnamefont
  {Xu}},\ \bibnamefont {and}\ \bibinfo {author} {\bibfnamefont {A.~F.}\
  \bibnamefont {Young}}.\EOS\
\newblock \emph {\bibinfo {title} {{Direct magnetic imaging of fractional Chern
  insulators in twisted MoTe$_2$}}}.\EOS\
\newblock \href {\doibase 10.1038/s41586-024-08153-x} {\bibfield  {journal}
  {\emph {\bibinfo  {journal} {Nature}}\ }\textbf {\bibinfo {volume} {635}},
  \bibinfo {pages} {584--589} (\bibinfo {year} {2024})}\BibitemShut {NoStop}%
\bibitem [{\citenamefont {Jia}\ \emph {et~al.}(2025)\citenamefont {Jia},
  \citenamefont {Song}, \citenamefont {Zheng}, \citenamefont {Cheng},
  \citenamefont {Uzan}, \citenamefont {Yu}, \citenamefont {Tang}, \citenamefont
  {Pollak}, \citenamefont {Onyszczak}, \citenamefont {Watanabe}, \citenamefont
  {Taniguchi}, \citenamefont {Lei}, \citenamefont {Yao}, \citenamefont
  {Schoop}, \citenamefont {Ong},\ and\ \citenamefont {Wu}}]{Jia2026}%
  \BibitemOpen
  \bibinfo {author} {\bibfnamefont {Y.}~\bibnamefont {Jia}}, \bibinfo {author}
  {\bibfnamefont {T.}~\bibnamefont {Song}}, \bibinfo {author} {\bibfnamefont
  {Z.~J.}\ \bibnamefont {Zheng}}, \bibinfo {author} {\bibfnamefont
  {G.}~\bibnamefont {Cheng}}, \bibinfo {author} {\bibfnamefont {A.~J.}\
  \bibnamefont {Uzan}}, \bibinfo {author} {\bibfnamefont {G.}~\bibnamefont
  {Yu}}, \bibinfo {author} {\bibfnamefont {Y.}~\bibnamefont {Tang}}, \bibinfo
  {author} {\bibfnamefont {F.}~\bibnamefont {Pollak}, \bibfnamefont {Connor
  J.and~Yuan}}, \bibinfo {author} {\bibfnamefont {M.}~\bibnamefont
  {Onyszczak}}, \bibinfo {author} {\bibfnamefont {K.}~\bibnamefont {Watanabe}},
  \bibinfo {author} {\bibfnamefont {T.}~\bibnamefont {Taniguchi}}, \bibinfo
  {author} {\bibfnamefont {S.}~\bibnamefont {Lei}}, \bibinfo {author}
  {\bibfnamefont {N.}~\bibnamefont {Yao}}, \bibinfo {author} {\bibfnamefont
  {L.~M.}\ \bibnamefont {Schoop}}, \bibinfo {author} {\bibfnamefont {N.~P.}\
  \bibnamefont {Ong}},\ \bibnamefont {and}\ \bibinfo {author} {\bibfnamefont
  {S.}~\bibnamefont {Wu}}.\EOS\
\newblock \emph {\bibinfo {title} {{Anomalous superconductivity in twisted
  MoTe$_2$ nanojunctions}}}.\EOS\
\newblock \href {\doibase 10.1126/sciadv.adq5712} {\bibfield  {journal} {\emph
  {\bibinfo  {journal} {Sci. Adv.}}\ }\textbf {\bibinfo {volume} {11}},
  \bibinfo {pages} {eadq5712} (\bibinfo {year} {2025})}\BibitemShut {NoStop}%
\bibitem [{\citenamefont {Pan}\ \emph {et~al.}(2026)\citenamefont {Pan},
  \citenamefont {Yang}, \citenamefont {Wang}, \citenamefont {Cai},
  \citenamefont {Wang}, \citenamefont {Zhao}, \citenamefont {Watanabe},
  \citenamefont {Taniguchi}, \citenamefont {Zhang}, \citenamefont {Liu},
  \citenamefont {Yang},\ and\ \citenamefont {Gao}}]{Pan2026}%
  \BibitemOpen
  \bibinfo {author} {\bibfnamefont {H.}~\bibnamefont {Pan}}, \bibinfo {author}
  {\bibfnamefont {S.}~\bibnamefont {Yang}}, \bibinfo {author} {\bibfnamefont
  {Y.}~\bibnamefont {Wang}}, \bibinfo {author} {\bibfnamefont {X.}~\bibnamefont
  {Cai}}, \bibinfo {author} {\bibfnamefont {W.}~\bibnamefont {Wang}}, \bibinfo
  {author} {\bibfnamefont {Y.}~\bibnamefont {Zhao}}, \bibinfo {author}
  {\bibfnamefont {K.}~\bibnamefont {Watanabe}}, \bibinfo {author}
  {\bibfnamefont {T.}~\bibnamefont {Taniguchi}}, \bibinfo {author}
  {\bibfnamefont {L.}~\bibnamefont {Zhang}}, \bibinfo {author} {\bibfnamefont
  {Y.}~\bibnamefont {Liu}}, \bibinfo {author} {\bibfnamefont {B.}~\bibnamefont
  {Yang}},\ \bibnamefont {and}\ \bibinfo {author} {\bibfnamefont
  {W.}~\bibnamefont {Gao}}.\EOS\
\newblock \emph {\bibinfo {title} {{Optical Signatures of
  $\ensuremath{-}\frac{1}{3}$ Fractional Quantum Anomalous Hall State in
  Twisted ${\mathrm{MoTe}}_{2}$}}}.\EOS\
\newblock \href {\doibase 10.1103/f4dj-7sts} {\bibfield  {journal} {\emph
  {\bibinfo  {journal} {Phys. Rev. Lett.}}\ }\textbf {\bibinfo {volume} {136}},
  \bibinfo {pages} {056601} (\bibinfo {year} {2026})}\BibitemShut {NoStop}%
\bibitem [{\citenamefont {Xu}\ \emph {et~al.}(2025)\citenamefont {Xu},
  \citenamefont {Sun}, \citenamefont {Li}, \citenamefont {Zheng}, \citenamefont
  {Xu}, \citenamefont {Gao}, \citenamefont {Jia}, \citenamefont {Su},
  \citenamefont {Watanabe}, \citenamefont {Taniguchi}, \citenamefont {Tong},
  \citenamefont {Lu}, \citenamefont {Jia}, \citenamefont {Shi}, \citenamefont
  {Jiang}, \citenamefont {Lin}, \citenamefont {Zhang}, \citenamefont {Zhang},
  \citenamefont {Lei}, \citenamefont {Liu},\ and\ \citenamefont
  {Li}}]{xu2025SC}%
  \BibitemOpen
  \bibinfo {author} {\bibfnamefont {F.}~\bibnamefont {Xu}}, \bibinfo {author}
  {\bibfnamefont {Z.}~\bibnamefont {Sun}}, \bibinfo {author} {\bibfnamefont
  {J.}~\bibnamefont {Li}}, \bibinfo {author} {\bibfnamefont {C.}~\bibnamefont
  {Zheng}}, \bibinfo {author} {\bibfnamefont {C.}~\bibnamefont {Xu}}, \bibinfo
  {author} {\bibfnamefont {J.}~\bibnamefont {Gao}}, \bibinfo {author}
  {\bibfnamefont {T.}~\bibnamefont {Jia}}, \bibinfo {author} {\bibfnamefont
  {Y.}~\bibnamefont {Su}}, \bibinfo {author} {\bibfnamefont {K.}~\bibnamefont
  {Watanabe}}, \bibinfo {author} {\bibfnamefont {T.}~\bibnamefont {Taniguchi}},
  \bibinfo {author} {\bibfnamefont {B.}~\bibnamefont {Tong}}, \bibinfo {author}
  {\bibfnamefont {L.}~\bibnamefont {Lu}}, \bibinfo {author} {\bibfnamefont
  {J.}~\bibnamefont {Jia}}, \bibinfo {author} {\bibfnamefont {Z.}~\bibnamefont
  {Shi}}, \bibinfo {author} {\bibfnamefont {S.}~\bibnamefont {Jiang}}, \bibinfo
  {author} {\bibfnamefont {J.}~\bibnamefont {Lin}}, \bibinfo {author}
  {\bibfnamefont {Y.}~\bibnamefont {Zhang}}, \bibinfo {author} {\bibfnamefont
  {Y.}~\bibnamefont {Zhang}}, \bibinfo {author} {\bibfnamefont
  {S.}~\bibnamefont {Lei}}, \bibinfo {author} {\bibfnamefont {X.}~\bibnamefont
  {Liu}},\ \bibnamefont {and}\ \bibinfo {author} {\bibfnamefont
  {T.}~\bibnamefont {Li}}.\EOS\
\newblock \emph {\bibinfo {title} {{Signatures of unconventional
  superconductivity near reentrant and fractional quantum anomalous Hall
  insulators}}}.\EOS\
\newblock \href {https://arxiv.org/abs/2504.06972} {\bibfield  {journal} {\emph
  {\bibinfo  {journal} {arXiv:2504.06972}}\ } (\bibinfo {year}
  {2025})}\BibitemShut {NoStop}%
\bibitem [{\citenamefont {Pichler}\ \emph {et~al.}(2026)\citenamefont
  {Pichler}, \citenamefont {Kuhlenkamp},\ and\ \citenamefont
  {Knap}}]{pichler2026}%
  \BibitemOpen
  \bibinfo {author} {\bibfnamefont {F.}~\bibnamefont {Pichler}}, \bibinfo
  {author} {\bibfnamefont {C.}~\bibnamefont {Kuhlenkamp}},\ \bibnamefont {and}\
  \bibinfo {author} {\bibfnamefont {M.}~\bibnamefont {Knap}}.\EOS\
\newblock \emph {\bibinfo {title} {False {{Vacuum Decay}} in {{Flat-Band
  Ferromagnets}}: {{Role}} of {{Quantum Geometry}} and {{Chiral Edge
  States}}}}.\EOS\
\newblock \href {\doibase 10.1103/q7ss-pvcq} {\bibfield  {journal} {\emph
  {\bibinfo  {journal} {Phys. Rev. Lett.}}\ }\textbf {\bibinfo {volume} {136}},
  \bibinfo {pages} {156502} (\bibinfo {year} {2026})}\BibitemShut {NoStop}%
\bibitem [{\citenamefont {Holtzmann}\ \emph {et~al.}(2026)\citenamefont
  {Holtzmann}, \citenamefont {Li}, \citenamefont {Anderson}, \citenamefont
  {Cai}, \citenamefont {Park}, \citenamefont {Hu}, \citenamefont {Taniguchi},
  \citenamefont {Watanabe}, \citenamefont {Chu}, \citenamefont {Xiao} \emph
  {et~al.}}]{holtzmann2026optical}%
  \BibitemOpen
  \bibinfo {author} {\bibfnamefont {W.}~\bibnamefont {Holtzmann}}, \bibinfo
  {author} {\bibfnamefont {W.}~\bibnamefont {Li}}, \bibinfo {author}
  {\bibfnamefont {E.}~\bibnamefont {Anderson}}, \bibinfo {author}
  {\bibfnamefont {J.}~\bibnamefont {Cai}}, \bibinfo {author} {\bibfnamefont
  {H.}~\bibnamefont {Park}}, \bibinfo {author} {\bibfnamefont {C.}~\bibnamefont
  {Hu}}, \bibinfo {author} {\bibfnamefont {T.}~\bibnamefont {Taniguchi}},
  \bibinfo {author} {\bibfnamefont {K.}~\bibnamefont {Watanabe}}, \bibinfo
  {author} {\bibfnamefont {J.-H.}\ \bibnamefont {Chu}}, \bibinfo {author}
  {\bibfnamefont {D.}~\bibnamefont {Xiao}}, \bibnamefont {et~al.}\EOS\
\newblock \emph {\bibinfo {title} {Optical control of integer and fractional
  chern insulators}}.\EOS\
\newblock \href {\doibase 10.1038/s41586-025-09777-3} {\bibfield  {journal}
  {\emph {\bibinfo  {journal} {Nature}}\ }\textbf {\bibinfo {volume} {649}},
  \bibinfo {pages} {1147--1152} (\bibinfo {year} {2026})}\BibitemShut {NoStop}%
\bibitem [{\citenamefont {Huber}\ \emph {et~al.}(2026)\citenamefont {Huber},
  \citenamefont {Kuhlbrodt}, \citenamefont {Anderson}, \citenamefont {Li},
  \citenamefont {Watanabe}, \citenamefont {Taniguchi}, \citenamefont {Kroner},
  \citenamefont {Xu}, \citenamefont {Imamo{\u{g}}lu},\ and\ \citenamefont
  {Smole{\'n}ski}}]{huber2026optical}%
  \BibitemOpen
  \bibinfo {author} {\bibfnamefont {O.}~\bibnamefont {Huber}}, \bibinfo
  {author} {\bibfnamefont {K.}~\bibnamefont {Kuhlbrodt}}, \bibinfo {author}
  {\bibfnamefont {E.}~\bibnamefont {Anderson}}, \bibinfo {author}
  {\bibfnamefont {W.}~\bibnamefont {Li}}, \bibinfo {author} {\bibfnamefont
  {K.}~\bibnamefont {Watanabe}}, \bibinfo {author} {\bibfnamefont
  {T.}~\bibnamefont {Taniguchi}}, \bibinfo {author} {\bibfnamefont
  {M.}~\bibnamefont {Kroner}}, \bibinfo {author} {\bibfnamefont
  {X.}~\bibnamefont {Xu}}, \bibinfo {author} {\bibfnamefont {A.}~\bibnamefont
  {Imamo{\u{g}}lu}},\ \bibnamefont {and}\ \bibinfo {author} {\bibfnamefont
  {T.}~\bibnamefont {Smole{\'n}ski}}.\EOS\
\newblock \emph {\bibinfo {title} {Optical control over topological chern
  number in moir{\'e} materials}}.\EOS\
\newblock \href {\doibase 10.1038/s41586-025-09851-w} {\bibfield  {journal}
  {\emph {\bibinfo  {journal} {Nature}}\ }\textbf {\bibinfo {volume} {649}},
  \bibinfo {pages} {1153--1158} (\bibinfo {year} {2026})}\BibitemShut {NoStop}%
\bibitem [{\citenamefont {Cai}\ \emph {et~al.}(2026)\citenamefont {Cai},
  \citenamefont {Pan}, \citenamefont {Wang}, \citenamefont {Rasmita},
  \citenamefont {Yang}, \citenamefont {Zhao}, \citenamefont {Wang},
  \citenamefont {Duan}, \citenamefont {He}, \citenamefont {Watanabe},
  \citenamefont {Taniguchi}, \citenamefont {Liu}, \citenamefont
  {{Z{\'u}{\~n}iga-P{\'e}rez}}, \citenamefont {Yang},\ and\ \citenamefont
  {Gao}}]{cai2026optical}%
  \BibitemOpen
  \bibinfo {author} {\bibfnamefont {X.}~\bibnamefont {Cai}}, \bibinfo {author}
  {\bibfnamefont {H.}~\bibnamefont {Pan}}, \bibinfo {author} {\bibfnamefont
  {Y.}~\bibnamefont {Wang}}, \bibinfo {author} {\bibfnamefont {A.}~\bibnamefont
  {Rasmita}}, \bibinfo {author} {\bibfnamefont {S.}~\bibnamefont {Yang}},
  \bibinfo {author} {\bibfnamefont {Y.}~\bibnamefont {Zhao}}, \bibinfo {author}
  {\bibfnamefont {W.}~\bibnamefont {Wang}}, \bibinfo {author} {\bibfnamefont
  {R.}~\bibnamefont {Duan}}, \bibinfo {author} {\bibfnamefont {R.}~\bibnamefont
  {He}}, \bibinfo {author} {\bibfnamefont {K.}~\bibnamefont {Watanabe}},
  \bibinfo {author} {\bibfnamefont {T.}~\bibnamefont {Taniguchi}}, \bibinfo
  {author} {\bibfnamefont {Z.}~\bibnamefont {Liu}}, \bibinfo {author}
  {\bibfnamefont {J.}~\bibnamefont {{Z{\'u}{\~n}iga-P{\'e}rez}}}, \bibinfo
  {author} {\bibfnamefont {B.}~\bibnamefont {Yang}},\ \bibnamefont {and}\
  \bibinfo {author} {\bibfnamefont {W.}~\bibnamefont {Gao}}.\EOS\
\newblock \emph {\bibinfo {title} {Optical switching of a moir\'e {{Chern}}
  ferromagnet}}.\EOS\
\newblock \href {\doibase 10.1038/s41586-025-10048-4} {\bibfield  {journal}
  {\emph {\bibinfo  {journal} {Nature}}\ }\textbf {\bibinfo {volume} {650}},
  \bibinfo {pages} {580--584} (\bibinfo {year} {2026})}\BibitemShut {NoStop}%
\bibitem [{\citenamefont {Fausti}\ \emph {et~al.}(2011)\citenamefont {Fausti},
  \citenamefont {Tobey}, \citenamefont {Dean}, \citenamefont {Kaiser},
  \citenamefont {Dienst}, \citenamefont {Hoffmann}, \citenamefont {Pyon},
  \citenamefont {Takayama}, \citenamefont {Takagi},\ and\ \citenamefont
  {Cavalleri}}]{fausti2011lightinduced}%
  \BibitemOpen
  \bibinfo {author} {\bibfnamefont {D.}~\bibnamefont {Fausti}}, \bibinfo
  {author} {\bibfnamefont {R.~I.}\ \bibnamefont {Tobey}}, \bibinfo {author}
  {\bibfnamefont {N.}~\bibnamefont {Dean}}, \bibinfo {author} {\bibfnamefont
  {S.}~\bibnamefont {Kaiser}}, \bibinfo {author} {\bibfnamefont
  {A.}~\bibnamefont {Dienst}}, \bibinfo {author} {\bibfnamefont {M.~C.}\
  \bibnamefont {Hoffmann}}, \bibinfo {author} {\bibfnamefont {S.}~\bibnamefont
  {Pyon}}, \bibinfo {author} {\bibfnamefont {T.}~\bibnamefont {Takayama}},
  \bibinfo {author} {\bibfnamefont {H.}~\bibnamefont {Takagi}},\ \bibnamefont
  {and}\ \bibinfo {author} {\bibfnamefont {A.}~\bibnamefont {Cavalleri}}.\EOS\
\newblock \emph {\bibinfo {title} {Light-induced superconductivity in a
  stripe-ordered cuprate}}.\EOS\
\newblock \href {\doibase 10.1126/science.1197294} {\bibfield  {journal} {\emph
  {\bibinfo  {journal} {Science}}\ }\textbf {\bibinfo {volume} {331}}, \bibinfo
  {pages} {189--191} (\bibinfo {year} {2011})}\BibitemShut {NoStop}%
\bibitem [{\citenamefont {Wang}\ \emph {et~al.}(2013)\citenamefont {Wang},
  \citenamefont {Steinberg}, \citenamefont {Jarillo-Herrero},\ and\
  \citenamefont {Gedik}}]{wang2013observation}%
  \BibitemOpen
  \bibinfo {author} {\bibfnamefont {Y.~H.}\ \bibnamefont {Wang}}, \bibinfo
  {author} {\bibfnamefont {H.}~\bibnamefont {Steinberg}}, \bibinfo {author}
  {\bibfnamefont {P.}~\bibnamefont {Jarillo-Herrero}},\ \bibnamefont {and}\
  \bibinfo {author} {\bibfnamefont {N.}~\bibnamefont {Gedik}}.\EOS\
\newblock \emph {\bibinfo {title} {{Observation of Floquet-Bloch States on the
  Surface of a Topological Insulator}}}.\EOS\
\newblock \href {\doibase 10.1126/science.1239834} {\bibfield  {journal} {\emph
  {\bibinfo  {journal} {Science}}\ }\textbf {\bibinfo {volume} {342}}, \bibinfo
  {pages} {453--457} (\bibinfo {year} {2013})}\BibitemShut {NoStop}%
\bibitem [{\citenamefont {Stojchevska}\ \emph {et~al.}(2014)\citenamefont
  {Stojchevska}, \citenamefont {Vaskivskyi}, \citenamefont {Mertelj},
  \citenamefont {Kusar}, \citenamefont {Svetin}, \citenamefont {Brazovskii},\
  and\ \citenamefont {Mihailovic}}]{stojchevska2014ultrafast}%
  \BibitemOpen
  \bibinfo {author} {\bibfnamefont {L.}~\bibnamefont {Stojchevska}}, \bibinfo
  {author} {\bibfnamefont {I.}~\bibnamefont {Vaskivskyi}}, \bibinfo {author}
  {\bibfnamefont {T.}~\bibnamefont {Mertelj}}, \bibinfo {author} {\bibfnamefont
  {P.}~\bibnamefont {Kusar}}, \bibinfo {author} {\bibfnamefont
  {D.}~\bibnamefont {Svetin}}, \bibinfo {author} {\bibfnamefont
  {S.}~\bibnamefont {Brazovskii}},\ \bibnamefont {and}\ \bibinfo {author}
  {\bibfnamefont {D.}~\bibnamefont {Mihailovic}}.\EOS\
\newblock \emph {\bibinfo {title} {{Ultrafast Switching to a Stable Hidden
  Quantum State in an Electronic Crystal}}}.\EOS\
\newblock \href {\doibase 10.1126/science.1241591} {\bibfield  {journal} {\emph
  {\bibinfo  {journal} {Science}}\ }\textbf {\bibinfo {volume} {344}}, \bibinfo
  {pages} {177--180} (\bibinfo {year} {2014})}\BibitemShut {NoStop}%
\bibitem [{\citenamefont {McIver}\ \emph {et~al.}(2020)\citenamefont {McIver},
  \citenamefont {Schulte}, \citenamefont {Stein}, \citenamefont {Matsuyama},
  \citenamefont {Jotzu}, \citenamefont {Meier},\ and\ \citenamefont
  {Cavalleri}}]{mciver2020Lightinduced}%
  \BibitemOpen
  \bibinfo {author} {\bibfnamefont {J.~W.}\ \bibnamefont {McIver}}, \bibinfo
  {author} {\bibfnamefont {B.}~\bibnamefont {Schulte}}, \bibinfo {author}
  {\bibfnamefont {F.-U.}\ \bibnamefont {Stein}}, \bibinfo {author}
  {\bibfnamefont {T.}~\bibnamefont {Matsuyama}}, \bibinfo {author}
  {\bibfnamefont {G.}~\bibnamefont {Jotzu}}, \bibinfo {author} {\bibfnamefont
  {G.}~\bibnamefont {Meier}},\ \bibnamefont {and}\ \bibinfo {author}
  {\bibfnamefont {A.}~\bibnamefont {Cavalleri}}.\EOS\
\newblock \emph {\bibinfo {title} {Light-induced anomalous {{Hall}} effect in
  graphene}}.\EOS\
\newblock \href {\doibase 10.1038/s41567-019-0698-y} {\bibfield  {journal}
  {\emph {\bibinfo  {journal} {Nat. Phys.}}\ }\textbf {\bibinfo {volume} {16}},
  \bibinfo {pages} {38--41} (\bibinfo {year} {2020})}\BibitemShut {NoStop}%
\bibitem [{\citenamefont {Kobayashi}\ \emph {et~al.}(2023)\citenamefont
  {Kobayashi}, \citenamefont {Heide}, \citenamefont {Johnson}, \citenamefont
  {Tiwari}, \citenamefont {Liu}, \citenamefont {Reis}, \citenamefont {Heinz},\
  and\ \citenamefont {Ghimire}}]{kobayashi2023floquet}%
  \BibitemOpen
  \bibinfo {author} {\bibfnamefont {Y.}~\bibnamefont {Kobayashi}}, \bibinfo
  {author} {\bibfnamefont {C.}~\bibnamefont {Heide}}, \bibinfo {author}
  {\bibfnamefont {A.~C.}\ \bibnamefont {Johnson}}, \bibinfo {author}
  {\bibfnamefont {V.}~\bibnamefont {Tiwari}}, \bibinfo {author} {\bibfnamefont
  {F.}~\bibnamefont {Liu}}, \bibinfo {author} {\bibfnamefont {D.~A.}\
  \bibnamefont {Reis}}, \bibinfo {author} {\bibfnamefont {T.~F.}\ \bibnamefont
  {Heinz}},\ \bibnamefont {and}\ \bibinfo {author} {\bibfnamefont
  {S.}~\bibnamefont {Ghimire}}.\EOS\
\newblock \emph {\bibinfo {title} {Floquet engineering of strongly driven
  excitons in monolayer tungsten disulfide}}.\EOS\
\newblock \href {\doibase https://doi.org/10.1038/s41567-022-01849-9}
  {\bibfield  {journal} {\emph {\bibinfo  {journal} {Nat. Phys.}}\ }\textbf
  {\bibinfo {volume} {19}}, \bibinfo {pages} {171--176} (\bibinfo {year}
  {2023})}\BibitemShut {NoStop}%
\bibitem [{\citenamefont {Mitra}\ \emph {et~al.}(2024)\citenamefont {Mitra},
  \citenamefont {{Jim{\'e}nez-Gal{\'a}n}}, \citenamefont {Aulich},
  \citenamefont {Neuhaus}, \citenamefont {Silva}, \citenamefont {Pervak},
  \citenamefont {Kling},\ and\ \citenamefont {Biswas}}]{mitra2024light}%
  \BibitemOpen
  \bibinfo {author} {\bibfnamefont {S.}~\bibnamefont {Mitra}}, \bibinfo
  {author} {\bibfnamefont {{\'A}.}~\bibnamefont {{Jim{\'e}nez-Gal{\'a}n}}},
  \bibinfo {author} {\bibfnamefont {M.}~\bibnamefont {Aulich}}, \bibinfo
  {author} {\bibfnamefont {M.}~\bibnamefont {Neuhaus}}, \bibinfo {author}
  {\bibfnamefont {R.~E.~F.}\ \bibnamefont {Silva}}, \bibinfo {author}
  {\bibfnamefont {V.}~\bibnamefont {Pervak}}, \bibinfo {author} {\bibfnamefont
  {M.~F.}\ \bibnamefont {Kling}},\ \bibnamefont {and}\ \bibinfo {author}
  {\bibfnamefont {S.}~\bibnamefont {Biswas}}.\EOS\
\newblock \emph {\bibinfo {title} {Light-wave-controlled {{Haldane}} model in
  monolayer hexagonal boron nitride}}.\EOS\
\newblock \href {\doibase 10.1038/s41586-024-07244-z} {\bibfield  {journal}
  {\emph {\bibinfo  {journal} {Nature}}\ }\textbf {\bibinfo {volume} {628}},
  \bibinfo {pages} {752--757} (\bibinfo {year} {2024})}\BibitemShut {NoStop}%
\bibitem [{\citenamefont {de~la Torre}\ \emph {et~al.}(2021)\citenamefont
  {de~la Torre}, \citenamefont {Kennes}, \citenamefont {Claassen},
  \citenamefont {Gerber}, \citenamefont {McIver},\ and\ \citenamefont
  {Sentef}}]{delatorre2021nonthermal}%
  \BibitemOpen
  \bibinfo {author} {\bibfnamefont {A.}~\bibnamefont {de~la Torre}}, \bibinfo
  {author} {\bibfnamefont {D.~M.}\ \bibnamefont {Kennes}}, \bibinfo {author}
  {\bibfnamefont {M.}~\bibnamefont {Claassen}}, \bibinfo {author}
  {\bibfnamefont {S.}~\bibnamefont {Gerber}}, \bibinfo {author} {\bibfnamefont
  {J.~W.}\ \bibnamefont {McIver}},\ \bibnamefont {and}\ \bibinfo {author}
  {\bibfnamefont {M.~A.}\ \bibnamefont {Sentef}}.\EOS\
\newblock \emph {\bibinfo {title} {{Colloquium: Nonthermal pathways to
  ultrafast control in quantum materials}}}.\EOS\
\newblock \href {\doibase 10.1103/RevModPhys.93.041002} {\bibfield  {journal}
  {\emph {\bibinfo  {journal} {Rev. Mod. Phys.}}\ }\textbf {\bibinfo {volume}
  {93}}, \bibinfo {pages} {041002} (\bibinfo {year} {2021})}\BibitemShut
  {NoStop}%
\bibitem [{\citenamefont {Bao}\ \emph {et~al.}(2022)\citenamefont {Bao},
  \citenamefont {Tang}, \citenamefont {Sun},\ and\ \citenamefont
  {Zhou}}]{bao2022Lightinduced}%
  \BibitemOpen
  \bibinfo {author} {\bibfnamefont {C.}~\bibnamefont {Bao}}, \bibinfo {author}
  {\bibfnamefont {P.}~\bibnamefont {Tang}}, \bibinfo {author} {\bibfnamefont
  {D.}~\bibnamefont {Sun}},\ \bibnamefont {and}\ \bibinfo {author}
  {\bibfnamefont {S.}~\bibnamefont {Zhou}}.\EOS\
\newblock \emph {\bibinfo {title} {Light-induced emergent phenomena in {{2D}}
  materials and topological materials}}.\EOS\
\newblock \href {\doibase 10.1038/s42254-021-00388-1} {\bibfield  {journal}
  {\emph {\bibinfo  {journal} {Nat. Rev. Phys.}}\ }\textbf {\bibinfo {volume}
  {4}}, \bibinfo {pages} {33--48} (\bibinfo {year} {2022})}\BibitemShut
  {NoStop}%
\bibitem [{\citenamefont {Tsui}\ \emph {et~al.}(1982)\citenamefont {Tsui},
  \citenamefont {Stormer},\ and\ \citenamefont {Gossard}}]{tsui1982}%
  \BibitemOpen
  \bibinfo {author} {\bibfnamefont {D.~C.}\ \bibnamefont {Tsui}}, \bibinfo
  {author} {\bibfnamefont {H.~L.}\ \bibnamefont {Stormer}},\ \bibnamefont
  {and}\ \bibinfo {author} {\bibfnamefont {A.~C.}\ \bibnamefont
  {Gossard}}.\EOS\
\newblock \emph {\bibinfo {title} {{Two-Dimensional Magnetotransport in the
  Extreme Quantum Limit}}}.\EOS\
\newblock \href {\doibase 10.1103/PhysRevLett.48.1559} {\bibfield  {journal}
  {\emph {\bibinfo  {journal} {Phys. Rev. Lett.}}\ }\textbf {\bibinfo {volume}
  {48}}, \bibinfo {pages} {1559--1562} (\bibinfo {year} {1982})}\BibitemShut
  {NoStop}%
\bibitem [{\citenamefont {Laughlin}(1983)}]{laughlin1983}%
  \BibitemOpen
  \bibinfo {author} {\bibfnamefont {R.~B.}\ \bibnamefont {Laughlin}}.\EOS\
\newblock \emph {\bibinfo {title} {{Anomalous quantum Hall effect: an
  incompressible quantum fluid with fractionally charged excitations}}}.\EOS\
\newblock \href {\doibase 10.1103/PhysRevLett.50.1395} {\bibfield  {journal}
  {\emph {\bibinfo  {journal} {Phys. Rev. Lett.}}\ }\textbf {\bibinfo {volume}
  {50}}, \bibinfo {pages} {1395} (\bibinfo {year} {1983})}\BibitemShut
  {NoStop}%
\bibitem [{\citenamefont {Hafezi}\ \emph {et~al.}(2007)\citenamefont {Hafezi},
  \citenamefont {S\o{}rensen}, \citenamefont {Demler},\ and\ \citenamefont
  {Lukin}}]{Hafezi_fractional_2007}%
  \BibitemOpen
  \bibinfo {author} {\bibfnamefont {M.}~\bibnamefont {Hafezi}}, \bibinfo
  {author} {\bibfnamefont {A.~S.}\ \bibnamefont {S\o{}rensen}}, \bibinfo
  {author} {\bibfnamefont {E.}~\bibnamefont {Demler}},\ \bibnamefont {and}\
  \bibinfo {author} {\bibfnamefont {M.~D.}\ \bibnamefont {Lukin}}.\EOS\
\newblock \emph {\bibinfo {title} {Fractional quantum hall effect in optical
  lattices}}.\EOS\
\newblock \href {\doibase 10.1103/PhysRevA.76.023613} {\bibfield  {journal}
  {\emph {\bibinfo  {journal} {Phys. Rev. A}}\ }\textbf {\bibinfo {volume}
  {76}}, \bibinfo {pages} {023613} (\bibinfo {year} {2007})}\BibitemShut
  {NoStop}%
\bibitem [{\citenamefont {Kapit}\ and\ \citenamefont
  {Mueller}(2010)}]{kapit_exact_2010}%
  \BibitemOpen
  \bibinfo {author} {\bibfnamefont {E.}~\bibnamefont {Kapit}}\ \bibnamefont
  {and}\ \bibinfo {author} {\bibfnamefont {E.}~\bibnamefont {Mueller}}.\EOS\
\newblock \emph {\bibinfo {title} {Exact {Parent} {Hamiltonian} for the
  {Quantum} {Hall} {States} in a {Lattice}}}.\EOS\
\newblock \href {\doibase 10.1103/PhysRevLett.105.215303} {\bibfield  {journal}
  {\emph {\bibinfo  {journal} {Phys. Rev. Lett.}}\ }\textbf {\bibinfo {volume}
  {105}}, \bibinfo {pages} {215303} (\bibinfo {year} {2010})}\BibitemShut
  {NoStop}%
\bibitem [{\citenamefont {Sheng}\ \emph {et~al.}(2011)\citenamefont {Sheng},
  \citenamefont {Gu}, \citenamefont {Sun},\ and\ \citenamefont
  {Sheng}}]{sheng2011}%
  \BibitemOpen
  \bibinfo {author} {\bibfnamefont {D.}~\bibnamefont {Sheng}}, \bibinfo
  {author} {\bibfnamefont {Z.-C.}\ \bibnamefont {Gu}}, \bibinfo {author}
  {\bibfnamefont {K.}~\bibnamefont {Sun}},\ \bibnamefont {and}\ \bibinfo
  {author} {\bibfnamefont {L.}~\bibnamefont {Sheng}}.\EOS\
\newblock \emph {\bibinfo {title} {{Fractional quantum Hall effect in the
  absence of Landau levels}}}.\EOS\
\newblock \href {\doibase 10.1038/ncomms1380} {\bibfield  {journal} {\emph
  {\bibinfo  {journal} {Nat. Comm.}}\ }\textbf {\bibinfo {volume} {2}},
  \bibinfo {pages} {389} (\bibinfo {year} {2011})}\BibitemShut {NoStop}%
\bibitem [{\citenamefont {Neupert}\ \emph {et~al.}(2011)\citenamefont
  {Neupert}, \citenamefont {Santos}, \citenamefont {Chamon},\ and\
  \citenamefont {Mudry}}]{neupert2011}%
  \BibitemOpen
  \bibinfo {author} {\bibfnamefont {T.}~\bibnamefont {Neupert}}, \bibinfo
  {author} {\bibfnamefont {L.}~\bibnamefont {Santos}}, \bibinfo {author}
  {\bibfnamefont {C.}~\bibnamefont {Chamon}},\ \bibnamefont {and}\ \bibinfo
  {author} {\bibfnamefont {C.}~\bibnamefont {Mudry}}.\EOS\
\newblock \emph {\bibinfo {title} {{Fractional Quantum Hall States at Zero
  Magnetic Field}}}.\EOS\
\newblock \href {\doibase 10.1103/PhysRevLett.106.236804} {\bibfield  {journal}
  {\emph {\bibinfo  {journal} {Phys. Rev. Lett.}}\ }\textbf {\bibinfo {volume}
  {106}}, \bibinfo {pages} {236804} (\bibinfo {year} {2011})}\BibitemShut
  {NoStop}%
\bibitem [{\citenamefont {Tang}\ \emph {et~al.}(2011)\citenamefont {Tang},
  \citenamefont {Mei},\ and\ \citenamefont {Wen}}]{tang2011}%
  \BibitemOpen
  \bibinfo {author} {\bibfnamefont {E.}~\bibnamefont {Tang}}, \bibinfo {author}
  {\bibfnamefont {J.-W.}\ \bibnamefont {Mei}},\ \bibnamefont {and}\ \bibinfo
  {author} {\bibfnamefont {X.-G.}\ \bibnamefont {Wen}}.\EOS\
\newblock \emph {\bibinfo {title} {{High-Temperature Fractional Quantum Hall
  States}}}.\EOS\
\newblock \href {\doibase 10.1103/PhysRevLett.106.236802} {\bibfield  {journal}
  {\emph {\bibinfo  {journal} {Phys. Rev. Lett.}}\ }\textbf {\bibinfo {volume}
  {106}}, \bibinfo {pages} {236802} (\bibinfo {year} {2011})}\BibitemShut
  {NoStop}%
\bibitem [{\citenamefont {Regnault}\ and\ \citenamefont
  {Bernevig}(2011)}]{regnault_fractional_2011}%
  \BibitemOpen
  \bibinfo {author} {\bibfnamefont {N.}~\bibnamefont {Regnault}}\ \bibnamefont
  {and}\ \bibinfo {author} {\bibfnamefont {B.~A.}\ \bibnamefont
  {Bernevig}}.\EOS\
\newblock \emph {\bibinfo {title} {Fractional {Chern} {Insulator}}}.\EOS\
\newblock \href {\doibase 10.1103/PhysRevX.1.021014} {\bibfield  {journal}
  {\emph {\bibinfo  {journal} {Phys. Rev. X}}\ }\textbf {\bibinfo {volume}
  {1}}, \bibinfo {pages} {021014} (\bibinfo {year} {2011})}\BibitemShut
  {NoStop}%
\bibitem [{\citenamefont {Lu}\ \emph {et~al.}(2024)\citenamefont {Lu},
  \citenamefont {Han}, \citenamefont {Yao}, \citenamefont {Reddy},
  \citenamefont {Yang}, \citenamefont {Seo}, \citenamefont {Watanabe},
  \citenamefont {Taniguchi}, \citenamefont {Fu},\ and\ \citenamefont
  {Ju}}]{lu2024}%
  \BibitemOpen
  \bibinfo {author} {\bibfnamefont {Z.}~\bibnamefont {Lu}}, \bibinfo {author}
  {\bibfnamefont {T.}~\bibnamefont {Han}}, \bibinfo {author} {\bibfnamefont
  {Y.}~\bibnamefont {Yao}}, \bibinfo {author} {\bibfnamefont {A.~P.}\
  \bibnamefont {Reddy}}, \bibinfo {author} {\bibfnamefont {J.}~\bibnamefont
  {Yang}}, \bibinfo {author} {\bibfnamefont {J.}~\bibnamefont {Seo}}, \bibinfo
  {author} {\bibfnamefont {K.}~\bibnamefont {Watanabe}}, \bibinfo {author}
  {\bibfnamefont {T.}~\bibnamefont {Taniguchi}}, \bibinfo {author}
  {\bibfnamefont {L.}~\bibnamefont {Fu}},\ \bibnamefont {and}\ \bibinfo
  {author} {\bibfnamefont {L.}~\bibnamefont {Ju}}.\EOS\
\newblock \emph {\bibinfo {title} {{Fractional quantum anomalous Hall effect in
  multilayer graphene}}}.\EOS\
\newblock \href {\doibase 10.1038/s41586-023-07010-7} {\bibfield  {journal}
  {\emph {\bibinfo  {journal} {Nature}}\ }\textbf {\bibinfo {volume} {626}},
  \bibinfo {pages} {759--764} (\bibinfo {year} {2024})}\BibitemShut {NoStop}%
\bibitem [{\citenamefont {Wang}\ \emph {et~al.}(2025)\citenamefont {Wang},
  \citenamefont {Choe}, \citenamefont {Anderson}, \citenamefont {Li},
  \citenamefont {Ingham}, \citenamefont {Arsenault}, \citenamefont {Li},
  \citenamefont {Hu}, \citenamefont {Taniguchi}, \citenamefont {Watanabe} \emph
  {et~al.}}]{wang2025hidden}%
  \BibitemOpen
  \bibinfo {author} {\bibfnamefont {Y.}~\bibnamefont {Wang}}, \bibinfo {author}
  {\bibfnamefont {J.}~\bibnamefont {Choe}}, \bibinfo {author} {\bibfnamefont
  {E.}~\bibnamefont {Anderson}}, \bibinfo {author} {\bibfnamefont
  {W.}~\bibnamefont {Li}}, \bibinfo {author} {\bibfnamefont {J.}~\bibnamefont
  {Ingham}}, \bibinfo {author} {\bibfnamefont {E.~A.}\ \bibnamefont
  {Arsenault}}, \bibinfo {author} {\bibfnamefont {Y.}~\bibnamefont {Li}},
  \bibinfo {author} {\bibfnamefont {X.}~\bibnamefont {Hu}}, \bibinfo {author}
  {\bibfnamefont {T.}~\bibnamefont {Taniguchi}}, \bibinfo {author}
  {\bibfnamefont {K.}~\bibnamefont {Watanabe}}, \bibnamefont {et~al.}\EOS\
\newblock \emph {\bibinfo {title} {Hidden states and dynamics of fractional
  fillings in twisted {MoTe}$_2$ bilayers}}.\EOS\
\newblock \href {\doibase https://doi.org/10.1038/s41586-025-08954-8}
  {\bibfield  {journal} {\emph {\bibinfo  {journal} {Nature}}\ }\textbf
  {\bibinfo {volume} {641}}, \bibinfo {pages} {1149--1155} (\bibinfo {year}
  {2025})}\BibitemShut {NoStop}%
\bibitem [{\citenamefont {Wu}\ \emph {et~al.}(2019)\citenamefont {Wu},
  \citenamefont {Lovorn}, \citenamefont {Tutuc}, \citenamefont {Martin},\ and\
  \citenamefont {MacDonald}}]{Wu_TopologicalInsulators_2019}%
  \BibitemOpen
  \bibinfo {author} {\bibfnamefont {F.}~\bibnamefont {Wu}}, \bibinfo {author}
  {\bibfnamefont {T.}~\bibnamefont {Lovorn}}, \bibinfo {author} {\bibfnamefont
  {E.}~\bibnamefont {Tutuc}}, \bibinfo {author} {\bibfnamefont
  {I.}~\bibnamefont {Martin}},\ \bibnamefont {and}\ \bibinfo {author}
  {\bibfnamefont {A.~H.}\ \bibnamefont {MacDonald}}.\EOS\
\newblock \emph {\bibinfo {title} {Topological insulators in twisted transition
  metal dichalcogenide homobilayers}}.\EOS\
\newblock \href {\doibase 10.1103/PhysRevLett.122.086402} {\bibfield  {journal}
  {\emph {\bibinfo  {journal} {Phys. Rev. Lett.}}\ }\textbf {\bibinfo {volume}
  {122}}, \bibinfo {pages} {086402} (\bibinfo {year} {2019})}\BibitemShut
  {NoStop}%
\bibitem [{\citenamefont {Devakul}\ \emph {et~al.}(2021)\citenamefont
  {Devakul}, \citenamefont {Crepel}, \citenamefont {Zhang},\ and\ \citenamefont
  {Fu}}]{Devakul2021}%
  \BibitemOpen
  \bibinfo {author} {\bibfnamefont {T.}~\bibnamefont {Devakul}}, \bibinfo
  {author} {\bibfnamefont {V.}~\bibnamefont {Crepel}}, \bibinfo {author}
  {\bibfnamefont {Y.}~\bibnamefont {Zhang}},\ \bibnamefont {and}\ \bibinfo
  {author} {\bibfnamefont {L.}~\bibnamefont {Fu}}.\EOS\
\newblock \emph {\bibinfo {title} {{Magic in twisted transition metal
  dichalcogenide bilayers}}}.\EOS\
\newblock \href {\doibase 10.1038/s41467-021-27042-9} {\bibfield  {journal}
  {\emph {\bibinfo  {journal} {Nat. Commun.}}\ }\textbf {\bibinfo {volume}
  {12}}, \bibinfo {pages} {6730} (\bibinfo {year} {2021})}\BibitemShut
  {NoStop}%
\bibitem [{\citenamefont {Reddy}\ \emph {et~al.}(2023)\citenamefont {Reddy},
  \citenamefont {Alsallom}, \citenamefont {Zhang}, \citenamefont {Devakul},\
  and\ \citenamefont {Fu}}]{Reddy2023}%
  \BibitemOpen
  \bibinfo {author} {\bibfnamefont {A.~P.}\ \bibnamefont {Reddy}}, \bibinfo
  {author} {\bibfnamefont {F.}~\bibnamefont {Alsallom}}, \bibinfo {author}
  {\bibfnamefont {Y.}~\bibnamefont {Zhang}}, \bibinfo {author} {\bibfnamefont
  {T.}~\bibnamefont {Devakul}},\ \bibnamefont {and}\ \bibinfo {author}
  {\bibfnamefont {L.}~\bibnamefont {Fu}}.\EOS\
\newblock \emph {\bibinfo {title} {{Fractional quantum anomalous Hall states in
  twisted bilayer ${\mathrm{MoTe}}_{2}$ and ${\mathrm{WSe}}_{2}$}}}.\EOS\
\newblock \href {\doibase 10.1103/PhysRevB.108.085117} {\bibfield  {journal}
  {\emph {\bibinfo  {journal} {Phys. Rev. B}}\ }\textbf {\bibinfo {volume}
  {108}}, \bibinfo {pages} {085117} (\bibinfo {year} {2023})}\BibitemShut
  {NoStop}%
\bibitem [{\citenamefont {Wang}\ \emph {et~al.}(2024)\citenamefont {Wang},
  \citenamefont {Zhang}, \citenamefont {Liu}, \citenamefont {He}, \citenamefont
  {Xu}, \citenamefont {Ran}, \citenamefont {Cao},\ and\ \citenamefont
  {Xiao}}]{Wang2024}%
  \BibitemOpen
  \bibinfo {author} {\bibfnamefont {C.}~\bibnamefont {Wang}}, \bibinfo {author}
  {\bibfnamefont {X.-W.}\ \bibnamefont {Zhang}}, \bibinfo {author}
  {\bibfnamefont {X.}~\bibnamefont {Liu}}, \bibinfo {author} {\bibfnamefont
  {Y.}~\bibnamefont {He}}, \bibinfo {author} {\bibfnamefont {X.}~\bibnamefont
  {Xu}}, \bibinfo {author} {\bibfnamefont {Y.}~\bibnamefont {Ran}}, \bibinfo
  {author} {\bibfnamefont {T.}~\bibnamefont {Cao}},\ \bibnamefont {and}\
  \bibinfo {author} {\bibfnamefont {D.}~\bibnamefont {Xiao}}.\EOS\
\newblock \emph {\bibinfo {title} {{Fractional Chern Insulator in Twisted
  Bilayer ${\mathrm{MoTe}}_{2}$}}}.\EOS\
\newblock \href {\doibase 10.1103/PhysRevLett.132.036501} {\bibfield  {journal}
  {\emph {\bibinfo  {journal} {Phys. Rev. Lett.}}\ }\textbf {\bibinfo {volume}
  {132}}, \bibinfo {pages} {036501} (\bibinfo {year} {2024})}\BibitemShut
  {NoStop}%
\bibitem [{\citenamefont {Yu}\ \emph {et~al.}(2024)\citenamefont {Yu},
  \citenamefont {Herzog-Arbeitman}, \citenamefont {Wang}, \citenamefont
  {Vafek}, \citenamefont {Bernevig},\ and\ \citenamefont {Regnault}}]{Yu2024}%
  \BibitemOpen
  \bibinfo {author} {\bibfnamefont {J.}~\bibnamefont {Yu}}, \bibinfo {author}
  {\bibfnamefont {J.}~\bibnamefont {Herzog-Arbeitman}}, \bibinfo {author}
  {\bibfnamefont {M.}~\bibnamefont {Wang}}, \bibinfo {author} {\bibfnamefont
  {O.}~\bibnamefont {Vafek}}, \bibinfo {author} {\bibfnamefont {B.~A.}\
  \bibnamefont {Bernevig}},\ \bibnamefont {and}\ \bibinfo {author}
  {\bibfnamefont {N.}~\bibnamefont {Regnault}}.\EOS\
\newblock \emph {\bibinfo {title} {{Fractional Chern insulators versus
  nonmagnetic states in twisted bilayer ${\mathrm{MoTe}}_{2}$}}}.\EOS\
\newblock \href {\doibase 10.1103/PhysRevB.109.045147} {\bibfield  {journal}
  {\emph {\bibinfo  {journal} {Phys. Rev. B}}\ }\textbf {\bibinfo {volume}
  {109}}, \bibinfo {pages} {045147} (\bibinfo {year} {2024})}\BibitemShut
  {NoStop}%
\bibitem [{\citenamefont {Coleman}(1977)}]{Coleman_FalseVacuum_1977}%
  \BibitemOpen
  \bibinfo {author} {\bibfnamefont {S.}~\bibnamefont {Coleman}}.\EOS\
\newblock \emph {\bibinfo {title} {Fate of the false vacuum: Semiclassical
  theory}}.\EOS\
\newblock \href {\doibase 10.1103/PhysRevD.15.2929} {\bibfield  {journal}
  {\emph {\bibinfo  {journal} {Phys. Rev. D}}\ }\textbf {\bibinfo {volume}
  {15}}, \bibinfo {pages} {2929--2936} (\bibinfo {year} {1977})}\BibitemShut
  {NoStop}%
\bibitem [{\citenamefont {Callan}\ and\ \citenamefont
  {Coleman}(1977)}]{Callan_FalseVacuum_1977}%
  \BibitemOpen
  \bibinfo {author} {\bibfnamefont {C.~G.}\ \bibnamefont {Callan}}\
  \bibnamefont {and}\ \bibinfo {author} {\bibfnamefont {S.}~\bibnamefont
  {Coleman}}.\EOS\
\newblock \emph {\bibinfo {title} {{Fate of the false vacuum. II. First quantum
  corrections}}}.\EOS\
\newblock \href {\doibase 10.1103/PhysRevD.16.1762} {\bibfield  {journal}
  {\emph {\bibinfo  {journal} {Phys. Rev. D}}\ }\textbf {\bibinfo {volume}
  {16}}, \bibinfo {pages} {1762--1768} (\bibinfo {year} {1977})}\BibitemShut
  {NoStop}%
\bibitem [{\citenamefont {Sidler}\ \emph {et~al.}(2017)\citenamefont {Sidler},
  \citenamefont {Back}, \citenamefont {Cotlet}, \citenamefont {Srivastava},
  \citenamefont {Fink}, \citenamefont {Kroner}, \citenamefont {Demler},\ and\
  \citenamefont {Imamoglu}}]{sidler2017fermi}%
  \BibitemOpen
  \bibinfo {author} {\bibfnamefont {M.}~\bibnamefont {Sidler}}, \bibinfo
  {author} {\bibfnamefont {P.}~\bibnamefont {Back}}, \bibinfo {author}
  {\bibfnamefont {O.}~\bibnamefont {Cotlet}}, \bibinfo {author} {\bibfnamefont
  {A.}~\bibnamefont {Srivastava}}, \bibinfo {author} {\bibfnamefont
  {T.}~\bibnamefont {Fink}}, \bibinfo {author} {\bibfnamefont {M.}~\bibnamefont
  {Kroner}}, \bibinfo {author} {\bibfnamefont {E.}~\bibnamefont {Demler}},\
  \bibnamefont {and}\ \bibinfo {author} {\bibfnamefont {A.}~\bibnamefont
  {Imamoglu}}.\EOS\
\newblock \emph {\bibinfo {title} {Fermi polaron-polaritons in charge-tunable
  atomically thin semiconductors}}.\EOS\
\newblock \href {\doibase https://doi.org/10.1038/nphys3949} {\bibfield
  {journal} {\emph {\bibinfo  {journal} {Nat. Phys.}}\ }\textbf {\bibinfo
  {volume} {13}}, \bibinfo {pages} {255--261} (\bibinfo {year}
  {2017})}\BibitemShut {NoStop}%
\bibitem [{\citenamefont {Smole\ifmmode~\acute{n}\else \'{n}\fi{}ski}\ \emph
  {et~al.}(2022)\citenamefont {Smole\ifmmode~\acute{n}\else \'{n}\fi{}ski},
  \citenamefont {Watanabe}, \citenamefont {Taniguchi}, \citenamefont {Kroner},\
  and\ \citenamefont {Imamo\ifmmode~\breve{g}\else
  \u{g}\fi{}lu}}]{Smolenski2022}%
  \BibitemOpen
  \bibinfo {author} {\bibfnamefont {T.}~\bibnamefont
  {Smole\ifmmode~\acute{n}\else \'{n}\fi{}ski}}, \bibinfo {author}
  {\bibfnamefont {K.}~\bibnamefont {Watanabe}}, \bibinfo {author}
  {\bibfnamefont {T.}~\bibnamefont {Taniguchi}}, \bibinfo {author}
  {\bibfnamefont {M.}~\bibnamefont {Kroner}},\ \bibnamefont {and}\ \bibinfo
  {author} {\bibfnamefont {A.}~\bibnamefont {Imamo\ifmmode~\breve{g}\else
  \u{g}\fi{}lu}}.\EOS\
\newblock \emph {\bibinfo {title} {{Spin-Valley Relaxation and Exciton-Induced
  Depolarization Dynamics of Landau-Quantized Electrons in MoSe$_2$
  Monolayer}}}.\EOS\
\newblock \href {\doibase 10.1103/PhysRevLett.128.127402} {\bibfield  {journal}
  {\emph {\bibinfo  {journal} {Phys. Rev. Lett.}}\ }\textbf {\bibinfo {volume}
  {128}}, \bibinfo {pages} {127402} (\bibinfo {year} {2022})}\BibitemShut
  {NoStop}%
\bibitem [{\citenamefont {Ciorciaro}\ \emph {et~al.}(2023)\citenamefont
  {Ciorciaro}, \citenamefont {Smole\'nski}, \citenamefont {Morera},
  \citenamefont {Kiper}, \citenamefont {Hiestand}, \citenamefont {Kroner},
  \citenamefont {Zhang}, \citenamefont {Watanabe}, \citenamefont {Taniguchi},
  \citenamefont {Demler},\ and\ \citenamefont {Imamoglu}}]{ciorciaro2023}%
  \BibitemOpen
  \bibinfo {author} {\bibfnamefont {L.}~\bibnamefont {Ciorciaro}}, \bibinfo
  {author} {\bibfnamefont {T.}~\bibnamefont {Smole\'nski}}, \bibinfo {author}
  {\bibfnamefont {I.}~\bibnamefont {Morera}}, \bibinfo {author} {\bibfnamefont
  {N.}~\bibnamefont {Kiper}}, \bibinfo {author} {\bibfnamefont
  {S.}~\bibnamefont {Hiestand}}, \bibinfo {author} {\bibfnamefont
  {M.}~\bibnamefont {Kroner}}, \bibinfo {author} {\bibfnamefont
  {Y.}~\bibnamefont {Zhang}}, \bibinfo {author} {\bibfnamefont
  {K.}~\bibnamefont {Watanabe}}, \bibinfo {author} {\bibfnamefont
  {T.}~\bibnamefont {Taniguchi}}, \bibinfo {author} {\bibfnamefont
  {E.}~\bibnamefont {Demler}},\ \bibnamefont {and}\ \bibinfo {author}
  {\bibfnamefont {A.}~\bibnamefont {Imamoglu}}.\EOS\
\newblock \emph {\bibinfo {title} {{Kinetic magnetism in triangular moir\'e
  materials}}}.\EOS\
\newblock \href {\doibase 10.1038/s41586-023-06633-0} {\bibfield  {journal}
  {\emph {\bibinfo  {journal} {Nature}}\ }\textbf {\bibinfo {volume} {623}},
  \bibinfo {pages} {509--513} (\bibinfo {year} {2023})}\BibitemShut {NoStop}%
\bibitem [{\citenamefont {Dey}\ \emph {et~al.}(2017)\citenamefont {Dey},
  \citenamefont {Yang}, \citenamefont {Robert}, \citenamefont {Wang},
  \citenamefont {Urbaszek}, \citenamefont {Marie},\ and\ \citenamefont
  {Crooker}}]{dey2017gate}%
  \BibitemOpen
  \bibinfo {author} {\bibfnamefont {P.}~\bibnamefont {Dey}}, \bibinfo {author}
  {\bibfnamefont {L.}~\bibnamefont {Yang}}, \bibinfo {author} {\bibfnamefont
  {C.}~\bibnamefont {Robert}}, \bibinfo {author} {\bibfnamefont
  {G.}~\bibnamefont {Wang}}, \bibinfo {author} {\bibfnamefont {B.}~\bibnamefont
  {Urbaszek}}, \bibinfo {author} {\bibfnamefont {X.}~\bibnamefont {Marie}},\
  \bibnamefont {and}\ \bibinfo {author} {\bibfnamefont {S.~A.}\ \bibnamefont
  {Crooker}}.\EOS\
\newblock \emph {\bibinfo {title} {{Gate-Controlled Spin-Valley Locking of
  Resident Carriers in WSe$_2$ Monolayers}}}.\EOS\
\newblock \href {\doibase 10.1103/PhysRevLett.119.137401} {\bibfield  {journal}
  {\emph {\bibinfo  {journal} {Phys. Rev. Lett.}}\ }\textbf {\bibinfo {volume}
  {119}}, \bibinfo {pages} {137401} (\bibinfo {year} {2017})}\BibitemShut
  {NoStop}%
\bibitem [{\citenamefont {Goryca}\ \emph {et~al.}(2019)\citenamefont {Goryca},
  \citenamefont {Wilson}, \citenamefont {Dey}, \citenamefont {Xu},\ and\
  \citenamefont {Crooker}}]{goryca2019_relaxation}%
  \BibitemOpen
  \bibinfo {author} {\bibfnamefont {M.}~\bibnamefont {Goryca}}, \bibinfo
  {author} {\bibfnamefont {N.~P.}\ \bibnamefont {Wilson}}, \bibinfo {author}
  {\bibfnamefont {P.}~\bibnamefont {Dey}}, \bibinfo {author} {\bibfnamefont
  {X.}~\bibnamefont {Xu}},\ \bibnamefont {and}\ \bibinfo {author}
  {\bibfnamefont {S.~A.}\ \bibnamefont {Crooker}}.\EOS\
\newblock \emph {\bibinfo {title} {{Detection of thermodynamic ``valley noise''
  in monolayer semiconductors: Access to intrinsic valley relaxation time
  scales}}}.\EOS\
\newblock \href {\doibase 10.1126/sciadv.aau4899} {\bibfield  {journal} {\emph
  {\bibinfo  {journal} {Sci. Adv.}}\ }\textbf {\bibinfo {volume} {5}}, \bibinfo
  {pages} {eaau4899} (\bibinfo {year} {2019})}\BibitemShut {NoStop}%
\bibitem [{\citenamefont {Li}\ \emph {et~al.}(2021)\citenamefont {Li},
  \citenamefont {Goryca}, \citenamefont {Yumigeta}, \citenamefont {Li},
  \citenamefont {Tongay},\ and\ \citenamefont {Crooker}}]{Li2021_relaxation}%
  \BibitemOpen
  \bibinfo {author} {\bibfnamefont {J.}~\bibnamefont {Li}}, \bibinfo {author}
  {\bibfnamefont {M.}~\bibnamefont {Goryca}}, \bibinfo {author} {\bibfnamefont
  {K.}~\bibnamefont {Yumigeta}}, \bibinfo {author} {\bibfnamefont
  {H.}~\bibnamefont {Li}}, \bibinfo {author} {\bibfnamefont {S.}~\bibnamefont
  {Tongay}},\ \bibnamefont {and}\ \bibinfo {author} {\bibfnamefont {S.~A.}\
  \bibnamefont {Crooker}}.\EOS\
\newblock \emph {\bibinfo {title} {{Valley relaxation of resident electrons and
  holes in a monolayer semiconductor: Dependence on carrier density and the
  role of substrate-induced disorder}}}.\EOS\
\newblock \href {\doibase 10.1103/PhysRevMaterials.5.044001} {\bibfield
  {journal} {\emph {\bibinfo  {journal} {Phys. Rev. Mater.}}\ }\textbf
  {\bibinfo {volume} {5}}, \bibinfo {pages} {044001} (\bibinfo {year}
  {2021})}\BibitemShut {NoStop}%
\bibitem [{\citenamefont {Glazov}\ \emph {et~al.}(2019)\citenamefont {Glazov},
  \citenamefont {Semina}, \citenamefont {Robert}, \citenamefont {Urbaszek},
  \citenamefont {Amand},\ and\ \citenamefont {Marie}}]{glazov2019intervalley}%
  \BibitemOpen
  \bibinfo {author} {\bibfnamefont {M.~M.}\ \bibnamefont {Glazov}}, \bibinfo
  {author} {\bibfnamefont {M.~A.}\ \bibnamefont {Semina}}, \bibinfo {author}
  {\bibfnamefont {C.}~\bibnamefont {Robert}}, \bibinfo {author} {\bibfnamefont
  {B.}~\bibnamefont {Urbaszek}}, \bibinfo {author} {\bibfnamefont
  {T.}~\bibnamefont {Amand}},\ \bibnamefont {and}\ \bibinfo {author}
  {\bibfnamefont {X.}~\bibnamefont {Marie}}.\EOS\
\newblock \emph {\bibinfo {title} {Intervalley polaron in atomically thin
  transition metal dichalcogenides}}.\EOS\
\newblock \href {\doibase 10.1103/PhysRevB.100.041301} {\bibfield  {journal}
  {\emph {\bibinfo  {journal} {Phys. Rev. B}}\ }\textbf {\bibinfo {volume}
  {100}}, \bibinfo {pages} {041301(R)} (\bibinfo {year} {2019})}\BibitemShut
  {NoStop}%
\bibitem [{\citenamefont {Hohenberg}\ and\ \citenamefont
  {Halperin}(1977)}]{HohenbergHalperin77}%
  \BibitemOpen
  \bibinfo {author} {\bibfnamefont {P.~C.}\ \bibnamefont {Hohenberg}}\
  \bibnamefont {and}\ \bibinfo {author} {\bibfnamefont {B.~I.}\ \bibnamefont
  {Halperin}}.\EOS\
\newblock \emph {\bibinfo {title} {Theory of dynamic critical phenomena}}.\EOS\
\newblock \href {\doibase 10.1103/RevModPhys.49.435} {\bibfield  {journal}
  {\emph {\bibinfo  {journal} {Rev. Mod. Phys.}}\ }\textbf {\bibinfo {volume}
  {49}}, \bibinfo {pages} {435--479} (\bibinfo {year} {1977})}\BibitemShut
  {NoStop}%
\bibitem [{\citenamefont {Bray}(1994)}]{bray1994theory}%
  \BibitemOpen
  \bibinfo {author} {\bibfnamefont {A.~J.}\ \bibnamefont {Bray}}.\EOS\
\newblock \emph {\bibinfo {title} {Theory of phase-ordering kinetics}}.\EOS\
\newblock \href {\doibase 10.1080/00018739400101505} {\bibfield  {journal}
  {\emph {\bibinfo  {journal} {Adv. in Phys.}}\ }\textbf {\bibinfo {volume}
  {43}}, \bibinfo {pages} {357--459} (\bibinfo {year} {1994})}\BibitemShut
  {NoStop}%
\bibitem [{\citenamefont {Park}\ \emph {et~al.}(2026)\citenamefont {Park},
  \citenamefont {Voinea}, \citenamefont {Tsui}, \citenamefont {Pu},
  \citenamefont {Watanabe}, \citenamefont {Taniguchi}, \citenamefont {Cooper},
  \citenamefont {Zaletel}, \citenamefont {Papi?},\ and\ \citenamefont
  {Yazdani}}]{park2026localspectroscopy}%
  \BibitemOpen
  \bibinfo {author} {\bibfnamefont {J.~M.}\ \bibnamefont {Park}}, \bibinfo
  {author} {\bibfnamefont {C.}~\bibnamefont {Voinea}}, \bibinfo {author}
  {\bibfnamefont {Y.-C.}\ \bibnamefont {Tsui}}, \bibinfo {author}
  {\bibfnamefont {S.}~\bibnamefont {Pu}}, \bibinfo {author} {\bibfnamefont
  {K.}~\bibnamefont {Watanabe}}, \bibinfo {author} {\bibfnamefont
  {T.}~\bibnamefont {Taniguchi}}, \bibinfo {author} {\bibfnamefont {N.~R.}\
  \bibnamefont {Cooper}}, \bibinfo {author} {\bibfnamefont {M.~P.}\
  \bibnamefont {Zaletel}}, \bibinfo {author} {\bibfnamefont {Z.}~\bibnamefont
  {Papi?}},\ \bibnamefont {and}\ \bibinfo {author} {\bibfnamefont
  {A.}~\bibnamefont {Yazdani}}.\EOS\
\newblock \emph {\bibinfo {title} {{Local spectroscopy of anyons bound to
  charge traps}}}.\EOS\
\newblock \href {https://arxiv.org/abs/2606.25024} {\  (\bibinfo {year}
  {2026})},\ \Eprint {http://arxiv.org/abs/2606.25024} {arXiv:2606.25024
  [cond-mat.mes-hall]} \BibitemShut {NoStop}%
\bibitem [{\citenamefont {Deng}\ \emph {et~al.}(2026)\citenamefont {Deng},
  \citenamefont {Sun}, \citenamefont {Pimenov}, \citenamefont {Taniguchi},
  \citenamefont {Watanabe}, \citenamefont {Mueller},\ and\ \citenamefont
  {Liu}}]{deng2026realspace}%
  \BibitemOpen
  \bibinfo {author} {\bibfnamefont {J.}~\bibnamefont {Deng}}, \bibinfo {author}
  {\bibfnamefont {Y.}~\bibnamefont {Sun}}, \bibinfo {author} {\bibfnamefont
  {D.}~\bibnamefont {Pimenov}}, \bibinfo {author} {\bibfnamefont
  {T.}~\bibnamefont {Taniguchi}}, \bibinfo {author} {\bibfnamefont
  {K.}~\bibnamefont {Watanabe}}, \bibinfo {author} {\bibfnamefont {E.~J.}\
  \bibnamefont {Mueller}},\ \bibnamefont {and}\ \bibinfo {author}
  {\bibfnamefont {X.}~\bibnamefont {Liu}}.\EOS\
\newblock \emph {\bibinfo {title} {{Real-space Imaging of Quantum Hall
  Quasiparticles}}}.\EOS\
\newblock \href {https://arxiv.org/abs/2606.25036} {\  (\bibinfo {year}
  {2026})},\ \Eprint {http://arxiv.org/abs/2606.25036} {arXiv:2606.25036
  [cond-mat.mes-hall]} \BibitemShut {NoStop}%
\bibitem [{\citenamefont {Li}\ \emph {et~al.}(2026)\citenamefont {Li},
  \citenamefont {Wang~Beach}, \citenamefont {Hu}, \citenamefont {Taniguchi},
  \citenamefont {Watanabe}, \citenamefont {Chu}, \citenamefont {Imamo{\u g}lu},
  \citenamefont {Cao}, \citenamefont {Xiao},\ and\ \citenamefont
  {Xu}}]{li2026}%
  \BibitemOpen
  \bibinfo {author} {\bibfnamefont {W.}~\bibnamefont {Li}}, \bibinfo {author}
  {\bibfnamefont {C.}~\bibnamefont {Wang~Beach}}, \bibinfo {author}
  {\bibfnamefont {C.}~\bibnamefont {Hu}}, \bibinfo {author} {\bibfnamefont
  {T.}~\bibnamefont {Taniguchi}}, \bibinfo {author} {\bibfnamefont
  {K.}~\bibnamefont {Watanabe}}, \bibinfo {author} {\bibfnamefont {J.-H.}\
  \bibnamefont {Chu}}, \bibinfo {author} {\bibfnamefont {A.}~\bibnamefont
  {Imamo{\u g}lu}}, \bibinfo {author} {\bibfnamefont {T.}~\bibnamefont {Cao}},
  \bibinfo {author} {\bibfnamefont {D.}~\bibnamefont {Xiao}},\ \bibnamefont
  {and}\ \bibinfo {author} {\bibfnamefont {X.}~\bibnamefont {Xu}}.\EOS\
\newblock \emph {\bibinfo {title} {{Signatures of fractional charges via
  anyon--trions in twisted MoTe$_2$}}}.\EOS\
\newblock \href {\doibase 10.1038/s41586-026-10101-w} {\bibfield  {journal}
  {\emph {\bibinfo  {journal} {Nature}}\ }\textbf {\bibinfo {volume} {651}},
  \bibinfo {pages} {48--53} (\bibinfo {year} {2026})}\BibitemShut {NoStop}%
\bibitem [{\citenamefont {Cao}\ \emph {et~al.}(2018{\natexlab{a}})\citenamefont
  {Cao}, \citenamefont {Fatemi}, \citenamefont {Demir}, \citenamefont {Fang},
  \citenamefont {Tomarken}, \citenamefont {Luo}, \citenamefont
  {Sanchez-Yamagishi}, \citenamefont {Watanabe}, \citenamefont {Taniguchi},
  \citenamefont {Kaxiras} \emph {et~al.}}]{cao2018correlated}%
  \BibitemOpen
  \bibinfo {author} {\bibfnamefont {Y.}~\bibnamefont {Cao}}, \bibinfo {author}
  {\bibfnamefont {V.}~\bibnamefont {Fatemi}}, \bibinfo {author} {\bibfnamefont
  {A.}~\bibnamefont {Demir}}, \bibinfo {author} {\bibfnamefont
  {S.}~\bibnamefont {Fang}}, \bibinfo {author} {\bibfnamefont {S.~L.}\
  \bibnamefont {Tomarken}}, \bibinfo {author} {\bibfnamefont {J.~Y.}\
  \bibnamefont {Luo}}, \bibinfo {author} {\bibfnamefont {J.~D.}\ \bibnamefont
  {Sanchez-Yamagishi}}, \bibinfo {author} {\bibfnamefont {K.}~\bibnamefont
  {Watanabe}}, \bibinfo {author} {\bibfnamefont {T.}~\bibnamefont {Taniguchi}},
  \bibinfo {author} {\bibfnamefont {E.}~\bibnamefont {Kaxiras}}, \bibnamefont
  {et~al.}\EOS\
\newblock \emph {\bibinfo {title} {Correlated insulator behaviour at
  half-filling in magic-angle graphene superlattices}}.\EOS\
\newblock \href {\doibase https://doi.org/10.1038/nature26154} {\bibfield
  {journal} {\emph {\bibinfo  {journal} {Nature}}\ }\textbf {\bibinfo {volume}
  {556}}, \bibinfo {pages} {80--84} (\bibinfo {year}
  {2018}{\natexlab{a}})}\BibitemShut {NoStop}%
\bibitem [{\citenamefont {Cao}\ \emph {et~al.}(2018{\natexlab{b}})\citenamefont
  {Cao}, \citenamefont {Fatemi}, \citenamefont {Fang}, \citenamefont
  {Watanabe}, \citenamefont {Taniguchi}, \citenamefont {Kaxiras},\ and\
  \citenamefont {Jarillo-Herrero}}]{cao2018unconventional}%
  \BibitemOpen
  \bibinfo {author} {\bibfnamefont {Y.}~\bibnamefont {Cao}}, \bibinfo {author}
  {\bibfnamefont {V.}~\bibnamefont {Fatemi}}, \bibinfo {author} {\bibfnamefont
  {S.}~\bibnamefont {Fang}}, \bibinfo {author} {\bibfnamefont {K.}~\bibnamefont
  {Watanabe}}, \bibinfo {author} {\bibfnamefont {T.}~\bibnamefont {Taniguchi}},
  \bibinfo {author} {\bibfnamefont {E.}~\bibnamefont {Kaxiras}},\ \bibnamefont
  {and}\ \bibinfo {author} {\bibfnamefont {P.}~\bibnamefont
  {Jarillo-Herrero}}.\EOS\
\newblock \emph {\bibinfo {title} {Unconventional superconductivity in
  magic-angle graphene superlattices}}.\EOS\
\newblock \href {\doibase https://doi.org/10.1038/nature26160} {\bibfield
  {journal} {\emph {\bibinfo  {journal} {Nature}}\ }\textbf {\bibinfo {volume}
  {556}}, \bibinfo {pages} {43--50} (\bibinfo {year}
  {2018}{\natexlab{b}})}\BibitemShut {NoStop}%
\bibitem [{\citenamefont {Zomer}\ \emph {et~al.}(2014)\citenamefont {Zomer},
  \citenamefont {Guimarães}, \citenamefont {Brant}, \citenamefont {Tombros},\
  and\ \citenamefont {van Wees}}]{Zomer2014}%
  \BibitemOpen
  \bibinfo {author} {\bibfnamefont {P.~J.}\ \bibnamefont {Zomer}}, \bibinfo
  {author} {\bibfnamefont {M.~H.~D.}\ \bibnamefont {Guimaraes}}, \bibinfo
  {author} {\bibfnamefont {J.~C.}\ \bibnamefont {Brant}}, \bibinfo {author}
  {\bibfnamefont {N.}~\bibnamefont {Tombros}},\ \bibnamefont {and}\ \bibinfo
  {author} {\bibfnamefont {B.~J.}\ \bibnamefont {van Wees}}.\EOS\
\newblock \emph {\bibinfo {title} {{Fast pick up technique for high quality
  heterostructures of bilayer graphene and hexagonal boron nitride}}}.\EOS\
\newblock \href {\doibase 10.1063/1.4886096} {\bibfield  {journal} {\emph
  {\bibinfo  {journal} {Appl. Phys. Lett.}}\ }\textbf {\bibinfo {volume}
  {105}}, \bibinfo {pages} {013101} (\bibinfo {year} {2014})}\BibitemShut
  {NoStop}%
\bibitem [{\citenamefont {Xu}\ \emph {et~al.}(2021)\citenamefont {Xu},
  \citenamefont {Horn}, \citenamefont {Zhu}, \citenamefont {Tang},
  \citenamefont {Ma}, \citenamefont {Li}, \citenamefont {Liu}, \citenamefont
  {Watanabe}, \citenamefont {Taniguchi}, \citenamefont {Hone} \emph
  {et~al.}}]{xu2021creation}%
  \BibitemOpen
  \bibinfo {author} {\bibfnamefont {Y.}~\bibnamefont {Xu}}, \bibinfo {author}
  {\bibfnamefont {C.}~\bibnamefont {Horn}}, \bibinfo {author} {\bibfnamefont
  {J.}~\bibnamefont {Zhu}}, \bibinfo {author} {\bibfnamefont {Y.}~\bibnamefont
  {Tang}}, \bibinfo {author} {\bibfnamefont {L.}~\bibnamefont {Ma}}, \bibinfo
  {author} {\bibfnamefont {L.}~\bibnamefont {Li}}, \bibinfo {author}
  {\bibfnamefont {S.}~\bibnamefont {Liu}}, \bibinfo {author} {\bibfnamefont
  {K.}~\bibnamefont {Watanabe}}, \bibinfo {author} {\bibfnamefont
  {T.}~\bibnamefont {Taniguchi}}, \bibinfo {author} {\bibfnamefont {J.~C.}\
  \bibnamefont {Hone}}, \bibnamefont {et~al.}\EOS\
\newblock \emph {\bibinfo {title} {Creation of moir{\'e} bands in a monolayer
  semiconductor by spatially periodic dielectric screening}}.\EOS\
\newblock \href {\doibase https://doi.org/10.1038/s41563-020-00888-y}
  {\bibfield  {journal} {\emph {\bibinfo  {journal} {Nat. Mat.}}\ }\textbf
  {\bibinfo {volume} {20}}, \bibinfo {pages} {645--649} (\bibinfo {year}
  {2021})}\BibitemShut {NoStop}%
\bibitem [{\citenamefont {Popert}\ \emph {et~al.}(2022)\citenamefont {Popert},
  \citenamefont {Shimazaki}, \citenamefont {Kroner}, \citenamefont {Watanabe},
  \citenamefont {Taniguchi}, \citenamefont {Imamo{\u g}lu},\ and\ \citenamefont
  {Smole{\'n}ski}}]{Popert2022}%
  \BibitemOpen
  \bibinfo {author} {\bibfnamefont {A.}~\bibnamefont {Popert}}, \bibinfo
  {author} {\bibfnamefont {Y.}~\bibnamefont {Shimazaki}}, \bibinfo {author}
  {\bibfnamefont {M.}~\bibnamefont {Kroner}}, \bibinfo {author} {\bibfnamefont
  {K.}~\bibnamefont {Watanabe}}, \bibinfo {author} {\bibfnamefont
  {T.}~\bibnamefont {Taniguchi}}, \bibinfo {author} {\bibfnamefont
  {A.}~\bibnamefont {Imamo{\u g}lu}},\ \bibnamefont {and}\ \bibinfo {author}
  {\bibfnamefont {T.}~\bibnamefont {Smole{\'n}ski}}.\EOS\
\newblock \emph {\bibinfo {title} {{Optical Sensing of Fractional Quantum Hall
  Effect in Graphene}}}.\EOS\
\newblock \href {\doibase 10.1021/acs.nanolett.2c02000} {\bibfield  {journal}
  {\emph {\bibinfo  {journal} {Nano Lett.}}\ }\textbf {\bibinfo {volume} {22}},
  \bibinfo {pages} {7363--7369} (\bibinfo {year} {2022})}\BibitemShut {NoStop}%
\bibitem [{\citenamefont {Smole{\'n}ski}\ \emph {et~al.}(2019)\citenamefont
  {Smole{\'n}ski}, \citenamefont {Cotlet}, \citenamefont {Popert},
  \citenamefont {Back}, \citenamefont {Shimazaki}, \citenamefont {Kn{\"u}ppel},
  \citenamefont {Dietler}, \citenamefont {Taniguchi}, \citenamefont {Watanabe},
  \citenamefont {Kroner} \emph {et~al.}}]{smolenski2019interaction}%
  \BibitemOpen
  \bibinfo {author} {\bibfnamefont {T.}~\bibnamefont {Smole{\'n}ski}}, \bibinfo
  {author} {\bibfnamefont {O.}~\bibnamefont {Cotlet}}, \bibinfo {author}
  {\bibfnamefont {A.}~\bibnamefont {Popert}}, \bibinfo {author} {\bibfnamefont
  {P.}~\bibnamefont {Back}}, \bibinfo {author} {\bibfnamefont {Y.}~\bibnamefont
  {Shimazaki}}, \bibinfo {author} {\bibfnamefont {P.}~\bibnamefont
  {Kn{\"u}ppel}}, \bibinfo {author} {\bibfnamefont {N.}~\bibnamefont
  {Dietler}}, \bibinfo {author} {\bibfnamefont {T.}~\bibnamefont {Taniguchi}},
  \bibinfo {author} {\bibfnamefont {K.}~\bibnamefont {Watanabe}}, \bibinfo
  {author} {\bibfnamefont {M.}~\bibnamefont {Kroner}}, \bibnamefont
  {et~al.}\EOS\
\newblock \emph {\bibinfo {title} {Interaction-induced shubnikov--de haas
  oscillations in optical conductivity of monolayer $\text{MoSe}_2$}}.\EOS\
\newblock \href {\doibase https://doi.org/10.1103/PhysRevLett.123.097403}
  {\bibfield  {journal} {\emph {\bibinfo  {journal} {Phys. Rev. Lett.}}\
  }\textbf {\bibinfo {volume} {123}}, \bibinfo {pages} {097403} (\bibinfo
  {year} {2019})}\BibitemShut {NoStop}%
\bibitem [{\citenamefont {St\v{r}eda}(1982)}]{streda1982theory}%
  \BibitemOpen
  \bibinfo {author} {\bibfnamefont {P.}~\bibnamefont {St\v{r}eda}}.\EOS\
\newblock \emph {\bibinfo {title} {Theory of quantised hall conductivity in two
  dimensions}}.\EOS\
\newblock \href {\doibase 10.1088/0022-3719/15/22/005} {\bibfield  {journal}
  {\emph {\bibinfo  {journal} {J. Phys. C: Solid State Phys.}}\ }\textbf
  {\bibinfo {volume} {15}}, \bibinfo {pages} {L717--L721} (\bibinfo {year}
  {1982})}\BibitemShut {NoStop}%
\bibitem [{\citenamefont {Bruno}\ \emph {et~al.}(1990)\citenamefont {Bruno},
  \citenamefont {Bayreuther}, \citenamefont {Beauvillain}, \citenamefont
  {Chappert}, \citenamefont {Lugert}, \citenamefont {Renard}, \citenamefont
  {Renard},\ and\ \citenamefont {Seiden}}]{bruno1990}%
  \BibitemOpen
  \bibinfo {author} {\bibfnamefont {P.}~\bibnamefont {Bruno}}, \bibinfo
  {author} {\bibfnamefont {G.}~\bibnamefont {Bayreuther}}, \bibinfo {author}
  {\bibfnamefont {P.}~\bibnamefont {Beauvillain}}, \bibinfo {author}
  {\bibfnamefont {C.}~\bibnamefont {Chappert}}, \bibinfo {author}
  {\bibfnamefont {G.}~\bibnamefont {Lugert}}, \bibinfo {author} {\bibfnamefont
  {D.}~\bibnamefont {Renard}}, \bibinfo {author} {\bibfnamefont {J.~P.}\
  \bibnamefont {Renard}},\ \bibnamefont {and}\ \bibinfo {author} {\bibfnamefont
  {J.}~\bibnamefont {Seiden}}.\EOS\
\newblock \emph {\bibinfo {title} {{Hysteresis properties of ultrathin
  ferromagnetic films}}}.\EOS\
\newblock \href {\doibase 10.1063/1.346944} {\bibfield  {journal} {\emph
  {\bibinfo  {journal} {J. Appl. Phys.}}\ }\textbf {\bibinfo {volume} {68}},
  \bibinfo {pages} {5759--5766} (\bibinfo {year} {1990})}\BibitemShut {NoStop}%
\bibitem [{\citenamefont {Qiu}\ and\ \citenamefont {Wu}(2025)}]{Qiu2025}%
  \BibitemOpen
  \bibinfo {author} {\bibfnamefont {W.-X.}\ \bibnamefont {Qiu}}\ \bibnamefont
  {and}\ \bibinfo {author} {\bibfnamefont {F.}~\bibnamefont {Wu}}.\EOS\
\newblock \emph {\bibinfo {title} {{Topological magnons and domain walls in
  twisted bilayer ${\mathrm{MoTe}}_{2}$}}}.\EOS\
\newblock \href {\doibase 10.1103/sl5k-c825} {\bibfield  {journal} {\emph
  {\bibinfo  {journal} {Phys. Rev. B}}\ }\textbf {\bibinfo {volume} {112}},
  \bibinfo {pages} {085132} (\bibinfo {year} {2025})}\BibitemShut {NoStop}%
\bibitem [{\citenamefont {Chen}\ \emph {et~al.}(2026)\citenamefont {Chen},
  \citenamefont {Li}, \citenamefont {Wang},\ and\ \citenamefont
  {Li}}]{chen2025fractionalcherninsulatorquantum}%
  \BibitemOpen
  \bibinfo {author} {\bibfnamefont {J.}~\bibnamefont {Chen}}, \bibinfo {author}
  {\bibfnamefont {Q.}~\bibnamefont {Li}}, \bibinfo {author} {\bibfnamefont
  {X.}~\bibnamefont {Wang}},\ \bibnamefont {and}\ \bibinfo {author}
  {\bibfnamefont {W.}~\bibnamefont {Li}}.\EOS\
\newblock \emph {\bibinfo {title} {{Fractional Chern insulator and quantum
  anomalous Hall crystal in twisted MoTe$_2$}}}.\EOS\
\newblock \href {\doibase https://doi.org/10.1016/j.scib.2026.01.014}
  {\bibfield  {journal} {\emph {\bibinfo  {journal} {Science Bulletin}}\
  }\textbf {\bibinfo {volume} {71}}, \bibinfo {pages} {1034--1042} (\bibinfo
  {year} {2026})}\BibitemShut {NoStop}%
\bibitem [{\citenamefont {Wang}\ \emph {et~al.}(2026)\citenamefont {Wang},
  \citenamefont {Liu},\ and\ \citenamefont
  {Moore}}]{wang2022structuredomainwallschiral}%
  \BibitemOpen
  \bibinfo {author} {\bibfnamefont {Y.-Q.}\ \bibnamefont {Wang}}, \bibinfo
  {author} {\bibfnamefont {C.}~\bibnamefont {Liu}},\ \bibnamefont {and}\
  \bibinfo {author} {\bibfnamefont {J.~E.}\ \bibnamefont {Moore}}.\EOS\
\newblock \emph {\bibinfo {title} {{Structure of domain walls in chiral spin
  liquids}}}.\EOS\
\newblock \href {\doibase 10.1073/pnas.2601093123} {\bibfield  {journal} {\emph
  {\bibinfo  {journal} {PNAS}}\ }\textbf {\bibinfo {volume} {123}}, \bibinfo
  {pages} {e2601093123} (\bibinfo {year} {2026})}\BibitemShut {NoStop}%
\end{thebibliography}

\newpage

\end{document}